\documentclass[12pt,preprint]{aastex}

\shorttitle{Stars In Galactic Center }
\shortauthors{Zhu et al.}

\begin{document}


\title{Radial Velocities of Stars in the Galactic Center \altaffilmark{1}}


\author{QINGFENG ZHU\altaffilmark{2}, ROLF P. KUDRITZKI\altaffilmark{3}, DONALD F. FIGER\altaffilmark{2}, FRANCISCO NAJARRO\altaffilmark{4} and DAVID MERRITT\altaffilmark{5}}

\altaffiltext{1}{Data presented here were obtained at the W. M. Keck
Observatory, which is operated as a scientific partnership among the
California Institute of Technology, the University of California,
and the National Aeronautics and Space Administration. The
Observatory was made available by the generous financial support of
the W. M. Keck Foundation.}

\altaffiltext{2}{Chester F. Carlson Center for Imaging Science,
Rochester Institute of Technology, 54 Lomb Memorial Drive,
Rochester, NY 14623-5604; zhuqf@cis.rit.edu; figer@cis.rit.edu}

\altaffiltext{3}{Institute for Astronomy, University of Hawaii, 2680
Woodlawn Drive, Honolulu, HI 96822; kud@ifa.hawaii.edu}

\altaffiltext{4}{Insituto de Estructura de la Materia, Consejo
Superior de Investigaciones Cientificas, Calle Serrano 121, 28006,
Madrid, Spain; najarro@damir.iem.csic.es}

\altaffiltext{5}{Department of Physics and Center for Computational
Relativity and Gravitation, 85 Lomb Memorial Drive, Rochester, NY
14623-5604; merritt@astro.rit.edu}



\begin{abstract}
We present results from K band slit scan observations of a
$\sim$20$''\times$20$''$ region of the Galactic center (GC) in two
separate epochs more than five years apart. The high resolution
(R~=~$\frac{\lambda}{\Delta\lambda}{\geq}14,000$) observations allow
the most accurate radial velocity and acceleration measurements of
the stars in the central parsec of the Galaxy. Detected stars can be
divided into three groups based on the CO absorption band heads at
$\sim$2.2935~$\mu$m and the He~I lines at $\sim$2.0581~$\mu$m and
$\sim$2.112, 2.113~$\mu$m: cool, narrow-line hot and broad-line hot.
The radial velocities of the cool, late-type stars have
approximately a symmetrical distribution with its center at
$\sim$-7.8$\pm$10.3~km~s$^{-1}$ and a standard deviation
$\sim$113.7$\pm$10.3~km~s$^{-1}$. Although our statistics are
dominated by the brightest stars, we estimate a central black hole
mass of 3.9$\pm$1.1$\times$10$^6$M$_{\odot}$, consistent with
current estimates from complete orbits of individual stars. Our
surface density profile and the velocity dispersion of the late type
stars support the existence of a low density region at the Galactic
center suggested by earlier observations. Many hot, early-type stars
show radial velocity changes higher than maximum values allowed by
pure circular orbital motions around a central massive object,
suggesting that the motions of these stars greatly deviate from
circular orbital motions around the Galactic center. The correlation
between the radial velocities of the early type He~I stars and their
declination offsets from Sagittarius~A$^*$ suggests that a
systematic rotation is present for the early-type population. No
figure rotation around the Galactic center for the late type stars
is supported by the new observations.
\end{abstract}





\section{Introduction}

The central half parsec in the Galactic center (GC) contains
$\sim$4$\times$10$^6$~M$_{\odot}$ of mass which dominates the
gravitational potential within several parsecs from the GC. This
fact raises many questions concerning the composition and origin of
the central mass and its effects on stellar objects in its vicinity.
Evidence strongly supports that a super massive black hole (SMBH) is
present at the GC \citep{gheMBTK00, schOGHLE02}. One of the most
important tasks is to determine the mass of the central object, or
equivalently, to measure the gravitational potential of the central
object. The gravitational potential can be probed by measuring bulk
motions of the ionized gas in the region \citep{lacTH82,serL85} or
by tracing a single star's 3D motion around the central mass
\citep{schOGHLE02,gheSHTL05}. These methods give similar estimates
of the central black mass, with $M\simeq3\times10^6~M_{\odot}$,
although a Sun-GC distance of 10~kpc was used in some of the earlier
works \citep{lacTH82,serL85}, which could result in an overestimate
of the enclosed mass. Both \cite{schOGHLE02} and \cite{gheSHTL05}
used a value of 8~kpc, which is also the value we assumed in the
current work.

We can also estimate the central mass by using statistical mass
estimators, which are based on the phase space distributions of
particles in dynamical systems. These mass estimators provide a
cross-check for other methods. We should note that not every star is
suitable for the statistical method. Many mass estimators ask for a
certain degree of velocity anisotropy for the dynamical system under
study. One such widely-used mass estimator, derived by \cite{bahT81}
(BT mass estimator), assumes that the velocities of stars in the
system are isotropic, which means that the stars have lived long
enough so that their motions are relaxed through close encounters.
Two-body relaxation time is related to the mean stellar mass, the
velocity dispersion and the stellar density \citep{binT87} and has a
value $\sim$6$\times$10$^9$~yrs at the distance of $\sim$0.7~pc from
a black hole with a mass of 3.7$\times$10$^6$M$_\odot$ for a
spherical velocity ellipsoid \citep{merS06}. Certainly, not all
stars in the GC have lived such a long time. Studies show that the
central parsec of the GC is populated by several stellar components,
including low mass red giants, hot He I emission line stars, some
luminous M-supergiant and asymptotic giant branch (AGB) stars, and
embedded near-IR featureless stars
\citep{allHH90,kraGDR91,tamRHCMR96,genTKKT96, bluSD96b,najKGLKH97,
genPEGO00,genSRE03}. Among these stars, only low mass red giant
branch stars have lifetimes long enough so that the two-body
relaxation can be effective \citep{davSRDC97}. Previous studies
suggest that the velocities of late type stars in the inner parsec
of the GC are close to being isotropic \citep{genTKKT96,genPEGO00}.
Thus it is possible to use mass estimators derived for isotropic
velocity distributions to estimate the central black mass, but it is
also important to try other estimators, like ones based on the
Virial Theorem, which are valid for any degree of anisotropy.

The GC is generally believed to be hostile to spontaneous star
formation through fragmentation and contraction of molecular clouds.
Extreme physical conditions, such as strong gravitational tidal
forces of the central object, large turbulent velocities inside
molecular clouds and strong magnetic fields can prevent star
formation in the region \citep{mor93}. However, discoveries of young
populations such as M supergiants, AGB stars, hot emission line
stars and main sequence stars \citep{genSRE03,gheSHTL05} pose a
great challenge to the star formation theory in the GC. Several
scenarios have been proposed to explain the observed young stellar
populations in the GC. These scenarios include the in situ star
formation following the cloud collapse triggered by compressing
shocks due to cloud-cloud collisions \citep{mor93} and the insertion
of young stars into the inner parsec of the Galaxy through mass
segregation after the stars formed elsewhere \citep{ger01,porMG03}.
Additionally, observations suggest two types of hot emission line
stars are present in the inner parsec of the Galaxy
\citep{pauMMR01,tanFNKGM06}. One type of stars show narrow
absorption doublet in the 2.112/2.113~$\mu$m He~I lines and narrow
emission in the 2.058~$\mu$m He~I line. The other type shows broad
emission in these He~I lines. These two types may represent two
different evolutionary stages of massive stars
\citep{pauMMR01,pauMS03}.

We carried out a set of high resolution spectroscopic survey
observations of the stars in the GC in order to acquire accurate
radial velocity measurements of the stars of both late (cool) and
early (hot) types in two different epochs separated by $\sim$5.8
years. The new observations allow us to estimate the mass of the
central black hole using the stellar velocity distribution and study
the velocity changes of these stars between the two epochs and other
properties of the stars in the region.

In \S~{\bf 2}, we introduce our observational method and data
reduction procedure. We present the results of the data analysis in
\S~{\bf 3}. We discuss the implications of the results in section
\S~{\bf 4} and summarize our findings in \S~{\bf 5}.

\section{Observations and Data Reduction}
\subsection{Observations}

\begin{figure}
\epsscale{1.0} \plotone{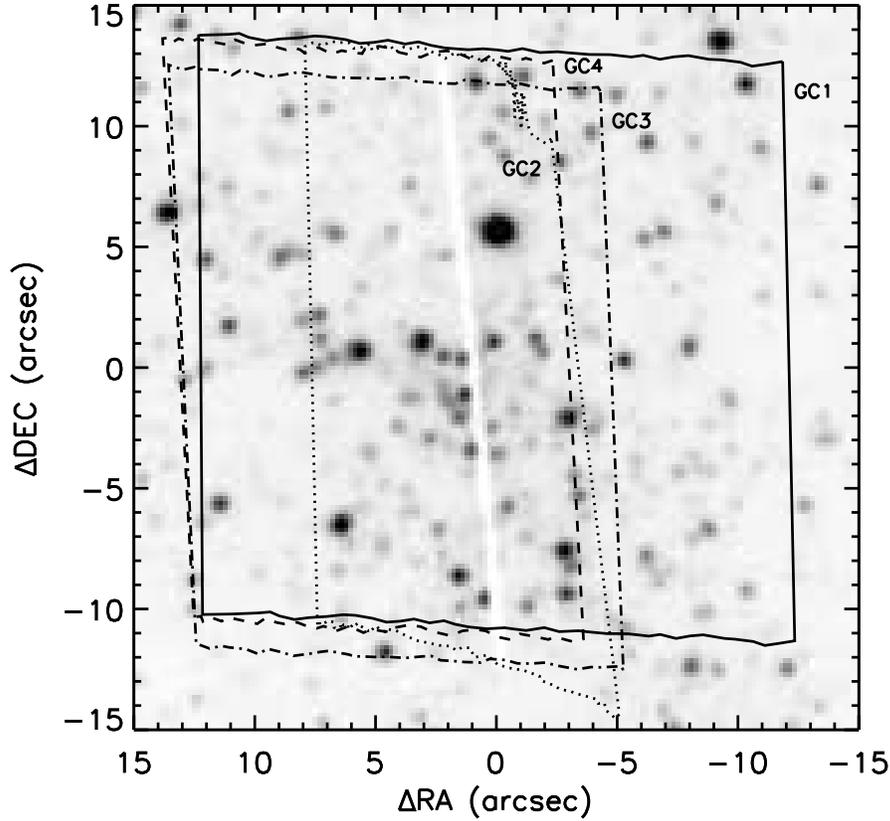} \caption{The slit-viewing camera
image with boxes indicating sky areas covered by observations.
Different types of lines are used to plot the boxes to avoid
confusion. Dates of the observations are indicated with names of the
slit scans (GC1: June 4th 1999, GC2: July 4th 1999, GC3: April 13th
2005, GC4: April 14th 2005). The nearly vertical white stripe at the
middle of the image is the shadow of the slit which is cut through
the mirror. The boundaries of boxes are computed based on slit
position parameters. \label{skybox}}
\end{figure}

Four slit scan observations were conducted over a
$\sim$20$''\times$20$''$ region in the GC with the near-infrared
spectrometer (NIRSPEC) on the Keck II telescope during June 4th
1999, July 4th 1999, April 13th 2005 and April 14th 2005. We refer
to these observations as ``GC1'', ``GC2'', ``GC3'' and ``GC4'' later
in this paper. The sky area covered by each scan is indicated in
Figure~\ref{skybox}. The corresponding observational parameters are
listed in Table~\ref{obspara}. The high resolution cross-dispersing
mode of the instrument was used to acquire spectra in the range of
2.04 to 2.38~$\mu$m to include the $^{12}$CO~(2-1) first overtone
band head features at $\sim$2.2935~$\mu$m and the Earth's
atmospheric features between 2.04 and 2.06~$\mu$m. Spatial-spectral
images of multiple echelle orders were obtained with the
1024$\times$1024 pixel InSb detector. The NIRSPEC-6 and NIRSPEC-7
filters were used during the slit scans. The spectral ranges of
echelle orders covered by each scan can be found in
Table~\ref{orderwave}. During each scan, the telescope was stepped
left to right across the field of the observations and an image
rotator controlled a long slit to keep slit orientation along the
north-south direction. Slit viewing camera (SCAM) images were taken
simultaneously with spectra. These images are used to determine the
location and orientation of the slit when the spectra were acquired.
The width of the slit was chosen to match seeing at the time when
the observations were carried out. The FWHM of seeing estimated
using SCAM images is $0.6''$, $0.4''$, $0.5''$, and $0.5''$ for data
obtained in GC1, GC2, GC3, and GC4, respectively. The estimated
resolving power (R$=\frac{\lambda}{\Delta\lambda}$) was
$\sim$~14,000 during GC1, $\sim$~23,300 during GC2 and $\sim$~16,700
during GC3 and GC4. Here, $\Delta\lambda$ is the width (FWHM) of arc
lamp lines in our calibration frames and is measured at
$\lambda$=2.1714~$\mu$m. Necessary flat, sky (dark spot) and
wavelength calibration frames were also acquired with the same
instrument setting for each night.

\subsection{Data Reduction}
To compare results from different epochs, it is necessary to reduce
data in a consistent way to avoid artifacts associated with
different methods. We developed an IDL data reduction code, which is
applied to all data sets. We removed background emission by
subtracting an average dark spot image from target images. However,
this procedure does not correct for sky variation during the scans,
which we have to tolerate. Additional contamination may come from
data which were taken before our observations. Unexpected lines or
stripes resulting from bright sources in the earlier observations
are seen in some of our images as ghost features. Fortunately, we
have those data, so we can determine the contaminating sources by
carefully examining our slit scan images and data taken immediately
before our observations. We subtract an optimal fraction of each
contaminating frame from our target images so that the resulting
images contain minimum amount of unwanted features (the correlation
coefficient between our data and the contaminating frame reaches
minimum for the region where the unwanted features were spotted).
This method removes most ghost features, such as the contamination
caused by bright arc lamp lines. After the subtractions, we divide
the resulting frames by the QTH lamp images (flat) to correct for
detector response variation. Finally, the data are treated with the
bad pixel cleaning procedure described in \cite{figGKMBM03}.

Three images of a bright star, Quintuplet star \#3 (Q3), taken with
the same instrument setting, are used as our telluric standard for
each night. The telescope was offset when these images were taken so
that Q3 is present at different locations along the slit in each
image. We subtract three images from each other to remove background
emission. Two resulting images with positive and negative stars are
flat-fielded and treated with the bad pixel cleaning procedure. The
arc, etalon lamp wavelength calibration frames are also flat-fielded
and cleaned of bad pixels. These frames are used in rectification
procedure to get accurate wavelength scale and de-wrapped orders for
target images.

To obtain an accurate wavelength scale, we use the follow procedure
to reduce observational data. First, we use Arc lamp lines
(including Ar, Ne, Kr and Xe lines) with known wavelengths to set
the wavelength scale across the etalon frame. The etalon line
wavelengths are measured based on this wavelength scale. The etalon
equation ($\lambda=\frac{2t}{N}$) is solved to find the etalon plate
thickness (t) and the orders (N) of the etalon lines. Theoretical
etalon line wavelengths are calculated using the derived etalon
thickness and order numbers. Finally, the new etalon line
wavelengths are used to derive de-wrapping matrices and calibration
reference wavelengths for the target images. Ideally, one etalon
thickness should be valid for all etalon orders present in the same
instrument setting. However, this is generally not the case. To
achieve the best wavelength calibration, we use different etalon
thicknesses for echelle orders. For echelle orders with at least two
Arc lamp lines bracketing a big portion of the orders, e.g. orders
\#32 and \#34-36, this method gives wavelengths with calibration
errors $\sim$3~km~s$^{-1}$. For the echelle order \#37, this method
gives bigger errors, around 7~km~s$^{-1}$. In the order \#33, which
contains the CO band head features, there are only two known Arc
lamp lines which are close to each other in the order. Using these
two lines to calibrate wavelengths will result in huge wavelength
errors for pixels far from these Arc lamp lines. In this case, we
carefully choose the etalon thickness so that wavelength
uncertainties of rectified Arc lamp lines in neighboring orders vary
monotonically. After doing this, two Arc lines have wavelength
errors $\sim$3~km~s$^{-1}$. These errors are estimated by measuring
the line wavelengths in the rectified Arc lamp line frames using the
de-wrapping matrices derived from the etalon wavelengths.

\begin{figure}
\epsscale{1.0} \plotone{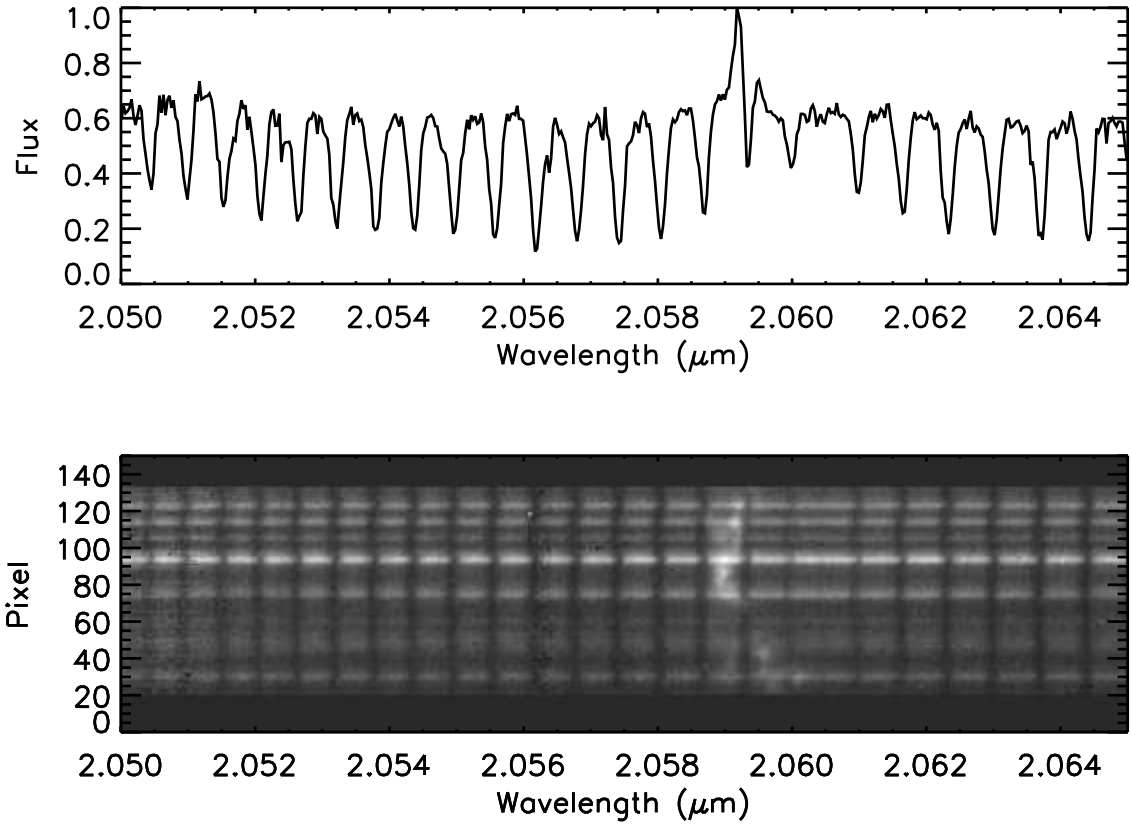} \caption{Telluric feature used to
find spectral shifts caused by star position in slit and an example
of rectified orders. Note the ripples, caused by star offset
position in the slit, in atmospheric CO$_2$ absorption features
(vertical dark lines in the image). Bright emission at
$\sim$2.06~$\mu$m is He~I~2.0581~$\mu$m line emission from ISM.
Image has been flipped so north is up. \label{atmo_order}}
\end{figure}

After rectification, target frames are divided into individual
orders. Stars are present as bright strips along the wavelength
direction in the rectified orders. An example of rectified orders is
shown in Figure~\ref{atmo_order} (bottom panel). Two prominent
spectral features in the image are vertical telluric (CO$_2$)
absorption lines and bright 2.058~$\mu$m He~I line emission from
local ionized gas in the GC. Spectra of individual stars are
extracted from each order with an extraction window of 5 pixels.
Local sky background at each wavelength is estimated from values of
pixels (1 pixel wide window) immediately above and below the pixels
belonging to stars and subtracted from the stellar spectra. We
sacrifice some s/n to avoid bad background extraction in the crowded
field. Spectrum of the telluric standard star is extracted in the
same way. Finally, the target spectra are divided by the spectrum of
the standard star to correct for absorption by the Earth's
atmosphere.

\begin{figure}
\epsscale{1.0} \plotone{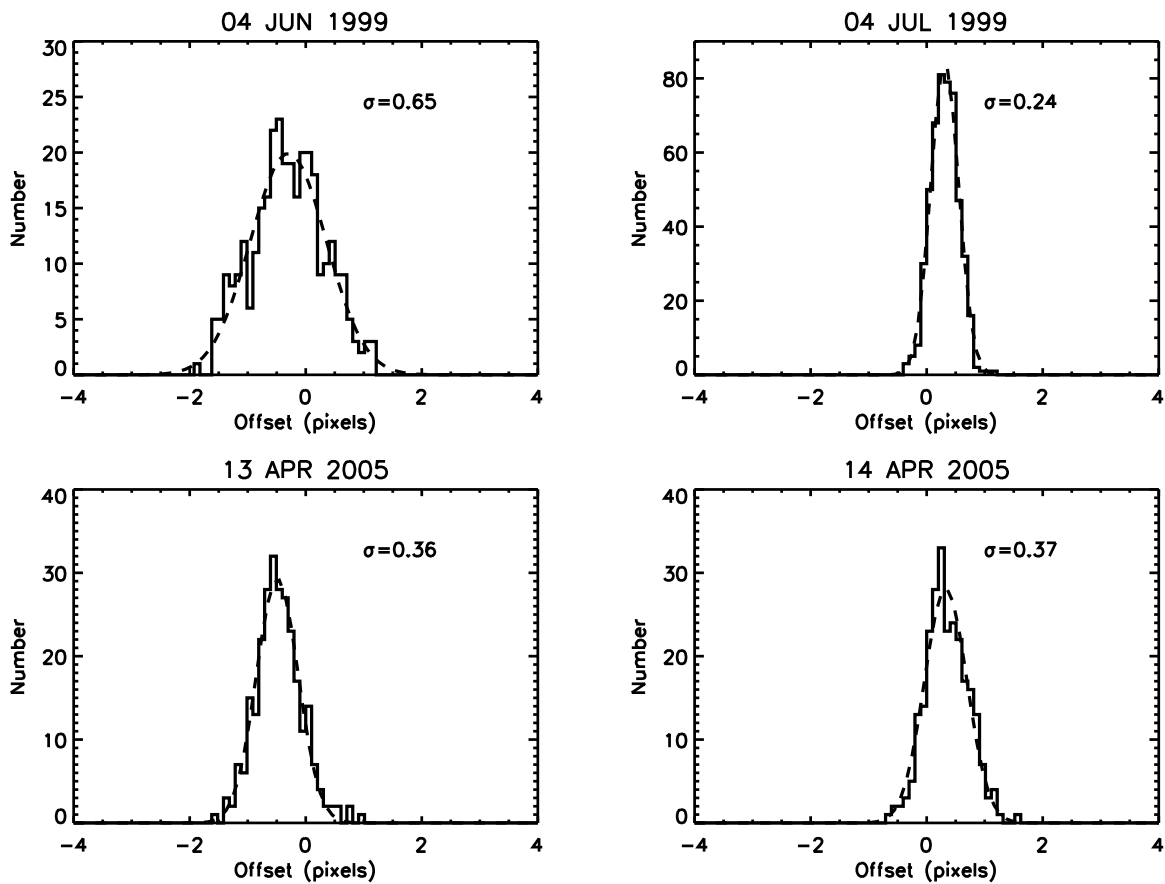} \caption{Distributions of star
positions within slits. Stars are generally offset from the center
of the slit. In order to correct for the spectral shift caused by
this effect, the offset is found by cross-correlating CO2 features
in the spectra of target stars and a reference star, which is
usually IRS7. Since IRS7 may also be offset from the center of the
slit, the distributions of the offset are not centered at 0 as shown
by these plots. The width of the distribution is proportional to the
width of the slit used for each scan. Widths of Gaussian fits to the
distributions are indicated in the plots. \label{specslitshift}}
\end{figure}

In rectified orders, stellar spectra are usually shifted along the
dispersion direction because stars are generally not centered in the
slit (Figure~\ref{atmo_order}). We can find the amount of shift by
cross-correlating the telluric features in stellar spectra with
those in the spectrum of the brightest object in our data sets
(usually IRS7). The cross-correlation produces the distribution of
the offset within the slit for four data sets
(Figure~\ref{specslitshift}). Because the location of a star in the
slit is random, it is not surprising that the offset has a normal
distribution. Gaussian fits of the distributions are also shown in
the figure and the width of the corresponding Gaussian fit is
indicated in each plot. These widths are consistent with the sizes
of slits used in four slit scans, which again proves that our
assumption of a normal distribution is correct. We can see from the
figure that centers of the distributions are usually offset from
zero, indicating that IRS7 is normally not centered in our slits.
The actual shift to which each extracted spectrum is subjected is
the difference between the calculated shift and the center of the
normal distribution. Finally, the correct shift is applied to each
spectrum.

Some observational parameters, such as telescope pointing and slit
orientation at the time when data were taken are stored in headers
of raw data files. However, the stored values are not accurate,
especially when the telescope tracking and/or image rotator control
fails, which happened during the observations of July, 1999. Thus,
instead of using the stored parameters, we use the sky images taken
by SCAM during the scan observations to calculate offsets of
telescope pointing and position angles of the slit by matching
positions of two bight stars (IRS11NE and IRS28) in the field.
Coordinates of the two stars are taken from \cite{figGKMBM03}. This
method allows better estimations of telescope pointing and slit
orientation. These parameters, together with offsets of stars from
the edge of the rectified orders, are used to calculate coordinates
of stars.

\begin{figure}
\epsscale{1.0} \plotone{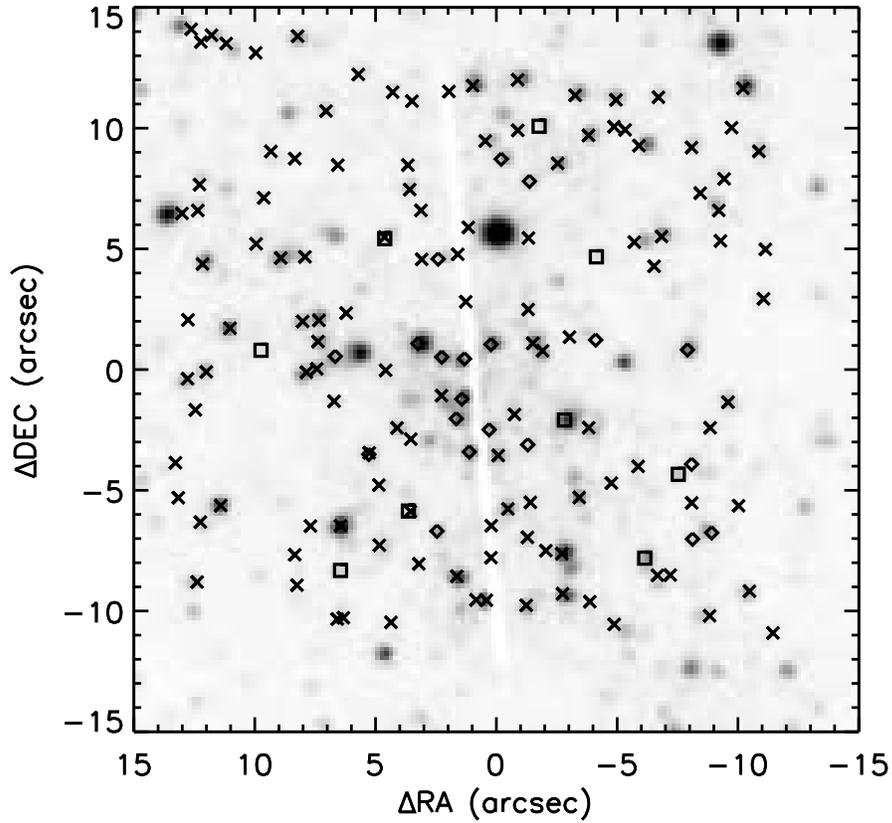} \caption{Positions of identified
stars in four slit scans. The grey scale image shows the proximate
sky area covered by four observations. The white strip in the middle
of the field is the shadow of the slit. Cool stars are indicated
with crosses, hot stars with He~I 2.112/2.113$\mu$m doublet
absorption feature are indicated with diamonds and hot stars showing
He~I 2.058$\mu$m broad emission are indicated with squares.
\label{star_position}}
\end{figure}

Depending on the width and tilting angle of the slit and the size of
telescope steps, the same star may be present in several slit
positions (different files). Spectra of same stars need to be
coadded before we make further analysis to avoid miscount of stars
and to improve s/n. We use stellar positions as the primary
criterion to identify different stars. At the same night, all
candidate stars with coordinate offsets less than 1.5$''$ are
probably the same star. However, if we extract more than one
spectrum from the same file, the spectra are considered from
different stars no matter how close the stars are in space. To
discriminate stars that are too close in coordinates, stellar
spectra are also used as the additional criterion. Spectra of the
same star are assumed to have similar line profiles and Doppler
shifts during the same night. The coadded spectra are then used to
find radial velocities for cool and hot stars based on different
spectral features. Spectra showing 2.2935~$\mu$m CO band head
absorption features are considered from relatively cool giant or
super giant stars. Absorption features at $\sim$2.112 and
2.113~$\mu$m and emission feature at $\sim$2.058~$\mu$m help us to
identify two types of hot stars (narrow-line and broad-line He~I
stars) described in \cite{tanFNKGM06} and \cite{pauMMR01}. The
positions of all stars of three types found in this work are shown
in Figure~\ref{star_position}.

\section{Results}
K band slit scan observations of the central parsec of the Galaxy
are carried out in two different epochs. Results from the first
epoch have been presented by \cite{figGKMBM03} for cool stars and by
\cite{tanFNKGM06} for hot stars. In the current work, we combine the
data from the two epochs to study the spatial distributions and the
accelerations of early and late populations in the inner parsec of
the Galaxy. By using an improved data reduction procedure, we have
identified 123 cool stars and 29 hot stars. We provide the most
accurate radial velocity measurements of these stars at two epochs.
Among the detected hot stars, 20 stars fall into the narrow-line
category and 9 stars fall into the broad-line category. In the
following, we present results from our analysis.

\subsection{Cool Stars}

The distinct $^{12}$CO (2-1) band head features with a sharp edge at
2.2935~$\mu$m are used to find radial velocities of cool stars. The
spectrum of Arcturus \citep{ramDFSB97} is used as the template. We
cross-correlate the coadded stellar spectra with the template
spectrum to find the best shifts and correlation coefficients. A big
coefficient ($\ge$0.6) indicates that the stellar spectrum under
examination shows evident CO band head features and that the star is
late-type. The criterion is picked based on visual inspection of the
results of cross-correlation. To understand the meaning of the
criterion, we simulate measured spectra by adding a certain amount
of white noise to the template spectrum. Our simulation shows that
the cross-correlation coefficient drops below 0.6 when s/n is
smaller than 1.7. The median value of s/n for these spectra is
$\sim$4.4. Stellar radial velocities are computed from wavelength
shifts of the CO band heads. We compute spectral types of these
stars using a linear relationship between the equivalent width of
the CO band heads and the surface temperature of a cool star
\citep{figGKMBM03}. The derived spectral type is used to make an
additional correction for the wavelength shift due to the stellar
surface temperature. This correction is typically less than
4~km~s$^{-1}$ \citep{figGKMBM03}. The wavelength shift caused by the
Earth's motion relative to the Sun is taken into account. The values
are computed using the macro, bary.pro, from the IDL Astronomy
User's Library. This correction is computed as +6.6, -8.1, 26.3, and
26.0~km~s$^{-1}$ for four data sets, respectively. Here a positive
velocity indicates a motion toward the GC.

\begin{figure}
\epsscale{1.0} \plottwo{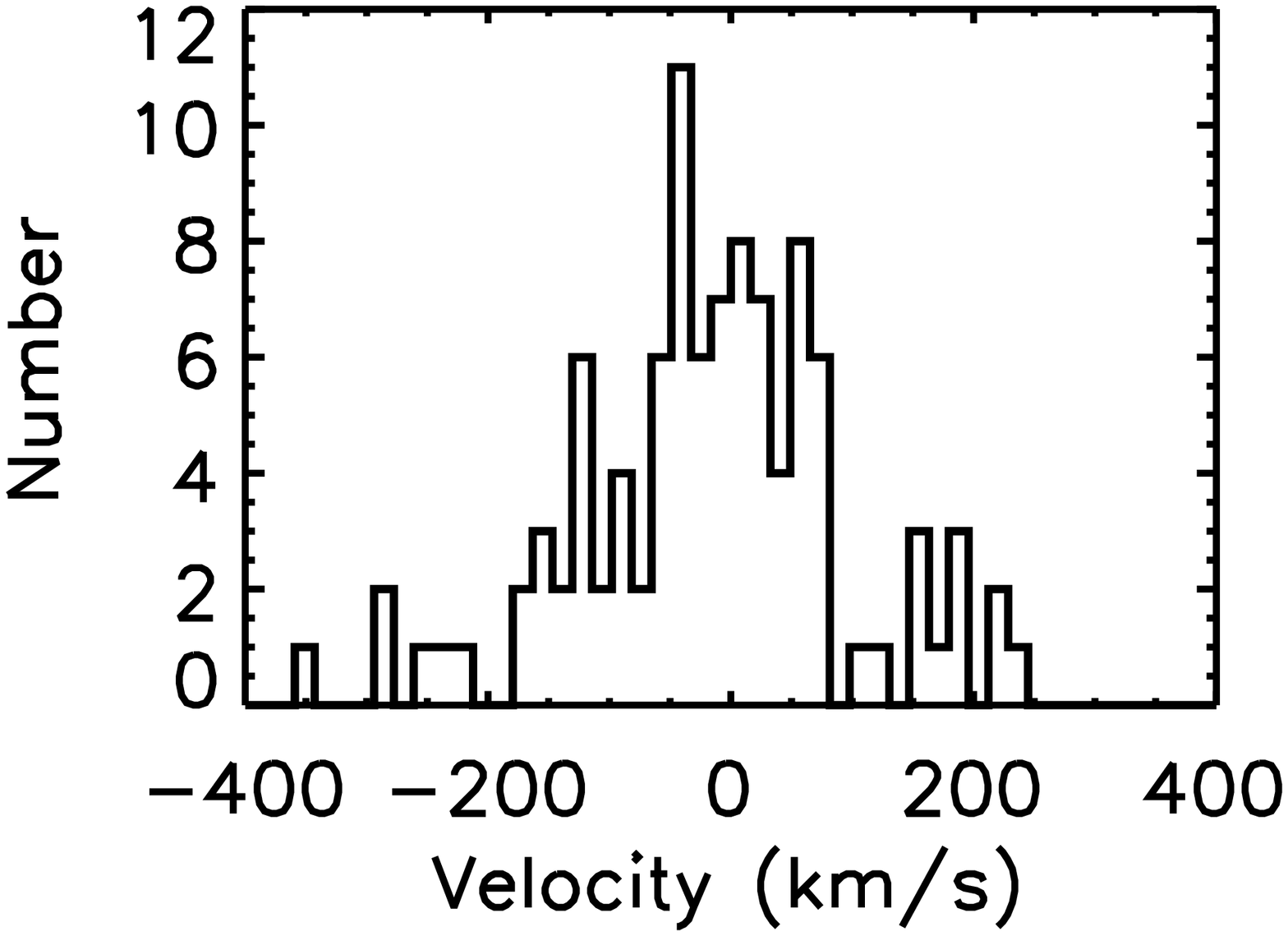}{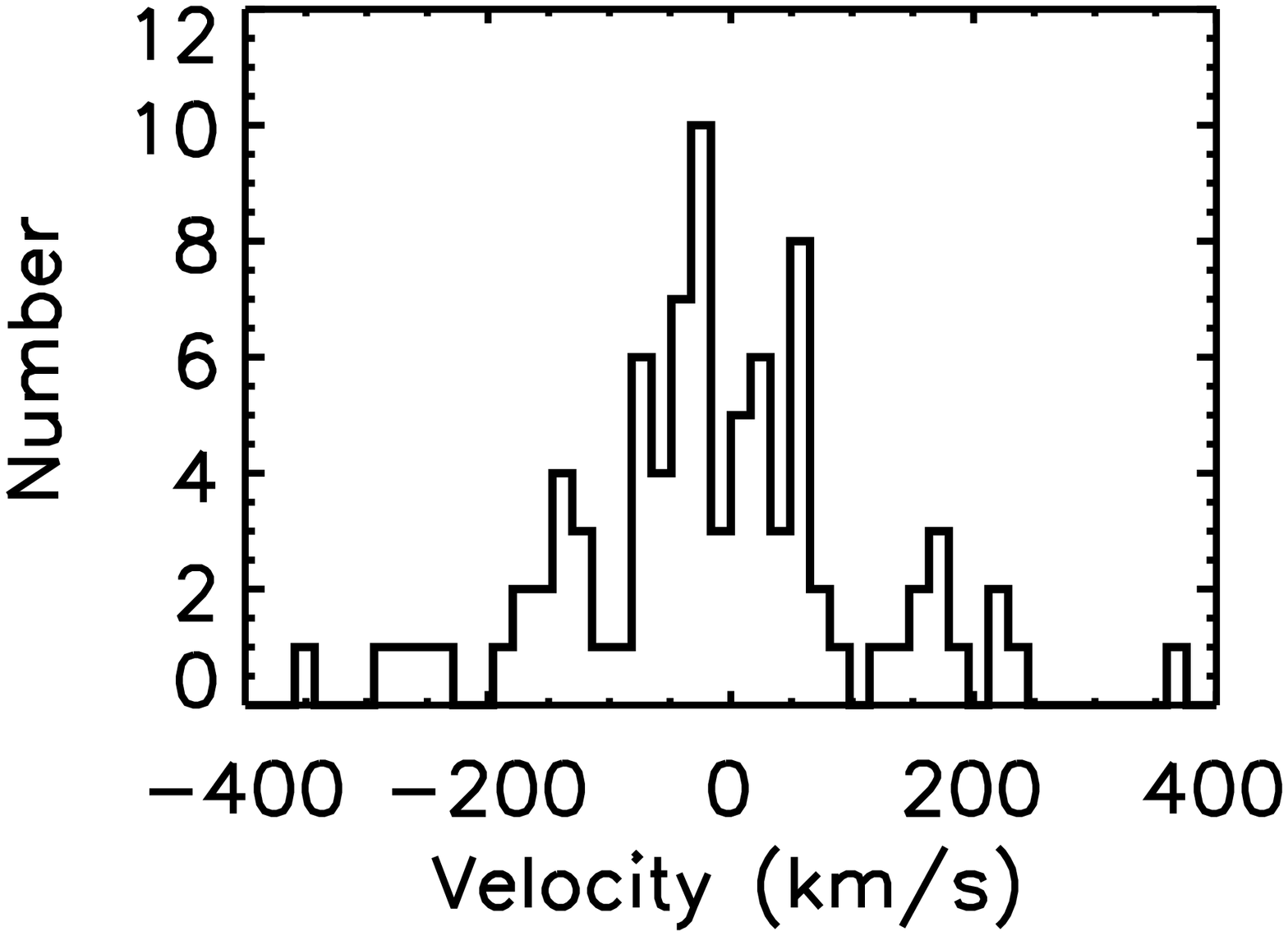} \caption{Distribution of
radial velocities of cool stars in combined 1999 (left) and 2005
(right) data sets. \label{coolstar_vdist}}
\end{figure}

The total numbers of cool stars identified in the four slit scans
are 98, 30, 68, and 68, respectively. In order to compare the
stellar velocities in the two epochs more than five years apart, we
combine two data sets in the year of 1999 (GC1 and GC2) and two data
sets in the year of 2005 (GC3 and GC4). A total of 102 cool stars
are included in the combined 1999 data set and 26 of these stars are
present in both GC1 and GC2. A total of 86 cool stars are included
in the combined 2005 data set and 50 of them are present in both GC3
and GC4. Figure~\ref{coolstar_vdist} shows the velocity
distributions for the two combined data sets. We list average radial
velocities of each epoch and estimated temperatures and spectral
types for these stars together with the velocities found by
\cite{figGKMBM03} in Table~\ref{coolstarvelo}. The mean and standard
deviation of the velocity differences for stars with two velocity
measurements in 1999 are 3.1~km~s$^{-1}$ and 2.5~km~s$^{-1}$.
Assuming that the errors of two velocity measurements are same but
uncorrelated, we can determine 1~$\sigma$ errors of the velocity
measurements to be 1.8 and 1.3~km~s$^{-1}$
($\frac{2.5^2}{2}$=$1.8^2$, $\frac{2.5^2}{4}$=$1.3^2$), in which the
bigger value is for stars with only one velocity measurement (the
uncertainty of individual measurements) and the smaller one is for
stars with two velocity measurements (the uncertainty of the mean of
two independent measurements). Applying the same method to data from
2005, we get a mean of 3.6~km~s$^{-1}$ and a standard deviation of
1.5~km~s$^{-1}$. Therefore, 1~$\sigma$ errors are 1.0 or
0.7~km~s$^{-1}$ for stars with one or two velocity measurements,
respectively. The uncertainties of our measurements for the first
epoch are bigger than those in \cite{figGKMBM03} although both
studies use same sets of data. We believe that this is due to the
difference in star and spectrum extraction. We used a smaller
(compared to that in \cite{figGKMBM03}) flux threshold to find
stars, and a smaller slit window to extract stellar spectra. A
bigger extracting window would result in more contamination of light
from near-by stars. Despite the different parameters used by two
studies, good agreement can be seen between the newly derived 1999
velocities and those found by \cite{figGKMBM03}. The differences
between the two measurements range from -1.9 to 12.3~km~s$^{-1}$
with a median difference of ~7.3~km~s$^{-1}$ and a standard
deviation of ~2.8~km~s$^{-1}$. We did not include IRS1NE(1) when we
computed these values. Our measurements and those in
\cite{figGKMBM03} for this source differ by ~240~km~s$^{-1}$. There
must be an error, e.g. erroneous source extraction, in the
measurements of \cite{figGKMBM03}. Measurements of \cite{figGKMBM03}
tend to have higher values than ours. This must be due to wavelength
calibration. Different calibration parameters were used to calibrate
wavelengths. We estimate that the systematic error caused by
wavelength calibration can be up to $\sim$7~km~s$^{-1}$. Finally, we
used 7.3~km~s$^{-1}$ as a safe estimate of the systematic error in
our measurements, which is the median difference between our
measurements and values in \cite{figGKMBM03} for stars observed in
the first epoch.

The medians and the standard deviations of stellar velocities in
these data sets are shown in Table~\ref{coolstarvelodis}. The
average heliocentric velocity (-7.8$\pm$10.3~km~s$^{-1}$,
10.3=7.3$\times\sqrt{2}$) corresponds to the projected motion of the
Sun with respect to the local standard of rest (LSR) and the
standard deviation (113.7$\pm$10.3~km~s$^{-1}$) is a measure of the
velocity dispersion of these stars. The skewness and kurtosis of
individual samples are also listed. Except for the data set from
GC2, values of other data sets are consistent with each other and
are also consistent with the values in \cite{dehB98} and
\cite{figGKMBM03}. The different values for the data set from GC2
are probably due to a small sky area coverage of GC2 and a small
size of the resulting sample.

\subsection{Hot Stars}
\subsubsection{Spectral Variability and Radial Velocity}
Based on data from the two epochs, we identified 29 hot stars. Among
them, 20 stars are identified as the narrow-line type and 9 as the
broad-line type He~I stars. We follow the classification suggested
by \cite{pauMMR01} and \cite{tanFNKGM06}. Stars showing broad
2.058$\mu$m line emission are considered as broad-line He~I stars.
Their 2.058~$\mu$m line full width at zero intensity (FWZI) is
usually over 1000~km~s$^{-1}$. Stars exhibiting 2.112/2.113~$\mu$m
absorption features are considered as the narrow-line He~I stars.
These lines are usually a few hundred km~s$^{-1}$ wide. Locations of
identified objects are indicated with squares for broad-line objects
and with diamonds for narrow-line objects in
Figure~\ref{star_position}. Correlation with previous identified
stars \citep{figGKMBM03,tanFNKGM06} is made based on calculated
stellar coordinates. The coordinates of a star are used as the name
when no published name is found.

We show multi-epoch spectra of the narrow-line stars with the He~I
2.112/2.113~$\mu$m absorption feature in
Figures~\ref{hotstar_spec2}. Figure~\ref{hotstar_spec3} shows
spectra of objects with broad He~I 2.058$\mu$m line emission. High
stellar density and large point spread functions make it difficult
to subtract background spectrum from stellar spectra. The extracted
sky spectra may contain emission from nearby stars, thus causing bad
sky subtraction. Weak component at 2.113~$\mu$m is especially not
always obvious. Emission from gas streamers \citep{pauMMR01} at the
GC sometimes makes it difficult to identify the stellar
photospherical features. In order to study the acceleration, we
assume that the shapes of the lines do not change on the time scale
of our observations. This assumption of time consistency is
supported by Figure~\ref{hotspec4}, which shows the profiles of
Br$\gamma$ line, He~I~2.112/2.113~$\mu$m and He~I~2.058$\mu$m lines
of two examples of the hot stars.

Of the 29 hot stars, many have radial velocity measurements by other
authors \citep{genPEGO00,pauMMR01,tanFNKGM06,pauGMNB06}. Our
improved spectral resolution and multi-epoch observations provide
more accurate velocity and acceleration measurements for these
stars. We fit the absorption and emission features with Gaussian
profiles to compute the Doppler shifts and the widths of the lines.
Examples of our best fits are shown in Figure~\ref{specfit}. The
continuum components are fitted with a second degree polynomial and
subtracted from the spectra before the Gaussian profile fitting. The
radial velocity of a narrow-line star can be derived from either the
2.112~$\mu$m line or the 2.113~$\mu$m line in most cases. For these
stars, the average of two velocities is used and the difference
between two measurements is listed as the uncertainty. Some
narrow-line objects show only one absorption trough in their
spectra, such as IRS16SW(W). In such a case, the rest wavelength of
2.112~$\mu$m is assumed and 5\% of the width (FWHM) of the Gaussian
profile is taken as the uncertainty.

It is difficult to estimate stellar velocities for the broad-line
stars because the emission lines of these stars generally do not
have a simple Gaussian profile. The observed 2.058~$\mu$m He~I line
often has a distorted P-Cygni shape. Some observed profiles can be
reasonably fitted with a Gaussian emission component plus a greatly
blue-shifted Gaussian absorption component. Because the blue-shifted
absorption component is presumably from the near side of the stellar
wind, radial velocities derived from this absorption feature would
significantly deviate from the real values. In order to get a good
estimate of the radial velocity of a broad line star, we first use
the emission component to calculate the radial velocity. Due to the
fact that the blue side of emission is pulled downward by the
blue-shifted absorption component, the actual line center before
absorption should be bluer than the center of the Gaussian fit. Thus
the derived radial velocity from the Gaussian fit serves as the
upper limit to the final estimate. We acquire a second estimate of
the radial velocity from the average line wavelength using the
continuum-subtracted fluxes as weights. For the absorption part of
the spectral line, the absolute values of negative fluxes are used
as the weights in the calculation. The resulting velocity serves as
the lower limit of the estimate of the radial velocity because the
presence of blue-shifted absorption makes the absolute flux weighted
wavelength bluer than if no absorption were present, considering the
blue-shifted absorption component is usually much weaker than the
emission component and is far from the emission peak. Finally, the
average value of the two estimates is taken as the final estimate of
the radial velocity and a half of the difference is taken as the
uncertainty.

The final radial velocities and the average equivalent widths (EW)
are listed in Table~\ref{hotstarvelo2} and \ref{hotstarvelo3}.
Unlike the cool star cases, radial velocities of some hot stars are
quite uncertain due to the variation of extracted line profiles from
night to night within one epoch. Big uncertainties are expected for
the measurements. Therefore, we list their spectra and velocities
individually. In the narrow line cases, the EW is computed from the
widths of Gaussian fits. In the broad line cases, the EW is simply
the integration of area under continuum-subtracted emission with the
continuum level being normalized to unity. Results show that both
radial velocities and line widths derived from different scans for
the same object are generally consistent with each other except for
in a few cases where spectral variability is important. We need to
point out that velocities estimated from different features are not
necessarily same. Theoretically, the Br$_\gamma$ line and the He~I
line at 2.058~$\mu$m can also be used to estimate the stellar
velocities of the narrow He~I line stars. The profiles of these
lines are usually complicated and result in different radial
velocity estimates because of nebula emission from streamers in the
region. It is also suggested that the widths and strengths of the
He~I emission lines are more connected with the density of the
stellar wind than with the terminal velocity \citep{naj07}. In the
cases of narrow He~I lines, the absorption feature at 2.112~$\mu$m
becomes prominent because emission in the 2.058~$\mu$m (and/or
2.112~$\mu$m) line is weak (also visually narrow). Thus the
velocities computed from the 2.112$\mu$m feature can be greatly
blue-shifted from the values computed from the emission features and
there would be systematic errors in the velocity measurements of the
narrow line objects. In some cases, this error can be up to
30~km~s$^{-1}$ \citep{naj07}. However, because estimating the
velocities of the narrow He~I line objects from the He~I emission
lines is not applicable due to the small strength of the lines, we
will leave out such an investigation in this work.

Because the sky area covered by our observations varies from time to
time (Figure~\ref{skybox}), some stars in our list are not covered
by all scans. \cite{pauGMNB06} obtained high spatial resolution
spectroimaging data of the region. We can find counterparts for many
of our sources in their list. In the following, we compare our
results with previous studies.

\subsubsection{Narrow-line stars}

We determine that 20 sources in our sample are narrow line type He~I
stars, including several stars which were considered featureless or
broad-lined in \cite{tanFNKGM06}. A better spectra extraction and
background subtraction procedure helps to reduce contamination by
light from neighboring stars. This contamination is significant in
\cite{tanFNKGM06}. The classifications of some broad line sources in
\cite{tanFNKGM06} are ambiguous due to the weak strength of the
lines.

-\textbf{IRS16NW}: Among all observed He~I emission line stars, this
one has the smallest projected distance from Sgr~A$^*$, only
$\sim$~1.0$''$ away. The star is identified as an Ofpe/WN9 star
\citep{pauGMNB06}. \cite{marGHEPG07} fit the observed K band
spectrum of the star and suggest that the star has just left the
main sequence and is in an early evolved stage. Our spectra show a
P-Cygni profile with strong blue-shifted absorption in the
2.058~$\mu$m line similar to that in \cite{pauMMR01} and
\cite{marGHEPG07}. The 2.112/2.113~$\mu$m absorption doublet is
prominent in our spectra. \cite{pauMMR01} derive a radial velocity
of -46~km~s$^{-1}$ for this object using the 2.058~$\mu$m line
observed in 1997. \cite{pauGMNB06} measure the radial velocity to be
-44~km~s$^{-1}$ during the time period of 2003-2004. Both values are
within the uncertainty range from our measurement.

-\textbf{IRS16C}: This star also has a spectral type of Ofpe/WN9
\citep{pauGMNB06}. Its spectrum resembles that of IRS16NW except
that more He~I absorption is present on the blue wing of the
Br$_{\gamma}$ line in IRS16NW \citep{marGHEPG07}. We detect the
2.112/2.113~$\mu$m absorption doublet in the spectrum of the object
from all four nights. Our measured velocity ($\sim$100~km~s$^{-1}$)
appears consistent within 1 sigma error bars from that in
\cite{pauGMNB06}, which is 125~km~s$^{-1}$.

-\textbf{IRS16SW(W)}: This object was identified as a short period
($\sim$19.4 days) eclipsing binary system
\citep{ottEG99,marTPOG06,peeBDSP07}. Its radial velocity varies in a
range between $\sim$300 and 677~km~s$^{-1}$
\citep{ottEG99,marTPOG06}. Its 2.112~$\mu$m line profile also varies
with time, depending on which phase the system is at. We detect the
2.112~$\mu$m absorption feature in all four data sets. Our derived
velocities are consistent with results in \cite{marTPOG06}, but are
different from those in \cite{tanFNKGM06} (247~km~s$^{-1}$) and
\cite{pauMMR01} (354~km~s$^{-1}$). High stellar density makes it
hard to extract spectrum for this star. Therefore, its spectrum may
be contaminated by light from neighboring stars, and result in big
uncertainties in the estimated velocities. Three sources (IRS16SW,
IRS16SSE1 and IRS16SSE2) in \cite{pauGMNB06} are close to IRS16SW(W)
based on their coordinates. However, none of them has an estimated
radial velocity close to our measurements.

-\textbf{IRS16CC}: We confirm the identification of this star as a
narrow-line star by \cite{tanFNKGM06}. The characteristic absorption
lines are observed in all four data sets. The derived radial
velocities are consistent with each other and with that in
\cite{tanFNKGM06}. 
No obvious velocity change over the 5.8 year time span is observed
for this star. \cite{pauMMR01} did not observe He~I emission from
this object. They claimed that the earlier identification of this
star as a helium star by \cite{bluSD96} was a mistake due to light
from a nearby source - He~I~N2, which is 0.6$''$ South and 0.3$''$
East of IRS16CC. Our derived velocities for IRS16CC are very
different from that of He~I~N2 in \cite{pauMMR01}, implying that
IRS16CC and He~I~N2 are probably two different sources. Our spatial
resolution does not allow us to separate IRS16CC and He~I~N2.
\cite{pauGMNB06} suggest that IRS16CC is an O9.5-B0.5 supergiant,
and they estimate the radial velocity of the star to be
241~km~s$^{-1}$, which is similar to our measurements.

-\textbf{MPE~1.0-7.4}: This star was classified as a B0.5 supergiant
in \cite{pauGMNB06}. The star belongs to IRS16 cluster of hot stars
and several bright He~I stars are in its vicinity. We detected the
star in the second scan. Our estimate of its radial velocity is
consistent with the value in \cite{pauGMNB06}.

-\textbf{IRS16NE}: \cite{tanFNKGM06} suggest that this star is a
spectroscopic binary system. We detect this star in all four data
sets. The radial velocity of this star changes from
$\sim$12.8~km~s$^{-1}$ in June 1999 to $\sim$-60~km~s$^{-1}$ in
2005, which supports the proposal by \cite{tanFNKGM06}. However, we
did not find evidence to support the change of the line shape (the
2.112~$\mu$m line width). \cite{pauGMNB06} measure the radial
velocity to be -10~km~s$^{-1}$ during 2003-2004, which is within the
uncertainty range of our measurements.

-\textbf{IRS33E}: This object is identified as a cool star (ID221)
in \cite{figGKMBM03}. However, no CO band head absorption feature is
observed for this object in our reduced data sets. The
misidentification may be due to contamination by the light from red
giants in the region and bad background subtraction. IRS33E was
identified as an Ofpe/WN9 star \citep{pauGMNB06}. \cite{marGHEPG07}
show that the K band spectrum of IRS33E resembles that of IRS34W,
IRS16NW and IRS16C except that the Br$_{\gamma}$ line of IRS33E has
a complicated line profile. The 2.112/2.113~$\mu$m doublet
absorption feature can be seen clearly in our spectra. Our
observations show that the radial velocity of this object ranges
$\sim$150~km~s$^{-1}$ and $\sim$110~km~s$^{-1}$ between 1999 and
2005, and does not change significantly over the 5.8 year time span.
\cite{pauMMR01} computed the radial velocity of the object (IRS33SE)
using the 2.058~$\mu$m emission line which is strongly affected by
diffuse emission from ISM. They had a much higher velocity
$\sim$258~km~s$^{-1}$. \cite{pauGMNB06} reports the radial velocity
of IRS33E to be 170$\pm$20~km~s$^{-1}$ between 2003 and 2005, still
higher than the values we found.

-\textbf{IRS34W}: This object was identified as a narrow-line type
He~I star by many investigators
\citep{pauMMR01,ottGES03,pauMS03,triMOPAE06}. \cite{pauGMNB06}
identified its spectral type to be Ofpe/WN9. The star is also
believed to be a LBV candidate \citep{pauMMR01,triMOPAE06} based on
its changing K band magnitude with $\Delta$K$\sim$1.5. Despite the
star's highly variable K magnitude, the spectrum of the star does
not show any variability. \cite{pauMMR01} point out that the source
consists of two stars (34E and 34 W) of similar brightness. Our
angular resolution is not high enough to separate them.
\cite{triMOPAE06} suggest that the star is not a LBV but a star with
a spectral type between O supergiant and LBV. \cite{triMOPAE06}
showed that the stellar spectrum has a clear P-Cygni profile in
2.058~$\mu$m and 2.112~$\mu$m He~I lines. A broad Br$_{\gamma}$ line
with a blue-shifted shoulder can also be seen in its spectrum. Our
spectra from GC1 and GC3 are similar to each other and to those in
\cite{triMOPAE06} and \cite{marGHEPG07}. Single peak absorption is
observed in the 2.112~$\mu$m line. \cite{tanFNKGM06} did not detect
the 2.112/2.113$\mu$m absorption feature, possibly due to bad sky
subtraction. The velocities derived from the absorption feature at
2.112~$\mu$m in our spectra are consistent with each other, although
they are different from that in \cite{pauMMR01}, which was derived
from the 2.058~$\mu$m He~I line. \cite{pauGMNB06} measure the radial
velocity of the star again and find a value similar to our result.

-\textbf{A21}: This star is located at the center of the eastern
cavity between the northern arm and the eastern arm. The 2.112$\mu$m
line absorption is quite strong. We detected the line in all four
data sets, although spectrum from GC2 is noisier than those from
other scans. Absorption of the 2.113~$\mu$m line is generally
missing in our spectra. The star does not show obvious radial
velocity shift over the 5-year time span.

-\textbf{ID308}: This star was identified as a cool star in
\cite{figGKMBM03}. Our multiple night observations show that it is a
narrow He~I line star. The CO absorption features present in the raw
spectra are greatly reduced after the subtraction of sky background
emission. This suggests that the CO feature is coming from the
background cluster of cool stars. Derived radial velocities are
consistent with each other. The position of the star matches that of
IRS1E (E67) in \cite{pauGMNB06}, which is classified as a B1-3
supergiant. \cite{pauGMNB06} found that the radial velocity of the
star is $\sim$8$\pm$20~km~s$^{-1}$, higher than our measurements.

-\textbf{MPE+1.41-12.2}: We detect the He~I absorption feature in
this object in all four data sets. Both 2.112 and 2.113$\mu$m lines
are detected in this object. Spectrum from GC1 is
$\sim$25~km~s$^{-1}$ more red shifted than those from GC3 and GC4.
The radial velocity from GC2 data set is very close to those from
GC3 and 4, which may be due to the uncertainty of our measurements.
Our measurements are within the uncertainty range of the value
($\sim$153$\pm$50~km~s$^{-1}$) by \cite{pauGMNB06}.

-\textbf{IRS6W}: This narrow-line object is only observed in our
first scan because it is outside of fields of other scans. It seems
that its coordinates in \cite{tanFNKGM06} are mixed with those of
6E. 6E, identified as a WC9 star by \cite{kraGENL95}, does not show
any obvious stellar photospheric feature in our observations.
\cite{pauMMR01} attribute observed broad 2.058~$\mu$m emission to
nearby helium streamers and suggest that the earlier WC9
classification was a misidentification. We do not observe broad
2.058~$\mu$m line mission, but we detect 2.112/2.113~$\mu$m
absorption doublet to confirm IRS6W to be a narrow He~I star. We do
not find this star in the list of \cite{pauGMNB06}.

-\textbf{IRS26}: The He~I absorption feature in the spectrum of this
star is detected in all four slit scans. Both 2.112 and 2.113~$\mu$m
lines can be seen in these spectra. The derived velocities are
consistent with each other and indicate no relative acceleration
during 1999-2005. \cite{pauGMNB06} estimate a similar radial
velocity for this star, and they suggest that the star is an O9 or
early B type supergiant.

-\textbf{AFNWB}: \cite{tanFNKGM06} identify this object as a broad
line He~I star. However, we did not observe the broad line emission
at the position of the star. We believe that the misclassification
is due to a nearby broad line source AFNW, which is $\sim$0.6$''$E
and 0.4$''$S of this star. AFNWB lies on the west edge of our
scanning field. We have only one measurement of its radial velocity.
\cite{pauGMNB06} do not include this star in their list.

-\textbf{G1138}: The same star is named as BSD~WC9B in
\cite{bluSD96}. In our spectrum, the star shows weak and narrow
absorption at $\sim$2.112~$\mu$m. We do not agree with
\cite{tanFNKGM06} on classifying this star as a broad line object.
No obvious 2.058~$\mu$m emission is observed at the location of the
star in our data set. Weak 2.058~$\mu$m line emission in the
\cite{tanFNKGM06} spectrum suggests that the confusion may be caused
by bad sky background subtraction.

-\textbf{-8.91-6.76}: \cite{tanFNKGM06} identified this star as a
narrow line He~I object. They mixed the star with BlumWC9, which was
identified as a WC9 Wolf-Rayet star by \cite{bluSD95}. We construct
the image for the entire field of our observations from our slit
scan data and compare the resulting image with that of \cite{mar07}.
The comparison shows that BlumWC9 and -8.91-6.76 are different
objects. BlumWC9 is the faint object $\sim$0.4$''$ NE of -8.91-6.76
and is probably buried in the light of -8.91-6.76 in our
observations. Our observations confirm that -8.91-6.76 is a narrow
line object. The 2.112/2.113~$\mu$m doublet absorption feature is
prominent in our K band spectrum of the star. Our derived stellar
velocity and line width agree with the results in \cite{tanFNKGM06}.

\subsubsection{Broad-line stars}
We identify 9 objects to be broad line He~I star candidates. Their
spectra show broad 2.058~$\mu$m or 2.112/2.113~$\mu$m line emission.

-\textbf{IRS13E}: IRS13E is a compact stellar cluster containing at
least three bright stars \citep{pauMMR01,schEIGO05,pauGMNB06}. It
has been suggested that IRS13E is the remnant of a massive cluster
falling into the GC from the outer region
\citep{maiPSR04,schEIGO05}. It is also suggested that IRS13E
contains an IMBH with a mass $\geq10^{4}$M$_{\odot}$ stabilizes the
inspiraling cluster and holds the massive cluster together against
the tidal force of Sgr~A$^*$ during the infall. \cite{pauGMNB06}
carefully analyze proper motions of stars that are assumed to be
members of IRS13E. They argue that the presence of an IMBH is
hindered by the ambiguous cluster membership of a few stars in the
analysis of \cite{maiPSR04} and \cite{schEIGO05}. Our data have
relatively low angular resolution compared to those AO observations.
IRS13E is present as one single broad line He~I source in our
observations. This classification is consistent with previous
studies \citep{najKGLKH97,pauMMR01}. Our spectra of four nights show
similar distorted P-Cygni profiles for this object. Blue-shifted
absorption components can be seen clearly. The wavelength shift
between profiles of the two epochs is obvious. The cross-correlation
fitting gives a $\sim$35~km~s$^{-1}$ difference for the radial
velocity at the two epochs. None of IRS13E members in
\cite{pauGMNB06} has a similar radial velocity as we observed.

-\textbf{IRS7W}: We detected strong and broad emission in He~I
2.058, 2.112~$\mu$m and Br$_{\gamma}$ lines in GC1, GC2 and GC3.
P-Cygni profiles with shallow blue-shifted absorption of the
2.058~$\mu$m line can be seen clearly in these spectra and resemble
each other. Our three measurements of the radial velocity indicate
that the line-of-sight acceleration of the star is small. This star
is classified as a WC9 star by \cite{pauGMNB06}. However, IRS7SW in
their list of stars seems closer to our source in coordinates.
IRS7SW has a type of WN8. Both two stars have a radial velocity
($\leq$-300~km~s$^{-1}$) more negative than our measurement
($\sim$-190~km~s$^{-1}$).

-\textbf{GCHe2}: GCHe2 \citep{tamRHCMR96} is also named as IRS~9W
and is classified as a broad He~I line star by \cite{pauMMR01}. It
is then identified as a WN8 star in \cite{pauGMNB06} and shows
strong He~II lines \citep{marGHEPG07}. The object shows broad
distorted P-Cygni profile 2.058~$\mu$m line emission in our spectra.
The overall line profiles and wavelength ranges of emission match
each other fairly well with a $\sim$50~km~s$^{-1}$ shift. Our
measurements of the radial velocity of the star have a big variance,
which can be due to the complexity of the line shape and the
uncertainty associated with the method to estimate the velocity in
the broad line cases. \cite{pauMMR01} estimate that the radial
velocity to be 221~km~s$^{-1}$, and the new measurement by
\cite{pauGMNB06} is $\sim$140~km~s$^{-1}$.

-\textbf{ID415}: We use the cool star designation from
\cite{figGKMBM03} for this object although IRS7E2 seems to be very
close to this star in location \citep{pauMMR01}. IRS7E2 was
identified to have a WN8 spectral type \citep{pauGMNB06}. We detect
broad 2.058~$\mu$m line emission from the star in three data sets.
The line profiles in these observations match each other very well
and show no relative velocity shift between them. Spectrum in
\cite{marGHEPG07} shows strong Br$_{\gamma}$ line emission with a
width similar to that of 2.058~$\mu$m He~I line, which is not shown
in our data. This object also shows strong CO absorption. It is
possible that light from multiple stars falls into our big beam
simultaneously. The CO and He~I features do not vary with time
indicating that the stars do not form a binary system.

-\textbf{AFNW}: AFNW is identified as a WN8 type star by
\cite{pauGMNB06} and shows broad emission in He~I lines (2.058 and
2.112~$\mu$m) and Br$\gamma$ line in our GC1 spectrum. This star is
also identified as a broad line object by \cite{pauMMR01}. Our
measurement of the radial velocity agrees with that in
\cite[][$\sim$159~km~s$^{-1}$]{pauMMR01}. The measurement by
\cite[][$\sim$70~km~s$^{-1}$]{pauGMNB06} is much smaller.

-\textbf{ID180}: ID180 is discovered by \cite{pauMMR01} as a broad
line object. The radial velocity was measured to be
$\sim$198~km~s$^{-1}$. \cite{pauGMNB06} measure the velocity again
and get a value $\sim$-230~km~s$^{-1}$. Our estimate of the velocity
is $\sim$-16~km~s$^{-1}$. It indicates that the measurements are
quite uncertain. The star is identified as a WC9 star by
\cite{pauGMNB06}.

-\textbf{AF}: \cite{pauMMR01} point out that this star is much
brighter than other broad line objects in their list and suggest
that a compact binary system may be present. Additional evidence is
provided by \cite{ottEG99}, who find that the K~band photometry of
this star shows some indication of time variability.
\cite{najHKKG94} modeled observed spectrum from the source with
unified atmosphere/wind models and concluded that the star was not a
classical Wolf-Rayet star because no other high excitation lines
were detected. They suggested that the star is an Ofpe/WN9 type blue
supergiant based on model parameters, which is consistent with the
classification by \cite{pauGMNB06}. \cite{marGHEPG07} point out that
AF has much broader lines compared with many other Ofpe/WN9 stars
and suggest that AF has a stronger wind and is in a more evolved
stage. In our data set, the AF star shows broad emission in
2.058~$\mu$m, 2.112~$\mu$m He~I and Br$\gamma$ lines that are not
related to ISM. The 2.058~$\mu$m line emission has a typical P-Cygni
profile with a strong emission peak and a shallow blue-shifted
absorption trough. We fit the emission peak with a single Gaussian
component to derived the stellar radial velocity, which is similar
to that in \cite{tanFNKGM06}, but $\sim$50~km~s$^{-1}$ higher than
in \cite{pauMMR01} and $\sim$80~km~s$^{-1}$ higher than in
\cite{pauGMNB06}.

-\textbf{IRS15SW}: IRS15SW is classified as WN8/WC9 by
\cite{pauGMNB06} because of C lines present in its K band spectrum
\citep{marGHEPG07}. Our observations show a broad emission line with
a regular P-Cygni profile at $\sim$2.058~$\mu$m. Radial velocity
measurements of this object at two epochs are similar to each other,
but less bluer than that in \cite[][$\sim$179~km~s$^{-1}$]{pauMMR01}
and \cite[][$\sim$-180~km~s$^{-1}$]{pauGMNB06}. The spectra indicate
a small line-of-sight acceleration during the time between two
epochs.

-\textbf{IRS9S}: We detected this star in all four data sets. Only
one spectrum shows weak 2.058~$\mu$m emission. All four spectra show
broad emission at $\sim$2.112~$\mu$m. Thus, we use this part of the
spectra to estimate the radial velocity of the star. All but one
estimated velocities are consistent with other, but much higher than
the measurement in \cite{pauGMNB06}. The star is identified as a WC9
star.

Due to the high stellar density in the region and the presence of
ISM streamers \citep{pauMMR01}, contamination from neighboring stars
and ISM is large. In our study, we carefully choose pixels for stars
and background during background subtraction and stellar spectrum
extraction to avoid as much contamination as possible. we use both
spatial and spectral information to identify different sources. This
helps us to identify many sources which were not found by
\cite{figGKMBM03} and \cite{tanFNKGM06}. However, the spatial
resolution of our observations is still low and is not able to
resolve many sources into individual stars in the crowded region. In
at least one case (ID415), we find characteristic spectral features
of cool and hot stars in the co-added spectrum at the same time.
Thus ID415 is included in both the cool and hot samples. We also
find one case (IRS13E) in which both narrow absorption and broad
emission features are present at the same time. These features must
come from different stars which lie closely on the sky. This
situation can only be improved by future high spatial resolution
observations. Finally, our radial velocity measurements of 7 of 20
narrow line stars
and 2 of 8 broad line stars 
detected in the first epoch have similar values as in
\cite{tanFNKGM06}. We also found that 11 narrow line stars
have radial velocities similar to the values in \cite{pauGMNB06}.
The differences between these measurements are less than
40~km~s$^{-1}$, roughly the median value of the uncertainties of the
radial velocities in \cite{pauGMNB06}. However, the differences
between our measurements and those in \cite{pauGMNB06} for broad
line stars are usually bigger than this value. Low spectral
resolution of data in \cite{pauGMNB06} and low spatial resolution of
our observations should be the major causes for the differences.
Different spectral resolutions of the observations and the large
intrinsic line widths of the stars can result in big differences and
uncertainties in the line centers found through Gaussian profile
fitting. The results suggest that large uncertainties are present in
the radial velocity measurements and high sensitivity and high
resolution observations are required to improve this situation.

\subsection{Distribution}

We show stellar surface densities and radial velocity dispersions of
the late type stars as functions of the projected distance (r$_P$)
from Sgr~A$^*$ in Figure~\ref{annuluscool}. The results from the
current data and from the published data in \cite{figGKMBM03} are
indicated with solid lines (filled circles) and dashed lines (open
circles), respectively. These plots only include those stars with a
projected Galactic radius r$_P$$<$0.4~pc ($\sim$10~arcsecs) so that
the resulting data set is complete in the sense that it covers a
circular region on the sky with its center at Sgr~A$^*$, and the
magnitude limit of the sample is K$\simeq$13. Same bins are used to
create the surface density plots and the corresponding velocity
dispersion plots. Our cool star surface density is consistent with
previous results \citep{figGKMBM03} by showing two bumps
(r$_P$~$\simeq$~0.07 and 0.3~pc) and one low density region at
r$_P$~$\sim$0.15~pc. The difference between two curves can be
attributed to the different sizes of the two samples. The low
surface density at r$_{P}\sim$0.15~pc shown by both curves indicates
the presence of a central ``hole'' of late type stars. Such a hole
is also supported by CO bandhead strength mapping and infrared
photometric observations \citep{selMBH90, halRRTCM96,
genTKKT96,schEAMGS07}. \cite{schEAMGS07} found an under-density of
stars at 5" and over-densities at 3" and 7" based on photometric
data. These results are similar to what we found. Beside the surface
density profiles, the velocity dispersions of late type stars are
quite flat in the observed region. No obvious velocity dispersion
increase toward the GC is supported by the data. Monte Carlo
simulations suggest that a central low stellar density region is
needed to explain the observed projected number density and flat
velocity dispersion curves \citep{figGKMBM03}. Mechanisms which have
been suggested to explain the formation of such a ``hole'' include
stellar collision \citep{lacTH82,genTKKT96,ale99,baiD99},
atmospheric striping \citep{aleL01}, atmospheric heating
\citep{aleM03} and dynamical ejection \citep{mirG00,figGKMBM03}. The
hump at the r$_P\sim$0.07~pc is a surprising result because one
would expect that these mechanisms causing the central ``hole''
would affect all stars in the region. This hump is probably due to
small number statistics and the stars observed at the small
projected radii are actually in front of or behind the central
``hole''. Due to the small area of the first bin, a small
fluctuation in the star count can result in a big surface density.
The velocity dispersions of late type stars are quite flat in the
observed region. No obvious velocity dispersion increase toward the
GC is supported by the data.

The surface density and the velocity dispersion of the early type
stars are shown as functions of the projected distance from
Sgr~A$^*$ in Figure~\ref{annulushot}. The surface density decreases
with the projected distance with a local peak at r$_P\sim$0.1~pc due
to the IRS16 cluster. Observations suggest that early type stars
between the projected distance r$_P$=1$''$ and 12$''$ from Sgr~A$^*$
belong to two disk-like structures rotating about the GC and at
least one disk is nearly edge-on \citep{levB03,genSRE03}.

The velocity distributions of both the late type and the early type
stars are quite symmetric as shown by Figure~\ref{star_vdist}. We
fit the velocity distributions with Gaussian functions, and the
fitted widths (standard deviations) are indicated with the
statistical quantities of the distributions. A 20~km~s$^{-1}$ bin
size is used to make the histograms before doing Gaussian fitting.
We note that the fitting results are not sensitive to the bin size
between 16 and 26~km~s$^{-1}$. The results show that standard
deviations from the two methods are different. For the late type
stars, the statistical standard deviation of the observed velocities
is close to 110~km~s$^{-1}$, while Gaussian fitting yields a value
of $\sim$87~km~s$^{-1}$. Velocities of the early type stars also
have significantly different values for statistical and fitted
standard deviations. Moreover, both velocity distributions skew
slightly to the left (more negative velocity). A positive kurtosis
of the late type star velocities indicates the number of the stars
with the smallest and/or highest velocities is greater than the
value expected by a normal distribution with the same mean and
variance. A negative kurtosis suggests a large number of early type
stars having intermediate velocities. Monte Carlo simulations of a
Gaussian-distributed velocity sample with a mean of -10~km~s$^{-1}$
and a standard deviation of 110~km~s$^{-1}$ show that there is
$\sim$7\% probability that the fitted standard deviation would fall
below 87~km~s$^{-1}$ when the sample size is equal to 97, and this
probability drops down to $\sim$4\% when the sample size is
increased to 123. The simulations suggest that the smaller standard
deviation found by our Gaussian fitting procedure is not entirely
due to the small size of our sample, and that the observed
velocities deviate from Gaussian distributions. The
D$'$Agostino-Pearson K$^2$ test based on the skewness and the
kurtosis \citep{dagBD90} also suggests that our Gaussian fittings
are poor, and that both velocity distributions deviate from Gaussian
distributions. The probability of the observed velocity distribution
being normal is $\sim$20\% ($\sim$60\%) for the late (early) type
stars.

The deviation from normality in the case of the late type stars is
probably due to our method of finding and identifying stars. Our
star-finding procedure tends to reject faint stars close to bright
ones and our star-identifying procedure also tends to throw away
stars with spectra of low s/n in order to compute correct wavelength
shifts. Therefore, our statistics are dominated by the brightest
stars in the region, which include many intermediate-mass
(3M$_{\odot}<$M$_*<8$M$_{\odot}$) AGB stars that are found in the
central parsec of the Galaxy \citep{genPEGO00,genSRE03}. These stars
have much shorter lifetimes (up to a few 10$^8$~yrs) compared with
low-mass (M$_*<3$M$_{\odot}$) stars, which lives for a few 10$^9$
years. These two populations of stars may have different velocity
distributions due to their different ages and relaxation times. The
two-body relaxation time for a solar mass star in the GC is of the
order $\sim$10$^9$~yrs, which is comparable to its age, while the
relaxation time scales as M$_*$$^{-1}$ and it is usually longer than
the lifetime of an intermediate-mass or massive star
\citep{genPEGO00,merS06}. We cross-correlate our late type stellar
sample with that provided by \cite{genTKKT96}, who divides the stars
into three K band magnitude groups: K$\leq$10.5, 10.5$\leq$K$\leq$12
and K$\geq$12. In total, 63 stars are present in both samples. The
comparison shows a substantial change in the luminosity function
from 25:53:120 in \cite{genTKKT96} to 19:18:26 in our case. A higher
percentage of bright stars are present in the sub-sample, supporting
our hypothesis. Thus our observed velocities of the late type stars
can have a non-Gaussian distribution even the velocities of the low
mass late type stars are normally distributed. In addition, the
``missing'' stars in the central hole may also drive the overall
velocity distribution of the late type stars away from a normal
distribution. In the case of the early type stars, the intrinsic
velocity distribution is probably non-Gaussian due to the short
ages.

\subsection{Rotation and Acceleration}

Radial velocities of the identified stars are plotted as a function
of the projected distance from the Sgr~A$^*$ in
Figure~\ref{star_vr}. Two plots are for cool (left) and hot (right)
populations, respectively. We also plot ideal circular orbital
velocities (solid lines) and escape velocities (dotted lines) around
a dominating central object with a mass of
3.5$\times$10$^6$~M$_{\odot}$ as comparison. Although several stars
have velocities beyond the limits of circular orbital motions,
radial velocities of most stars are within the range allowed by
circular orbits. The speed of IRS9 exceeds the escape velocity at
its projected distance. The star is probably ejected from the
central cluster, as suggested by \cite{reiMTOG07}.

We plot radial velocities of the cool stars against their Right
Ascension and Declination offsets from Sgr~A$^*$
(Figure~\ref{coolstar_disk}). No correlation is indicated by these
plots, which leads us to conclude that there is no, or a very small,
systematic rotation for the late-type population.
Figure~\ref{hotstar_disk} shows the average radial velocities of hot
stars verse their Right Ascension and Declination offsets from
Sgr~A$^*$. The plots include both the narrow- (filled circles) and
broad-line (open circles) stars. A correlation between the radial
velocity and the Declination offset can be seen for the hot stars
(right plot). This correlation is especially evident for the broad
emission line stars. Most hot stars distribute along a diagonal from
the upper left quadrant to the lower right quadrant through
Sgr~A$^*$, and the stars north of Sgr~A$^*$ have red-shifted
velocities and the stars south of Sgr~A$^*$ have blue-shifted
velocities. This result confirms a systematic rotation for hot
stars, consistent with other observations \citep{genPEGO00,
pauMMR01, pauGMNB06}. It has been suggested that hot stars at the GC
form one or two disk-like structures
\citep{genPEGO00,pauMMR01,levB03,genSRE03,pauGMNB06,luGHMBM07}. Our
observations support the existence of disk-like structure, however
additional information about stellar proper motions is needed to
construct the 3-D distribution of hot stars and study other
properties of the disk-like structures.

Using data sets from the two epochs, we examine acceleration of
stars in the region over this period. The changes of the radial
velocities of the stars that are observed in both 1999 and 2005 are
plotted against the projected Galactic distances in
Figure~\ref{star_vaccl}. No correlation is apparent between the
velocity changes and the projected distances from the Galactic
center. To compute the velocity changes of the observed hot stars,
we shift and cross-correlate spectra of the hot stars from two
epochs to compute the best velocity shifts to avoid big
uncertainties caused by low quality of spectra and large line
widths. The pixel size of our re-sampling grid is used as
uncertainties. In a few cases, when the relative wavelength shift
found by cross-correlation is not good due to the variation of the
line shape, and line features are obviously misaligned, we adjust
the spectral shift manually so that the overall line profiles from
different observations are aligned with each other, and the
difference between the shift found by cross-correlation and the
manual shift is used as uncertainties.

It can be shown (Appendix A) that for edge-on circular orbits the
velocity change of a star at a projected radius R$_0$ follows a
relation:
\begin{eqnarray}
\Delta\upsilon=-[s_1\frac{GM}{R}\frac{\sqrt{R^2-R_0^2}}{R^2}\tau+s_2\sqrt{(\frac{GM}{R})^3}\frac{R_0}{2R^3}\tau^2],
\end{eqnarray}
where $G$ is the gravitational constant, $M$ the mass of the central
object and $\tau$ the time between two observations. The signs of
two terms ($s_1$ and $s_2$) are determined by the direction of
rotation and the location of the star with respect to the tangent
point. This relation assumes that all stars are in circular orbits.
This assumption is not necessarily true since the stellar orbits are
generally not circular ($e$=0), but tend to have high eccentricities
\citep{bahW76}. Nevertheless, the derived formula helps us gain some
insights about the stellar system we observe. It can be proved that
for each projected distance there is a maximum radial velocity
change allowed by the formula. The maximum velocity change is
plotted as a function of the projected distance in
Figure~\ref{star_vaccl}. The curve is for a central object with a
mass of 3.5$\times$10$^{6}M_{\odot}$ and a time separation of 5.8
years between two velocity measurements. The plots show that at
least one cool star and many hot stars, in our sample, show a
velocity change larger than the maximum value allowed by circular rotations. 
IRS7 shows the biggest velocity change in our plot for cool stars
and the value exceeds the maximum velocity change allowed at its
projected distance from the GC. This was not reported by previous
observations because of relatively low spectral resolutions.
\cite{ottEG99} discussed the magnitude change of IRS7 and found that
the stellar light curve reached a temporal minimum in 1998. These
authors suggested
that IRS7 is a red supergiant on the asymptotic giant branch (AGB). 
Thus the large velocity change we observed may be related to the
stellar surface activity during the AGB phase. Follow-up
observations are needed to investigate such a possibility. Among all
observed hot stars, IRS16W(SW) and IRS16NE stand out with large
velocity changes over the 5.8 year period. This is consistent with
their identities as binary systems. GCHe2 also has a large velocity
change which may be an indication of the spectral variability of the
star. Our data show that the radial velocity change generally is
bigger for hot stars than for cool stars, suggesting that the
motions of the early-type stars deviate greatly from circular
orbital motions.

\subsection{Enclosed Mass}

We try to estimate the enclosed mass in the inner parsec of the GC
using different projected mass estimators. The classical
Bahcall-Tremaine (BT) estimator assumes that the velocities of the
particles in the system are isotropic and all the mass of the system
is in a central point \citep{bahT81}.
It predicts that the mass enclosed within a radius $R_0$ is: %
\begin{eqnarray}
M_{BT}=\frac{16}{{\pi}GN}\sum{\upsilon}_r^2R_P.
\label{enclosedmasseq}
\end{eqnarray}
Where G is the gravitational constant, N the number of stars,
$\upsilon_r$ the line-of-sight velocity, R$_P$ the projected radius
from the center, and the summation is over all stars within the
projected radius, R$_0$. Equally weighted mass estimators are used
in our study because, as pointed out by \cite{bahT81}, theoretically
every detected star carries the same amount of information about the
enclosed mass. The weighted mass estimators implicitly assume some
stars carry more information than others thus the result is biased
by those stars with bigger weights. Such as in \cite{figGKMBM03},
stars with bigger weights are from the overlapped region which was
covered by two scans (GC1 and GC2) in 1999. Since the region covered
by GC2 is much smaller than that covered by GC1 and is asymmetric to
Sgr~A$^*$ (Figure~\ref{skybox}), the calculated enclosed masses are
biased by stars in the smaller region. The resulting enclosed mass
is displayed as a function of the projected radius from the
Sgr~A$^*$ in Figure~\ref{enclosedmass} (left panel). In the
calculation, R$_0$ is the size of a circular region which is
centered at Sgr~A$^*$ and is entirely contained inside the area
covered by the combined data set. The summation only includes the
stars inside the circle to avoid the bias caused by the asymmetry of
the entire observed region. The largest possible R$_0$ in our case
is $\sim$0.4~pc. During the calculation, we use the average velocity
of each star in the sample for $\upsilon_r$ if multiple velocity
measurements are available. We do not include two stars, IRS9 and
-3.03+1.35, in the calculation because of their extremely large
radial velocities ($\upsilon_r{>}$~300~km~s$^{-1}$). The median
velocity ($\sim$~-10~km~s$^{-1}$) of the resulting sample of 76
stars is subtracted from $\upsilon_r$ to remove effects of the
systematic velocity shift before computing the enclosed masses. Our
calculation indicates that the mass enclosed within a projected
radius of 0.4~pc is $\sim$3.3$\pm$0.6$\times$10$^6$M$_{\odot}$. An
alternate version of the BT estimator assumes that the mass is
distributed like the light, but still assumes isotropic motions
\citep{heiTB85}. In this case, exactly twice amount of mass is
needed to bind the system.

When we have no knowledge about the velocity anisotropy of a system,
we can use a more traditional mass estimator based on the Virial
Theorem (VT), which also assumes that the mass of the system is at
the center \citep{bahT81,mer87}, but it does not assume any degree
of velocity anisotropy. The VT mass estimator predicts that the
central BH mass of a spherical system is:
\begin{eqnarray}
M_{VT}=\frac{3\pi}{2G}\frac{{\sum}{\upsilon}_r^2}{{\sum}R_P^{-1}}.
\label{enclosedmasseqvt1}
\end{eqnarray}
Placing all the mass in the center is equivalent to deriving the
minimum mass that can bind the system \citep{mer87}. Therefore, the
VT mass estimator puts a lower limit on the BH mass. An alternative
form of the VT mass estimator is \citep{limM60}:
\begin{eqnarray}
M^{\prime}_{VT}=\frac{3{\pi}N}{2G}\frac{{\sum}{\upsilon}_r^2}{{\sum}R_{ij}^{-1}},
(i<j). \label{enclosedmasseqvt2}
\end{eqnarray}
Here $R_{ij}$ is the projected distance between individual stars in
the region. This mass estimator assumes that the mass distributes in
the same way as stars in the system. It requires more mass to bond a
system if the mass distributes evenly, thus $M^{\prime}_{VT}$
equivalently puts an upper limit on the estimated BH mass. The
results from these two equations are shown in
Figure~\ref{enclosedmass} (right panel) with the filled circles for
the case with all mass in the central point and the open circles for
the case in which mass follows light. The ratios between two sets of
estimated masses range between $\sim$2.0 and $\sim$3.0. The best
estimate would be a value between
2.6$\pm$0.4$\times$10$^6$~M$_\odot$ and
5.8$\pm$1.7$\times$10$^6$~M$_\odot$, which is consistent the result
using the classical BT estimator. Using these values, we estimate
that the enclosed black mass within 0.4~pc of Sgr~A$^*$ is
3.9$\pm$1.1$\times$10$^6$~M$_\odot$. This value is consistent with
previous results using a single star's 3D orbit
\citep[3.7$\pm$1.5$\times$10$^6$~M$_\odot$][]{schOGHLE02,gheDMHT03,gheSHTL05}
or using the proper motions of late type stars
\citep[3.0$\pm$0.5$\times$10$^6$~M$_\odot$][]{genPEGO00}, so the BT
estimator gives a good estimate after all. We discard the value
estimated for the case of uniformly distributed mass (the alternate
version of the BT estimator) because the value is outside of the
range set by the Virial Theorem.

\section{Discussion}

\subsection{Projected mass estimators and stellar velocity distributions}
We estimate the mass of the central object using different projected
mass estimators. Although our estimate is consistent with the values
using the stellar orbits \citep{schOGHLE02,gheSHTL05}, we should be
aware of the weakness of such a statistical method. Projected mass
estimators largely rely on the theoretical work which usually
assumes a certain degree of anisotropy. In the Virial Theorem case,
the estimators are derived from systems which are well sampled in
both space and phase. Observations are inevitably incomplete in this
sense, particularly in the GC case, where observations are limited
by the size of sky coverage and non-uniform extinction across the
field of observations. The observed stars will not make a complete
sample of the dynamical system. This under-sampling will result in
an under-estimate of the enclosed mass.

The velocity distribution of a stellar population contains
information about how the corresponding sample evolves under
gravitational influences of the surrounding environment or whether
the population is dynamically relaxed. The velocity distributions of
late and early type stars in the inner parsec of the GC both are
approximately symmetric in shape (skewness$\simeq$0.0), but they
both deviate from a Gaussian distribution. Although we expect that
the late type sample has a more relaxed velocity distribution
because of the comparable lifetimes and relaxation times of these
stars, even relaxed populations can have non-Gaussian velocity
distributions, particularly in the vicinity of a black hole
\citep[e.g.][]{van94}. Moreover, previous studies suggest that
multiple populations of late type stars are present in the observed
region \citep[][and references therein]{genSRE03}. The overall
velocity distribution of the late type stars may not be a consistent
one. Depending on the nature of the central ``hole'' of late type
stars, the cluster of late type stars in the GC may deviate from a
spheroid and the observed radial velocity distribution deviates from
the real distribution.

\subsection{Spectral types of GC hot stars}
We show that stellar He~I features at 2.112/2.113~$\mu$m and
2.0581~$\mu$m can be used to divide hot He~I stars in the GC into
two groups: narrow-line and broad-line. This method is more accurate
than the one using only the 2.0581~$\mu$m feature because the
absorption feature at 2.112~$\mu$m is less affected by emission from
diffuse interstellar gas. However, we are not very clear about the
connection between this morphological classification and the
spectral types of He~I stars. Many works have been devoted to
identifying the spectral types of GC He~I stars
\citep{hanC94,hanCR96,morEHCB96,tamRHCMR96,figMN97,najKGLKH97,pauMMR01,pauGMNB06}.
These studies show that stars of a large variety of spectral types
emit He~I~2.058~$\mu$m line. These spectral types include WC, late
WN, Ofpe/WN9, LBV, Oe/Be. Despite the complexity of the situation, a
correlation seems to exist between the spectral types and line
widths of He~I stars. \cite{pauMMR01} suggested that GC He~I stars
can be loosely identified as LBV (narrow line objects) or WR (broad
line objects) based on their spectral morphologies and K band
magnitudes. Table~\ref{hotspty} lists spectral types of some He~I
stars from literatures \citep{pauGMNB06} and the table clearly shows
a tendency of identifying narrow line objects to be Ofpe/WN9 stars
and broad line objects to be late type Wolf-Rayet stars. The
Ofpe/WN9 and LBV stages are considered to be close to each other on
the evolutionary track of massive stars because stellar spectra at
these two stages have many similarities \citep{marGHEPG07}. Such a
tendency is consistent with evolutionary models of massive stars in
which stars lose a large amount of mass during the early stages of
evolution. More evolved stars tend to have higher surface gravity
and stronger radiation field because of their smaller sizes and high
surface temperatures due to exposed helium-burning cores. Such
models predict that massive stars of later types generally have
relatively faster winds. However, this simple solution does not
stand in some cases. Our spectrum of AF shows strong and broad
emission in multiple lines, but it is identified as an Ofpe/WN9 by
\cite{pauGMNB06}. Similar discrepancy also happens in the case of
ID415. It seems that both stages (Ofpe/WN9 and WC9) can have either
fast ($>$1000~km~s$^{-1}$) or slow ($<$1000~km~s$^{-1}$) winds.
Further studies are needed to better constrain the spectral types of
these stars.

\subsection{Origin of GC hot stars}
The presence of a large number of He~I stars in the GC poses a
challenge to understanding the origin of these stars and the star
formation history in the GC. The correlation between the radial
velocities of the hot stars and their projected distances
(Declination offsets) from Sgr~A$^*$ supports that all GC He~I stars
belong to one or two disk like structures which rotate about the
gravitational center. Previous studies of stellar dynamics at the GC
suggest that hot stars belong to two counter-rotating disks in two
planes that are at a large angle from each other \citep{pauGMNB06}.
New analysis indicates that there is only one disk \cite{luGHMBM07}.
Our observations do not allow us to make conclusive remarks on where
one or two such disks are present. Such conclusions need additional
information, such as proper motion velocities, about the stars.

In any case, the similar ages of the two types of stars suggest that
hot stars have formed in one or two star formation events
$\sim$6~Myr ago
\citep{genTKKT96,genPEGO00,pauMMR01,pauGMNB06,luGHMBM07}.
\cite{pauMMR01} suggest that the stars formed from the same disk of
gas and dust around the dynamical center. Due to tidal forces or/and
other unknown effects, radial structures developed in the disk
during the star formation event with the most massive stars in the
outer ring, less massive stars in the middle cluster. This scenario
can explain why all these stars are roughly along the same diagonal
in Figure~\ref{hotstar_disk} (right) and have similar ages. The
model is also consistent with the spatial distribution of broad line
objects and narrow line objects. However, it cannot explain why
planes of two disks form a big angle between them as shown by
\cite{pauGMNB06} since we would expect two disks to have similar
orientations if these structures have formed from the same disk of
gas and dust. Alternatively, stars could form from the different
dynamical structures of gas and dust, such as multiple molecular
clouds originally orbiting the GC. This model can avoid difficulties
associated with differences between two stellar disks, but it needs
to explain why stars were forming at about the same time in multiple
structures. Another competitive formation scenario is the
inspiraling cluster model in which stars are formed elsewhere
outside the central region of the GC and migrate inward due to
dynamical friction. Such a model has been proved to be unlikely
because the model predicted surface density and total mass of the
disk differ significantly from observations \citep{pauGMNB06}. It is
also very unlikely that two separate cluster inspiraling events
happen at the same time.

\section{Summary}
We report slit scan observations toward the GC in the two epochs
separated by $\sim$5.8 years. We identify stars based on
spectroscopic features at 2.2395~$\mu$m CO band heads,
2.112/2.112~$\mu$m He~I line and 2.058~$\mu$m He~I line. As a
result, 123 cool, late type (mostly red giants), 20 narrow line and
9 broad line early type He~I stars are detected in our combined data
set. We measure the radial velocities of these stars in the two
epochs and their velocity changes between the two epochs. The
velocity dispersion and the surface density distribution of the late
type stars support that a cavity of low stellar density is present
at the center of the system, which has been suggested by previous IR
photometries and model simulations. Using projected mass estimators,
we estimate that the mass of the central black object is
$\sim$3.9$\pm$1.1$\times$10$^{6}$~M$_{\odot}$. No systematic
rotation about the central black hole is indicated by our
observations for the late type stars. A systematic rotation exists
for the early type stars with the stars north to Sgr~A$^*$
relatively blue-shifted and the stars south to Sgr~A$^*$ relatively
red-shifted. The measured velocity changes of many early type stars
exceed the maximum values allowed by edge-on circular orbital
motions around a central object with
3.5$\times$10$^{6}$~M$_{\odot}$, suggesting that these stars may
move in high eccentricity orbits.

\acknowledgments The material in this paper is based on work
supported by NASA under award NNG 05-GC37G, through the Long Term
Space Astrophysics program. F. N. acknowledges PNAYA 2003-02785-E
and AYA 2004-08271-C02-02 grants and the Ramon y Cajal program. This
research was performed in the Rochester Imaging Detector Laboratory
with support from a NYSTAR Faculty Development Program grant. We
thank Alan Johnson, Christine Trombley, and Daniel Smialek for
proof-reading our paper and giving us suggestions.




\appendix

\section{The radial velocity change for a circular orbit system}

A star of mass $m$ moves at a speed $\upsilon$ in an edge-on
circular orbit with a radius $R$ around a central object of mass
$M$. We have:
\begin{eqnarray}
\upsilon=\sqrt{\frac{GM}{R}}
\end{eqnarray}
If the projected distance is $R_0$, we have:
\begin{eqnarray}
cos(\phi)&=&\frac{R_0}{R}, \qquad sin(\phi)=\frac{L}{R}.
\end{eqnarray}
Where $L=\sqrt{R^2-R_0^2}$ is the distance between the star's
location to the tangent point. The radial velocity of the star is:
\begin{eqnarray}
\upsilon_r={\upsilon}cos(\phi).
\end{eqnarray}
After a small time span $\tau$, the star moves $\Delta\phi$ in the
orbit. We have:
\begin{eqnarray}
\upsilon_r^\prime={\upsilon}cos(\phi+\Delta\phi).
\end{eqnarray}
with
\begin{eqnarray}
\Delta\phi=\frac{\upsilon\tau}{R}=\sqrt{\frac{GM}{R^3}}{\tau}.
\end{eqnarray}

\begin{figure}[!ht]
\epsscale{0.6} \plotone{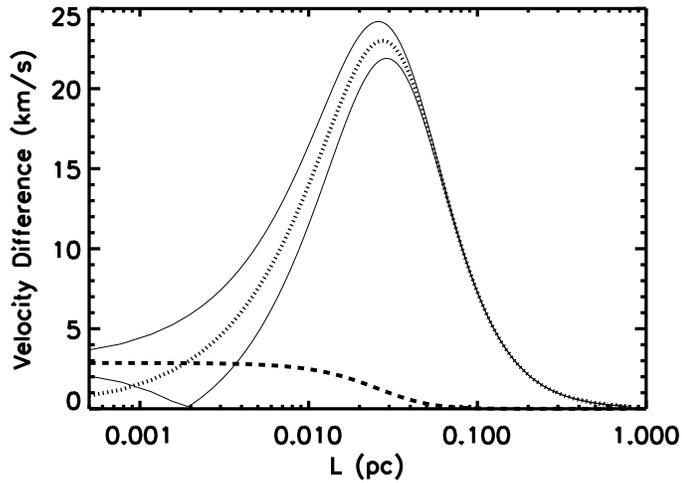} \caption{Absolute stellar radial
velocity change, over a period of 5.8 years, as a function of the
star's line-of-sight distance ($L$) from the tangent point of an
edge-on circular orbit around the galactic center. Values are
calculated using Eq.~\ref{eq_acce}. Absolute values of two terms in
the equation are shown with the dashed and dotted lines,
respectively. The absolute values of the sum and difference of the
two terms are shown with solid lines. The mass of the central black
hole is assumed to be 3.5${\times}10^{6}M_{\odot}$ and the radius of
the orbit (R$_0$) is 1~arcsec at the distance of 8~kpc. Note that
the value of the first term reaches its maximum when
L$^2$=$\frac{1}{2}$R$^2_0$.\label{vdifflos}}
\end{figure}

The change of the radial velocity is:
\begin{eqnarray}
\label{eq_acce}
\Delta\upsilon_{r}(R_0, L)&=&\upsilon_{r}^\prime-\upsilon_{r}={\upsilon}[cos(\phi+\Delta\phi)-cos(\phi)]\\
&\simeq&-\upsilon[sin(\phi)\Delta\phi+\frac{1}{2}cos(\phi)\Delta\phi^2]\\
&=&-[s_1\frac{GM}{R}\frac{\sqrt{R^2-R_0^2}}{R^2}\tau+s_2\sqrt{(\frac{GM}{R})^3}\frac{R_0}{2R^3}\tau^2].
\end{eqnarray}
Since all $\phi$, $\Delta\phi$ and $\upsilon$ can be positive or
negative, two terms in Eq.~\ref{eq_acce} can be either positive or
negative. Thus $s_1$ and $s_2$ are used to represent the signs of
two terms. The absolute change of the radial velocity will be the
absolute value of either the sum or the difference of two terms.
Note that the first term dominates the overall change of the radial
velocity except for at the tangent point, where the second term
dominates. The results are shown in Figure~\ref{vdifflos}.

Although we only derived the projected velocity change for a case of
an edge-on circular orbit, the same maximum velocity change is also
valid when we include orbit inclinations. Because the radius of an
actual orbit must be bigger than the observed projected distance,
the projected radial velocity of an inclined orbit must be smaller
than that of an edge-on orbit when other orbital parameters stay the
same. If we consider cases of a constant inclination angle
($\theta$), we will have similar formula. The only difference will
be a cosine factor (cos$\theta$) in the velocity terms. Therefore,
we will have a new curve which is under the curve of the edge-on
case, although the peak of the new curve will be at different
positions for different inclination angles.

\newpage



\newpage

\clearpage
\begin{figure}
\epsscale{0.8} \plotone{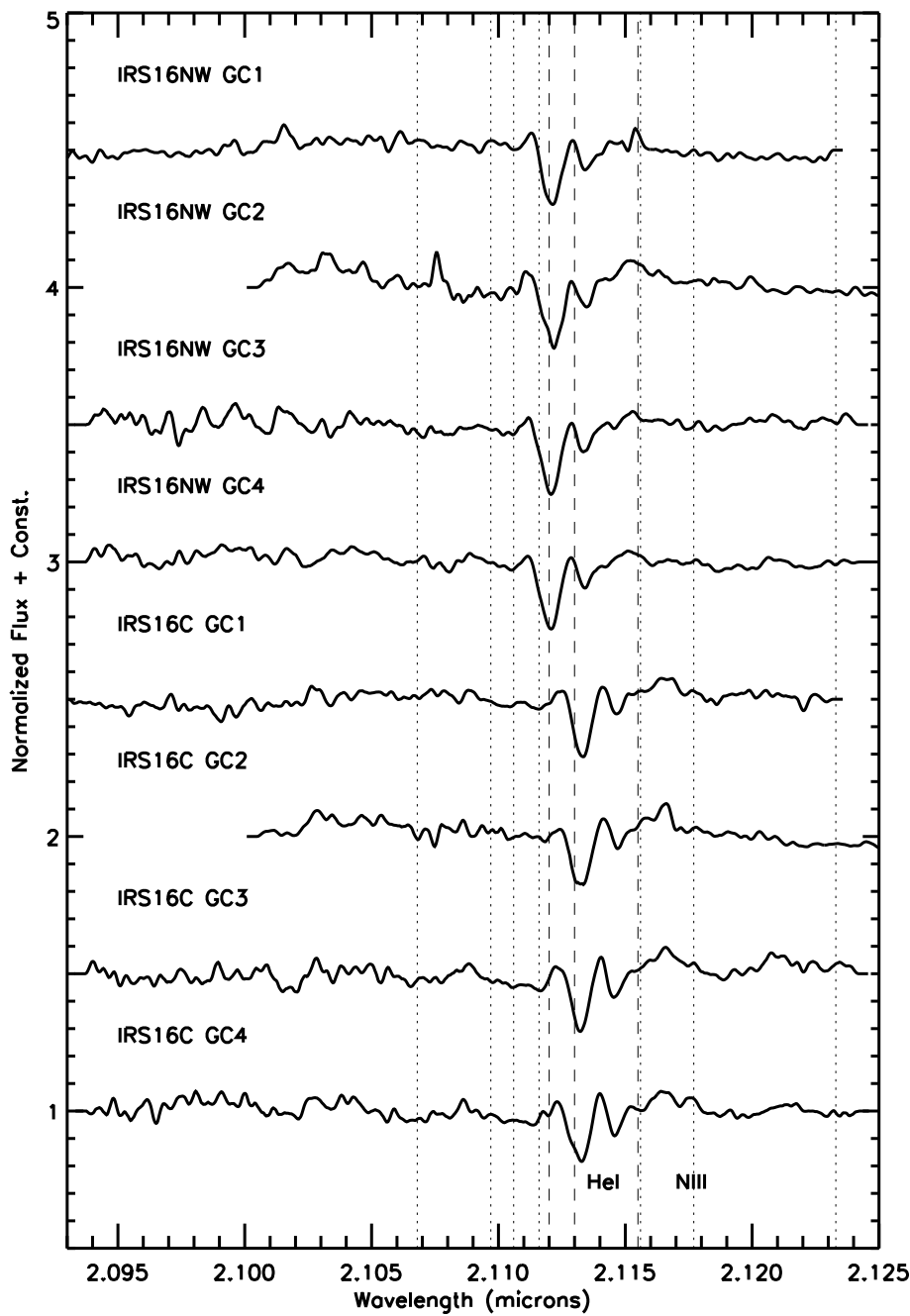} \caption{Spectra of He~I stars
showing the He~I~2.112/2.113$\mu$m absorption doublet. Each spectrum
is indicated with the name of the object followed by the name of the
slit scan (GC1: June 4th 1999, GC2: July 4th 1999, GC3: April 13th
2005, GC4: April 14th 2005). Vertical dashed lines in the plots
indicate the rest wavelengths of the photospherical lines and dotted
lines indicate the rest wavelengths of telluric OH lines. The
coadded spectra have been properly shifted to compensate the
barycentric motions of the target star in each data set.
\label{hotstar_spec2}}
\end{figure}

\addtocounter{figure}{-1}
\begin{figure}
\epsscale{0.8} \plotone{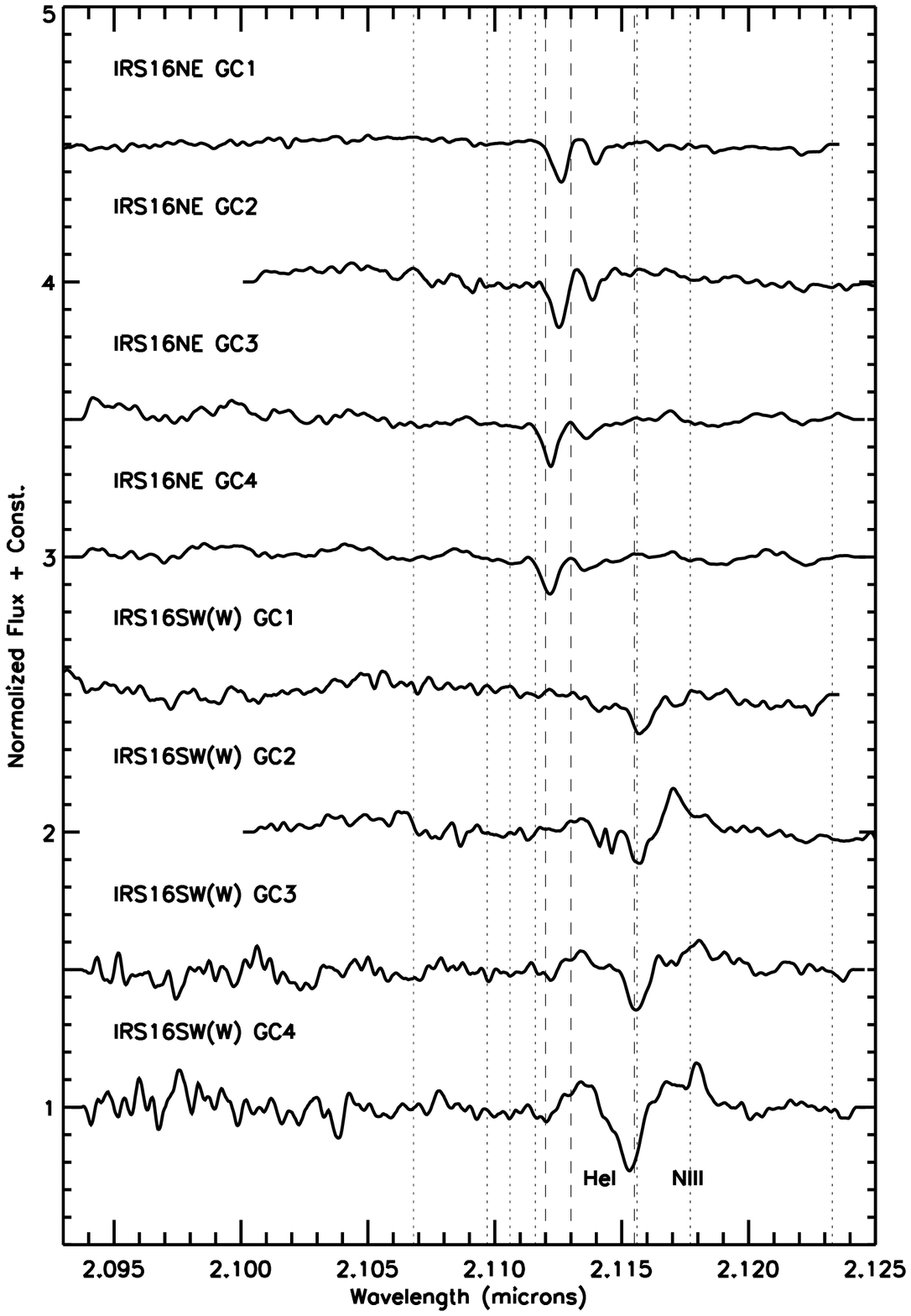} \caption{continued
\label{hotstar_spec22}}
\end{figure}

\addtocounter{figure}{-1}
\begin{figure}
\epsscale{0.8} \plotone{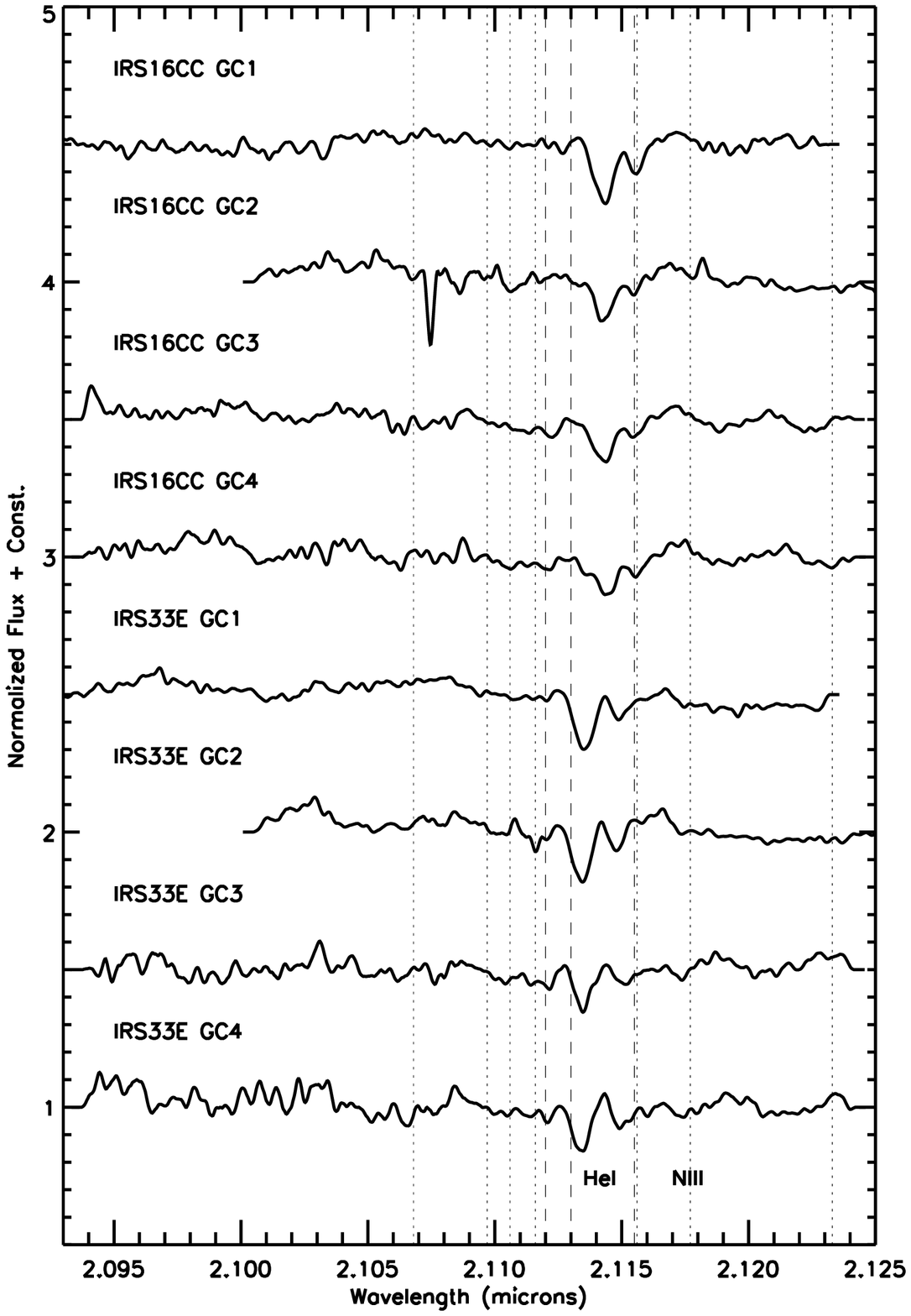} \caption{continued
\label{hotstar_spec23}}
\end{figure}

\addtocounter{figure}{-1}
\begin{figure}
\epsscale{0.8} \plotone{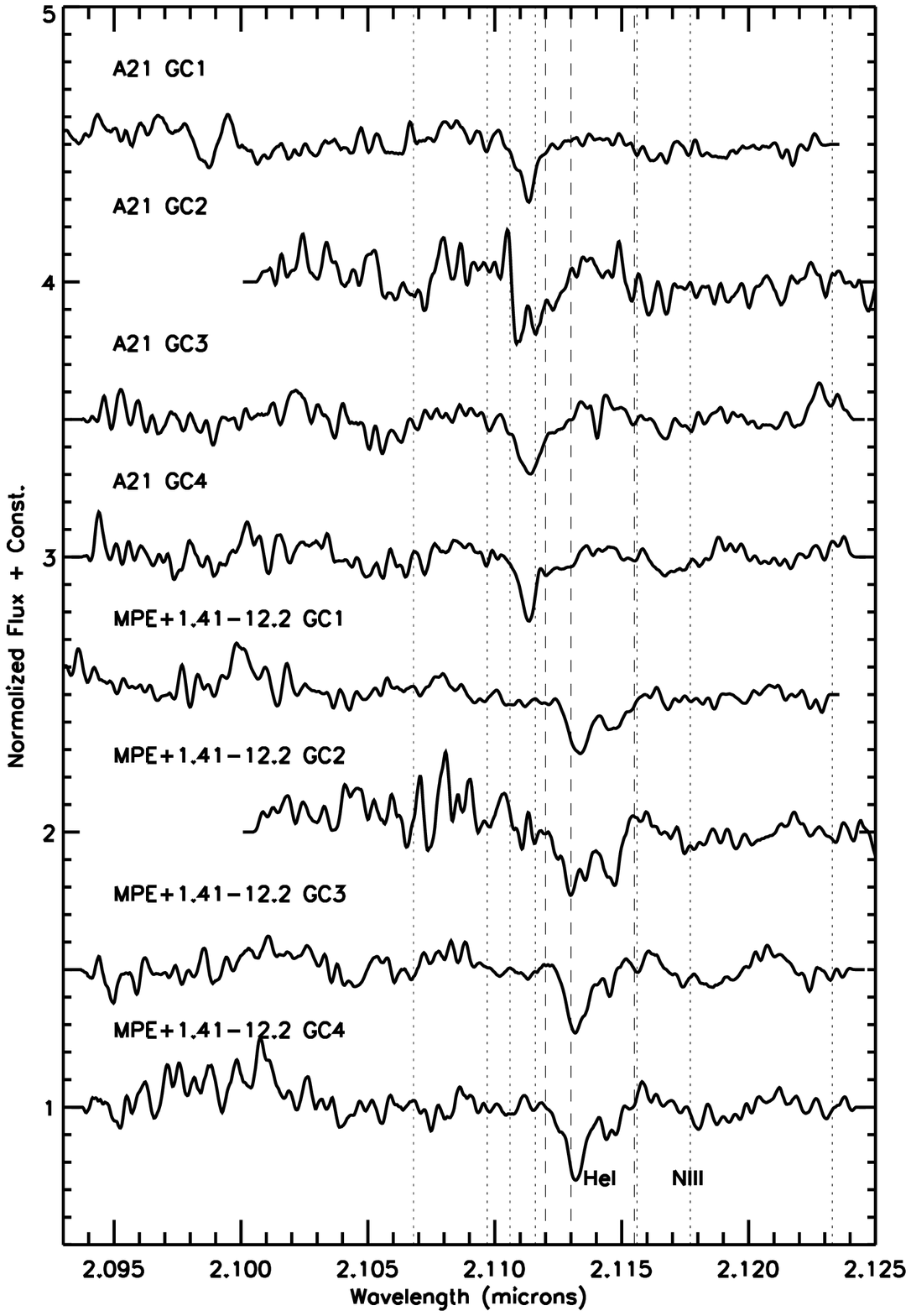} \caption{continued
\label{hotstar_spec24}}
\end{figure}

\addtocounter{figure}{-1}
\begin{figure}
\epsscale{0.8} \plotone{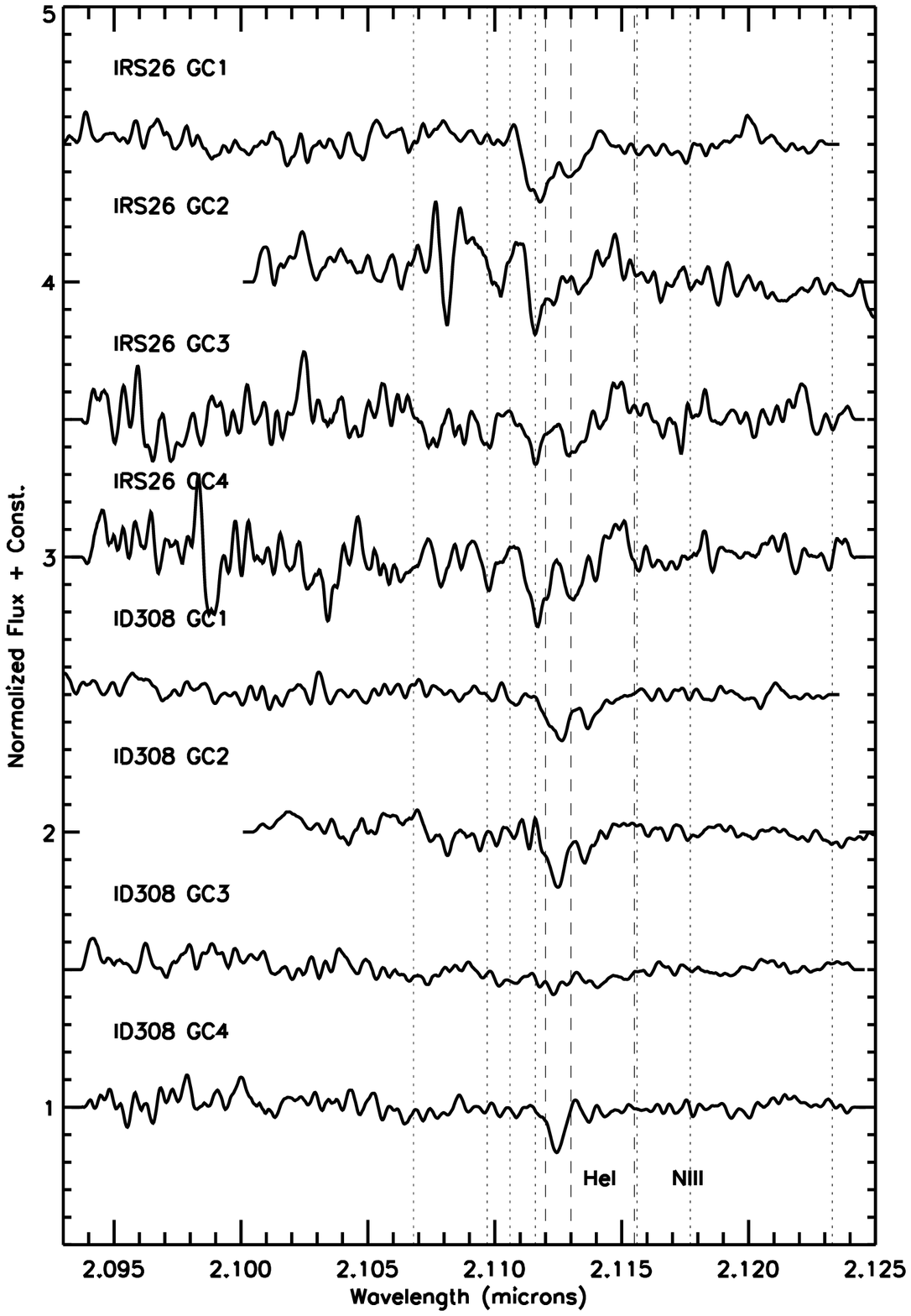} \caption{continued
\label{hotstar_spec25}}
\end{figure}

\addtocounter{figure}{-1}
\begin{figure}
\epsscale{0.8} \plotone{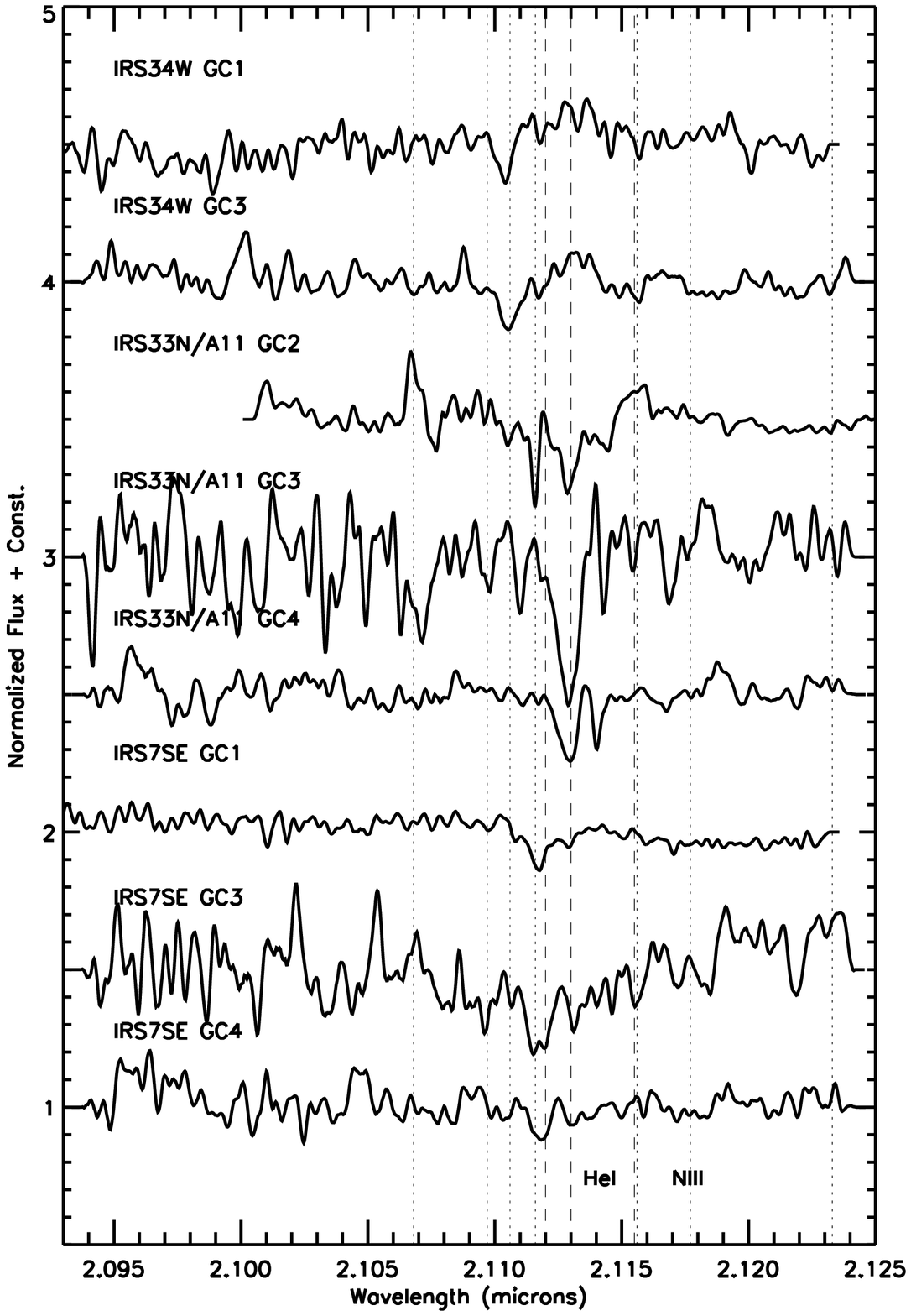} \caption{continued
\label{hotstar_spec26}}
\end{figure}

\addtocounter{figure}{-1}
\begin{figure}
\epsscale{0.8} \plotone{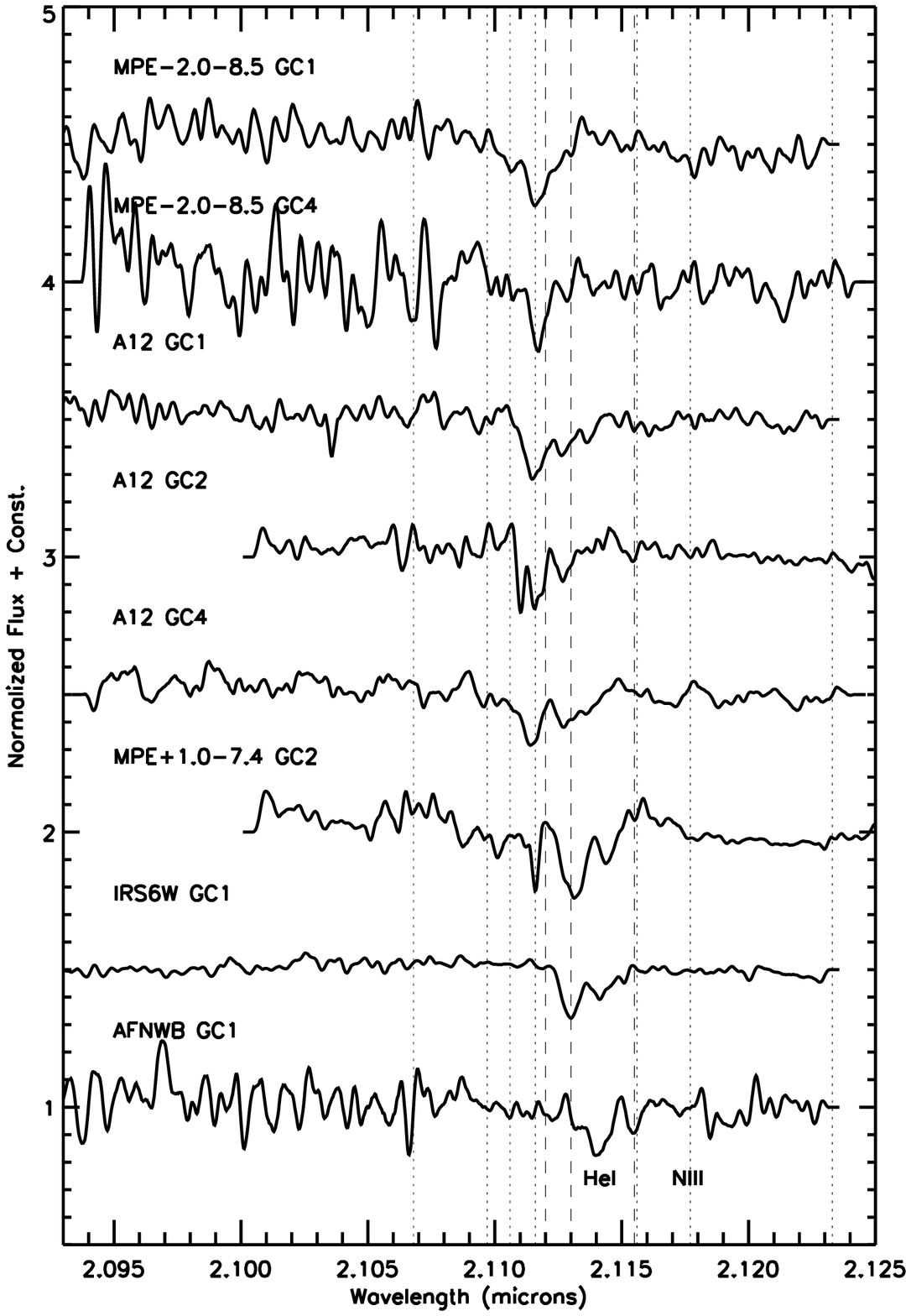} \caption{continued
\label{hotstar_spec27}}
\end{figure}

\addtocounter{figure}{-1}
\begin{figure}
\epsscale{0.8} \plotone{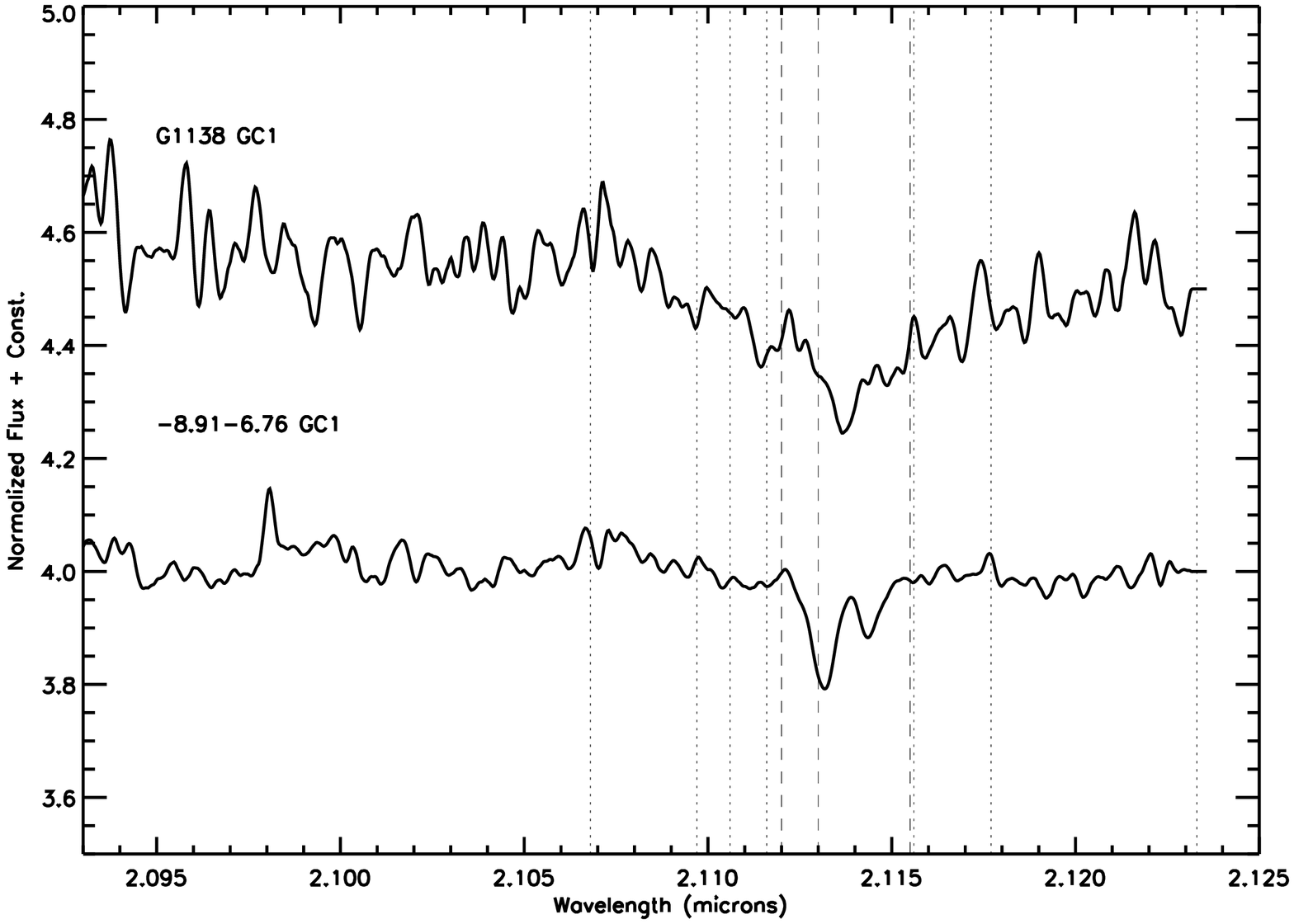} \caption{continued
\label{hotstar_spec2last}}
\end{figure}

\begin{figure}
\epsscale{1.0} \plotone{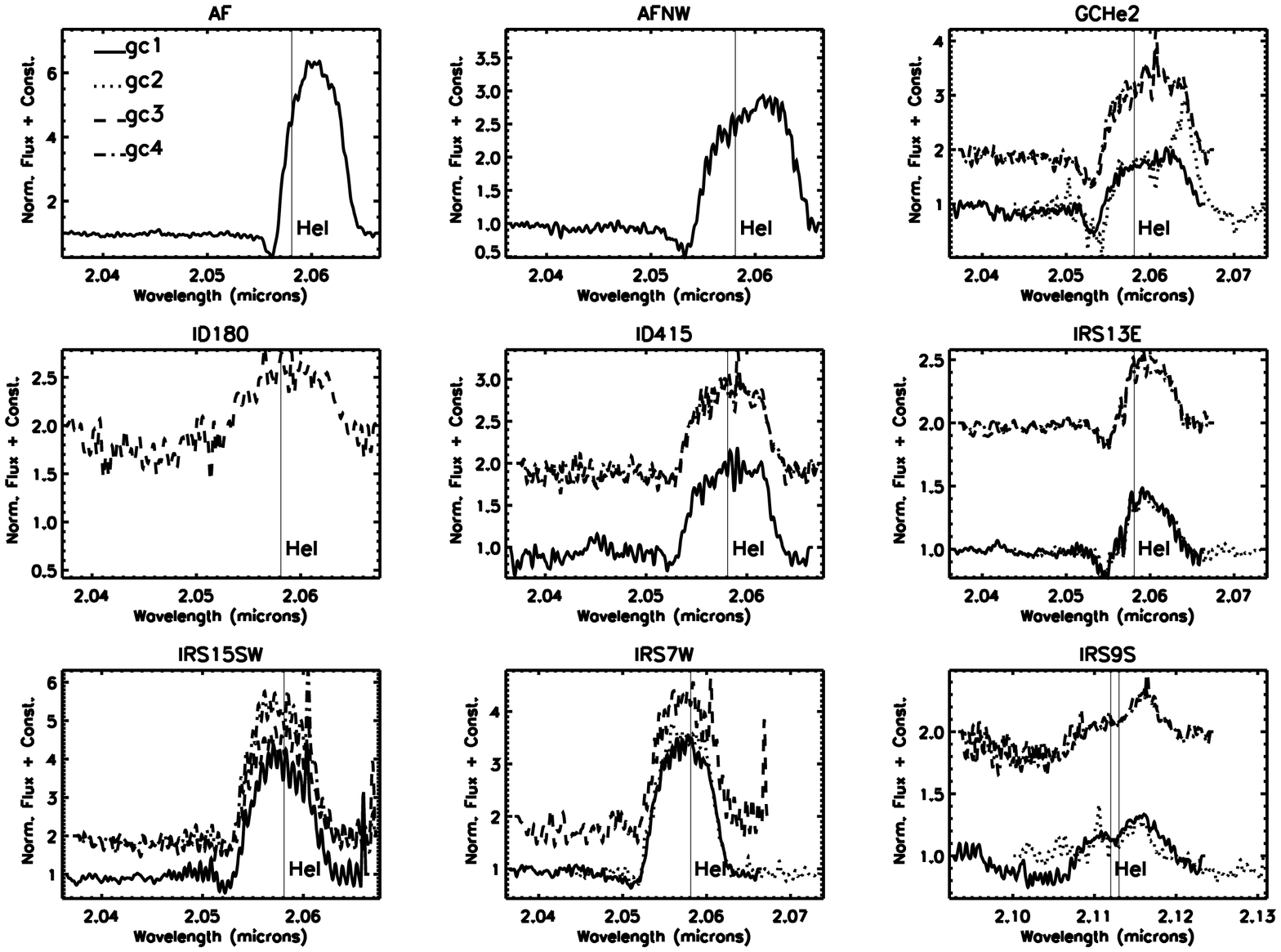} \caption{Spectra of He~I stars
showing the He~I~2.058$\mu$m broad emission line. Spectra of
different scans are plotted with different line types to avoid
confusion. Spectra from two slit scans in 2005 are offset positively
so that they are well separated from the spectra from two scans in
1999. The corresponding line styles are indicated by the names of
slit scans at the upper left corner of the first panel. The vertical
solid lines in the plots indicate the rest wavelengths of the He~I
lines. The coadded spectra have been properly shifted to compensate
the barycentric motions of the target star in each data set. Strong
repeating absorption features (zigzags) present in many spectra are
the results of incomplete removal of telluric CO2 lines
(Figure~\ref{atmo_order}). Imperfect background subtraction leaves
residual nebula emission in some spectra, such as in that of GCHe2
from GC2. \label{hotstar_spec3}}
\end{figure}

\begin{figure}
\epsscale{1.0} \plottwo{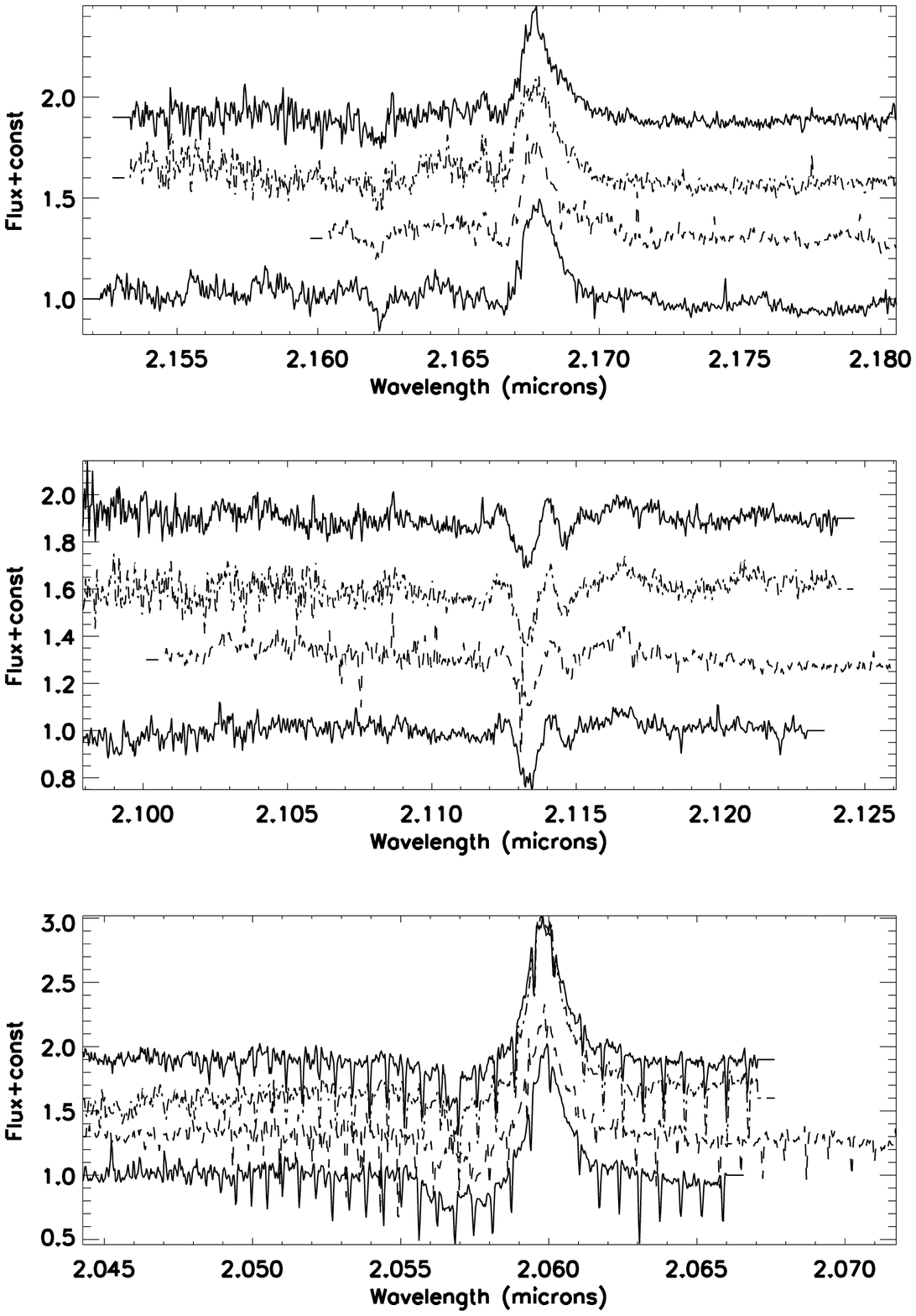}{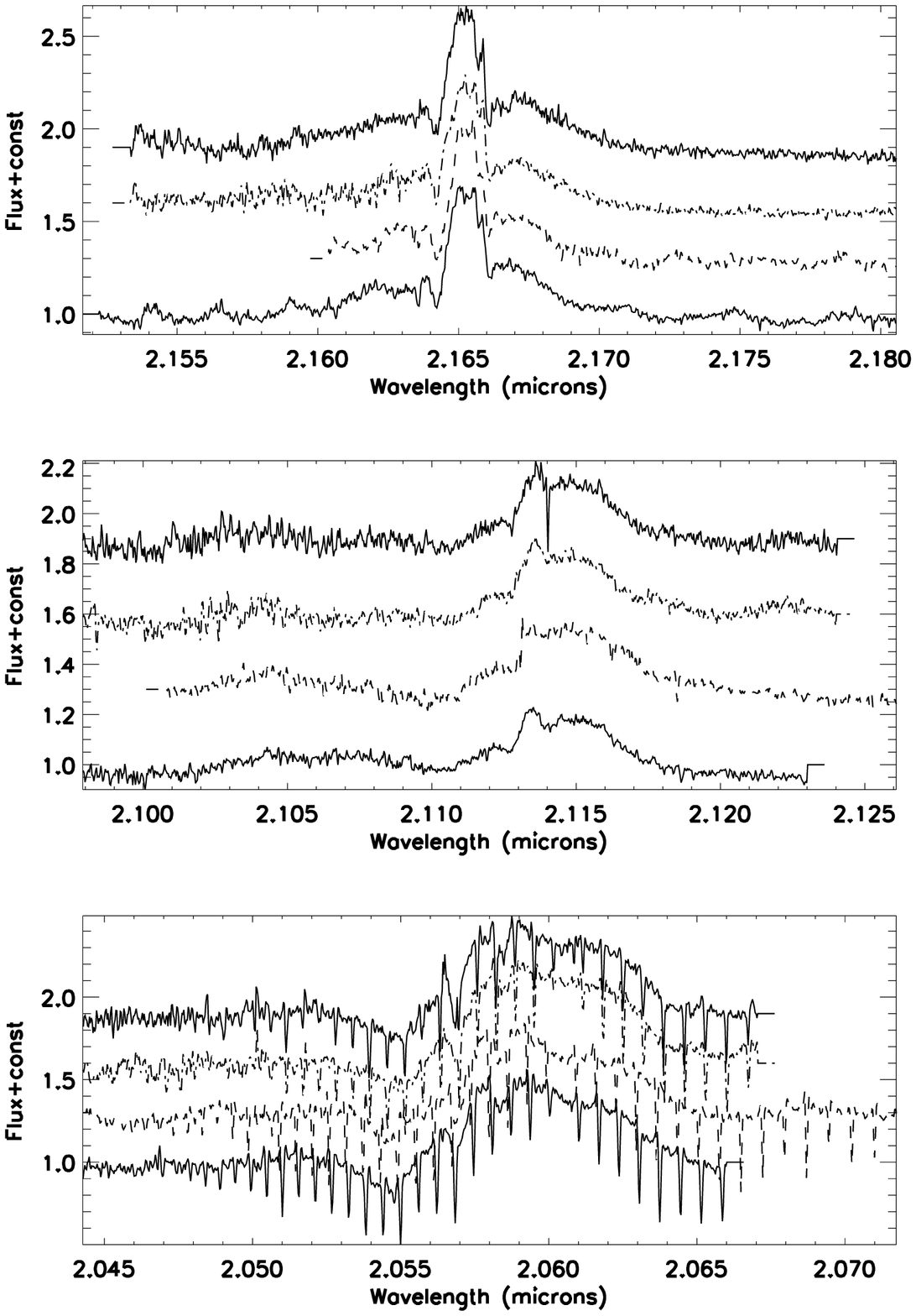} \caption{Examples of
multiple night spectra of He~I stars. Panels on the left are for
IRS16C, a typical narrow line object. Panels on the right are for
IRS13E, which is a broad line He~I star. Three panels on each side
show spectra centered at Br$\gamma$, He~I~2.112~$\mu$m and
He~I~2.058~$\mu$m lines, respectively. In each panel, spectra from
GC1, GC2, GC3 and GC4 are plotted from the bottom to the top. These
figures demonstrate that the shapes of these lines do not change
with time. Note that strong repeating absorption features seen in
the spectra in the bottom panels are the results of incomplete
removal of telluric CO2 lines. \label{hotspec4}}
\end{figure}

\begin{figure}
\epsscale{1.0} \plottwo{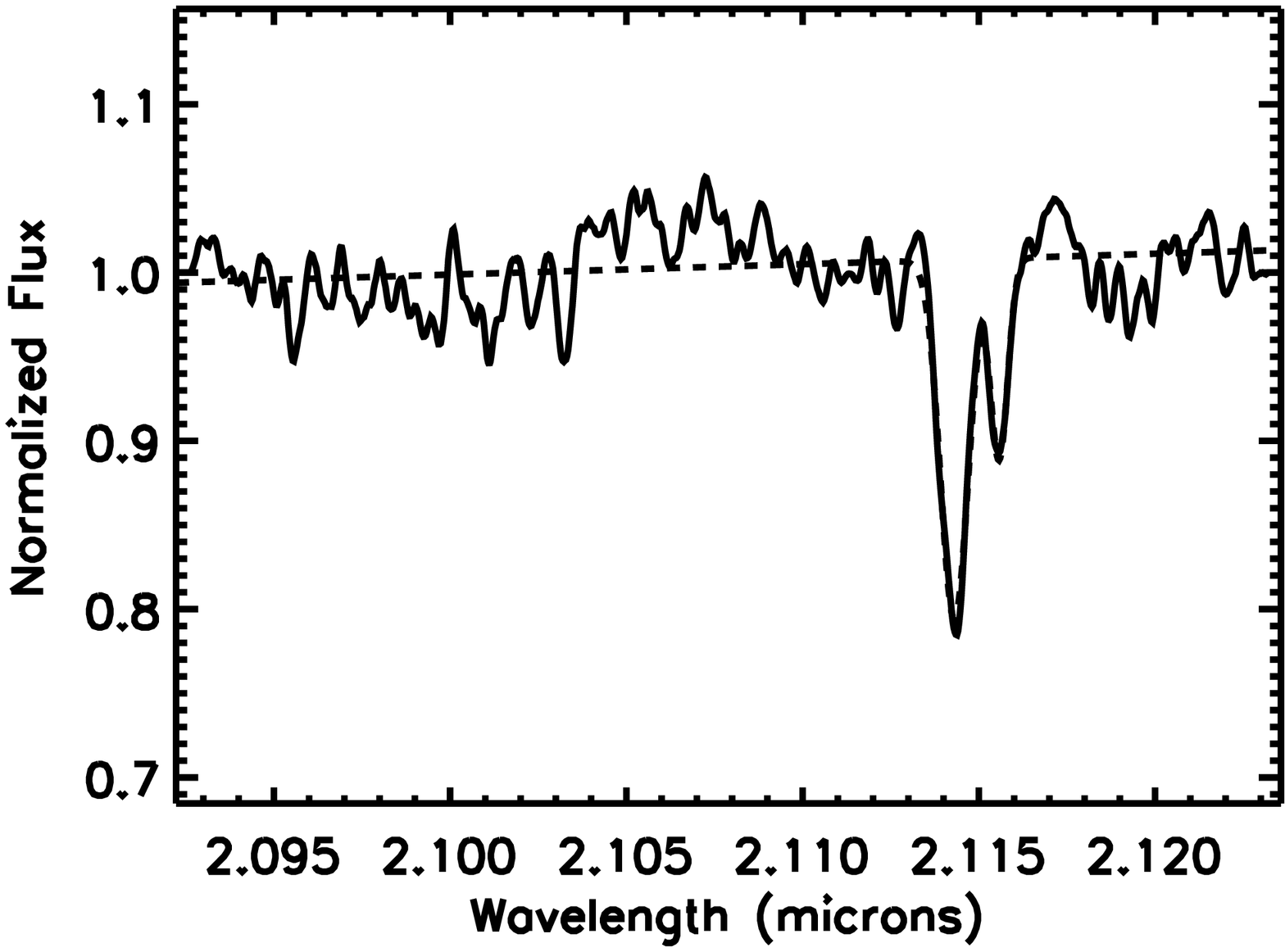}{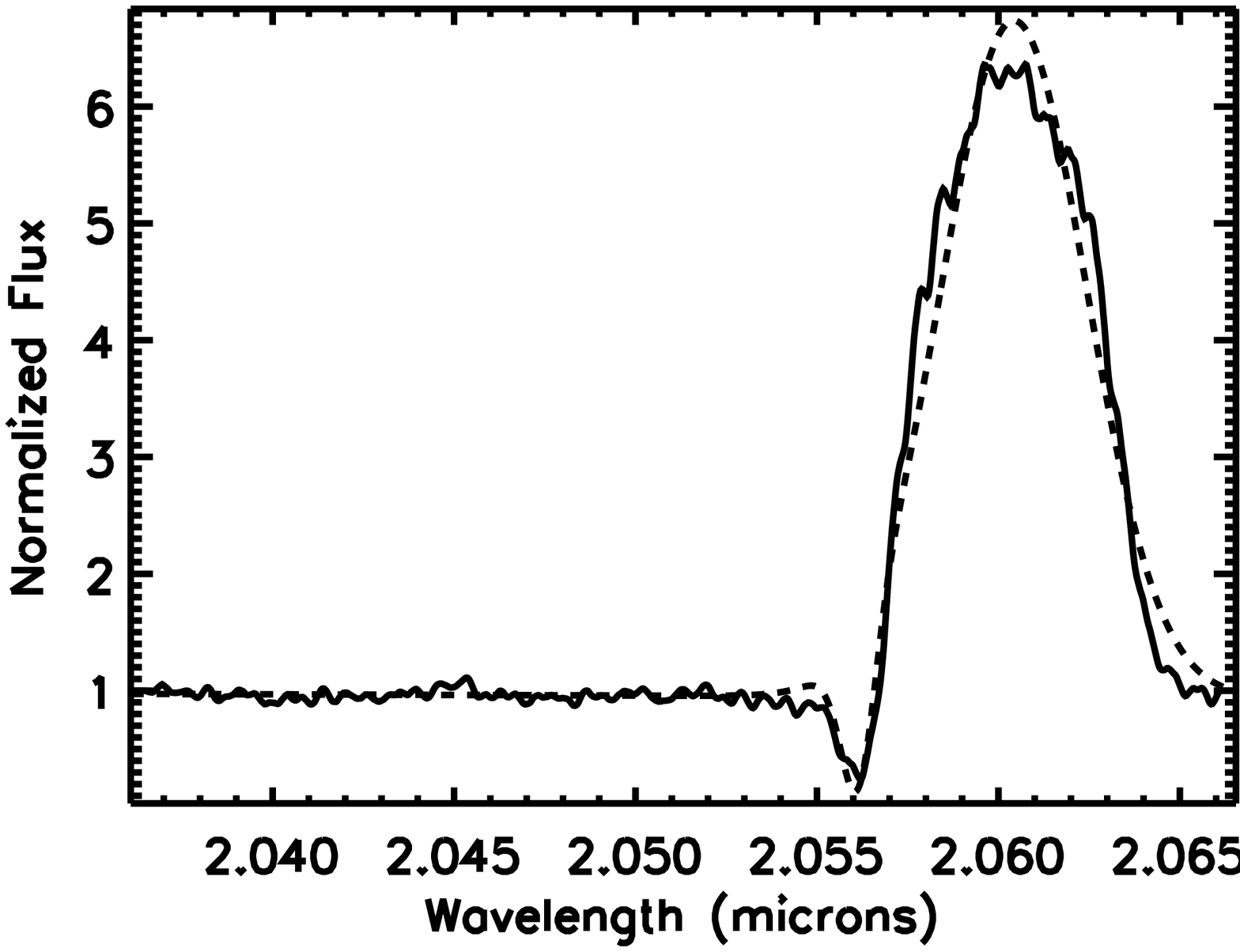} \caption{Examples of
Gaussian fitting to the spectral lines of He~I stars. The left panel
shows an example of fitting the absorption doublet from a
narrow-line object, IRS16CC. The right panel shows an example of
fitting the P-Cygni profile emission line of a broad-line star, AF.
Both spectra are acquired in GC1. In each plot, the continuum is
fitted with a second degree polynomial. \label{specfit}}
\end{figure}

\begin{figure}
\epsscale{1.0} \plottwo{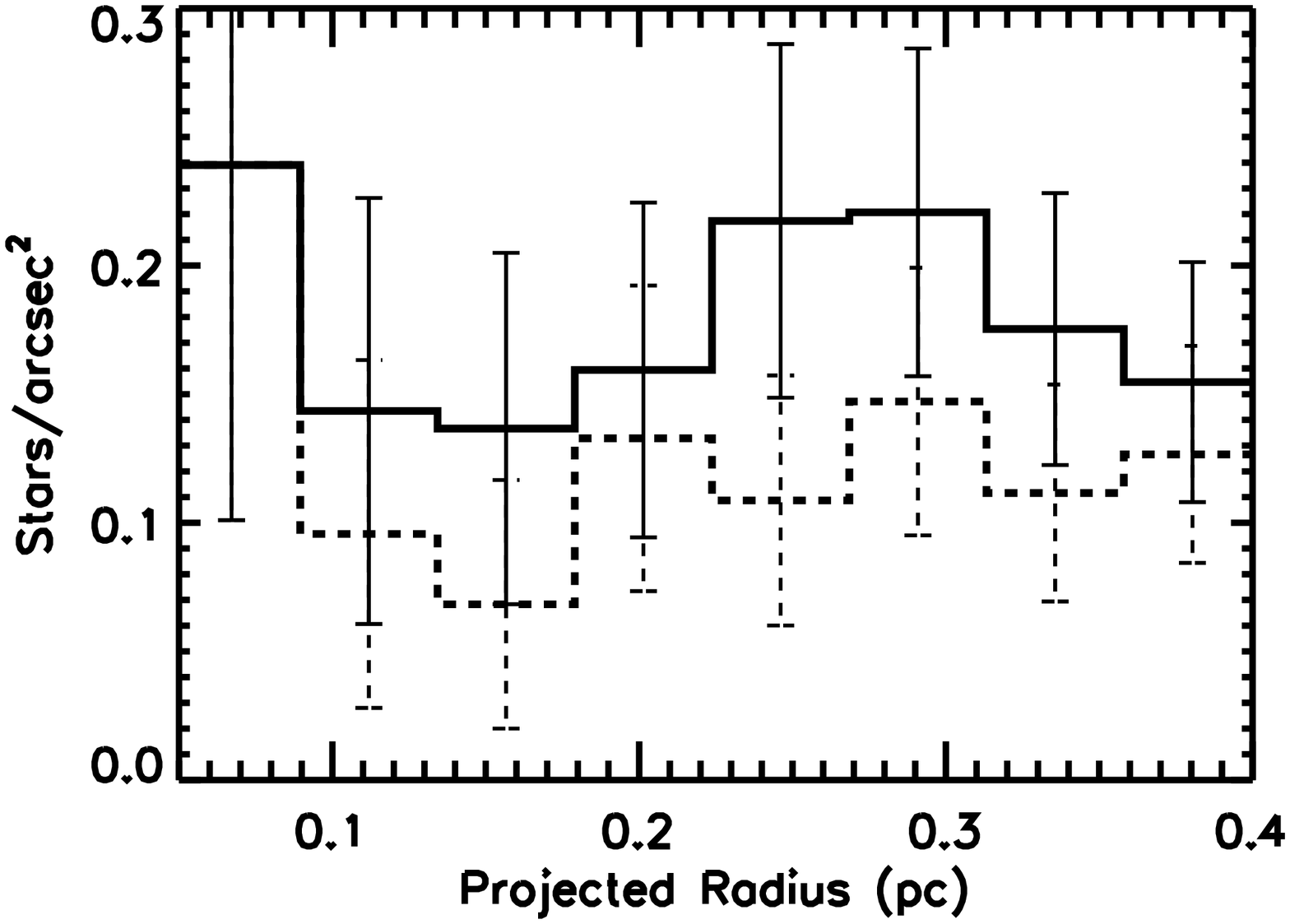}{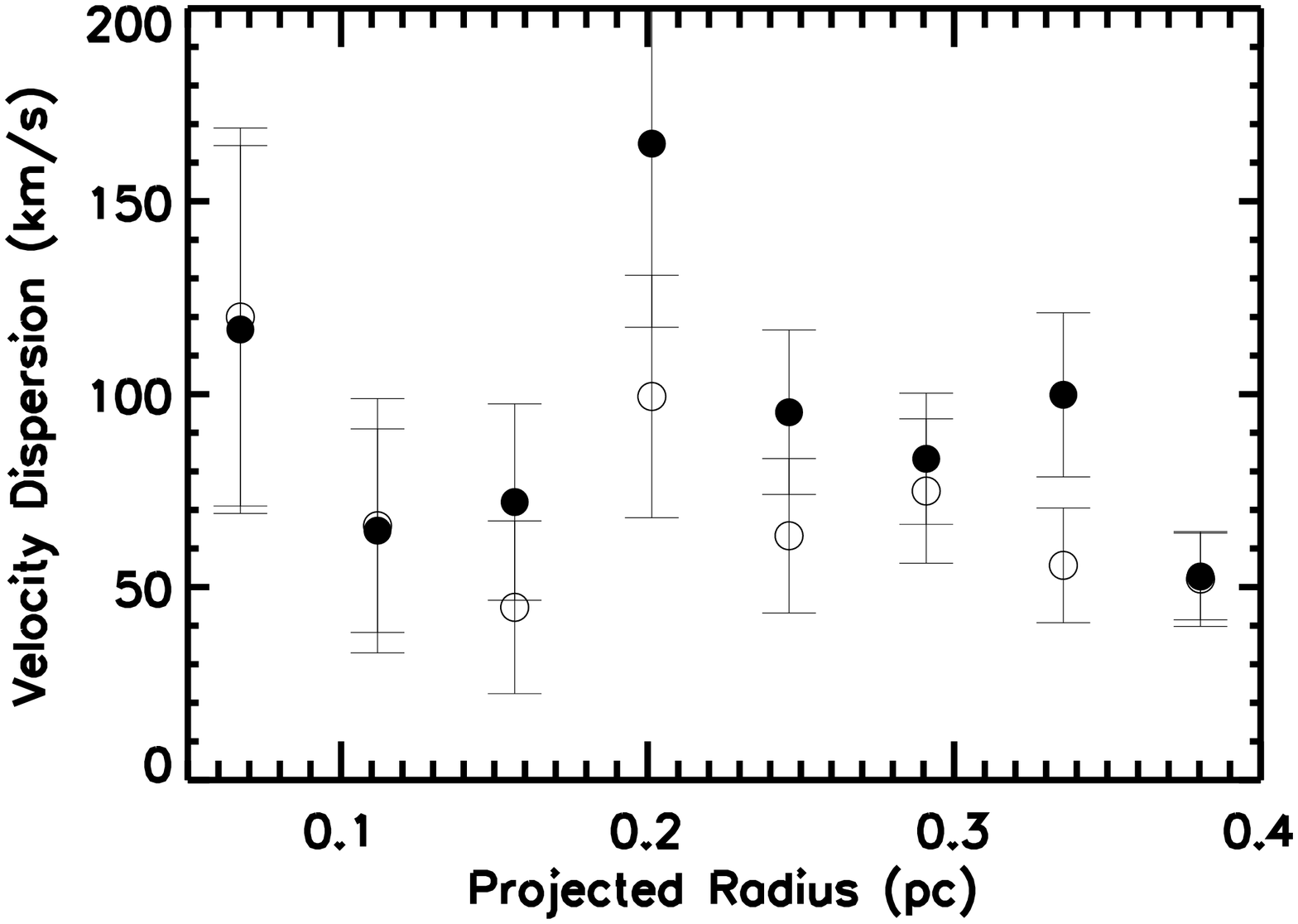} \caption{Surface number
density (left) and radial velocity dispersion (right) of late type
stars as functions of the projected distance from Sgr~A$^*$. Solid
lines (filled circles) are for the current data and the dashed lines
(open circles) are for the published data in \cite{figGKMBM03}.
\label{annuluscool}}
\end{figure}

\begin{figure}
\epsscale{1.0} \plottwo{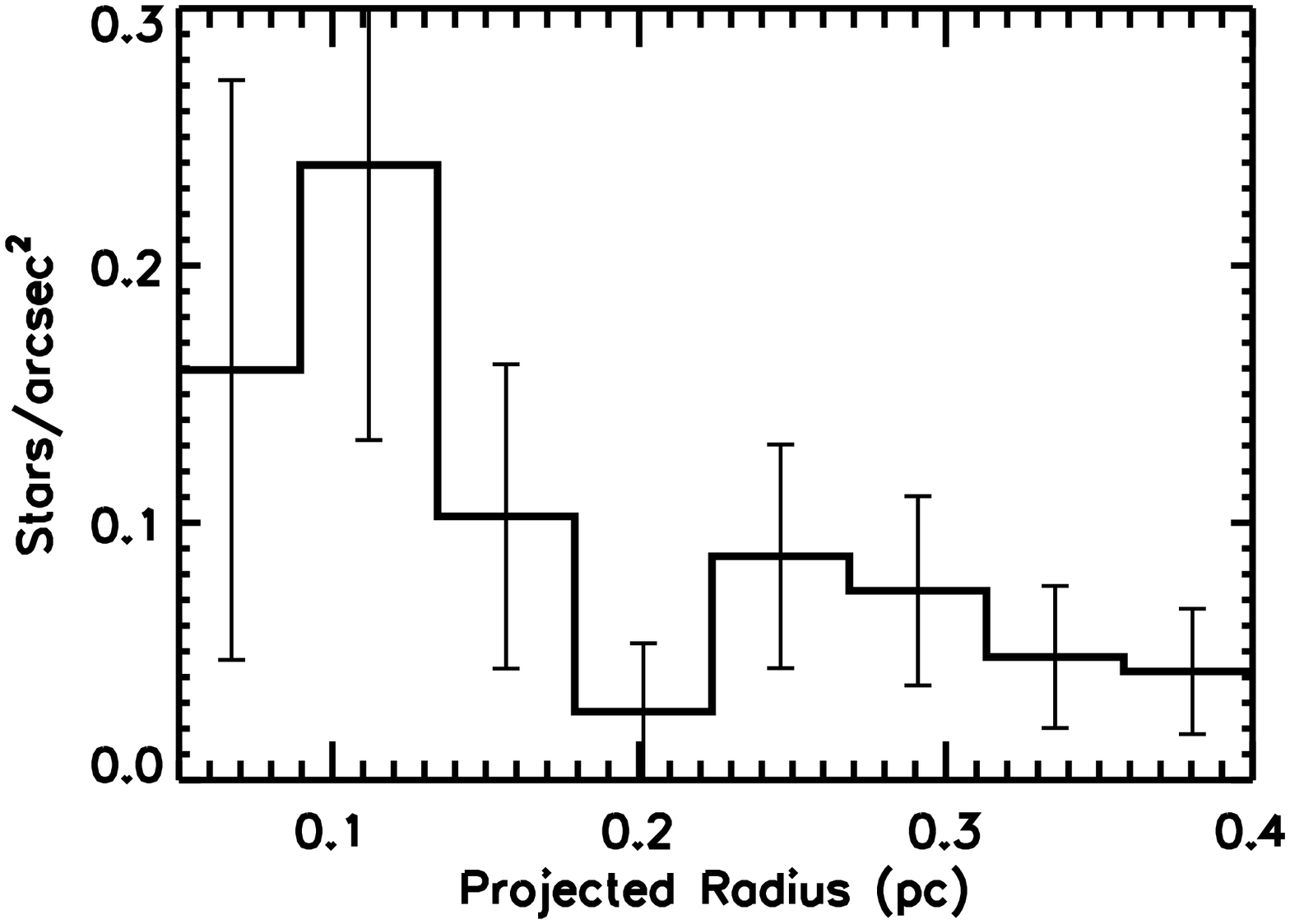}{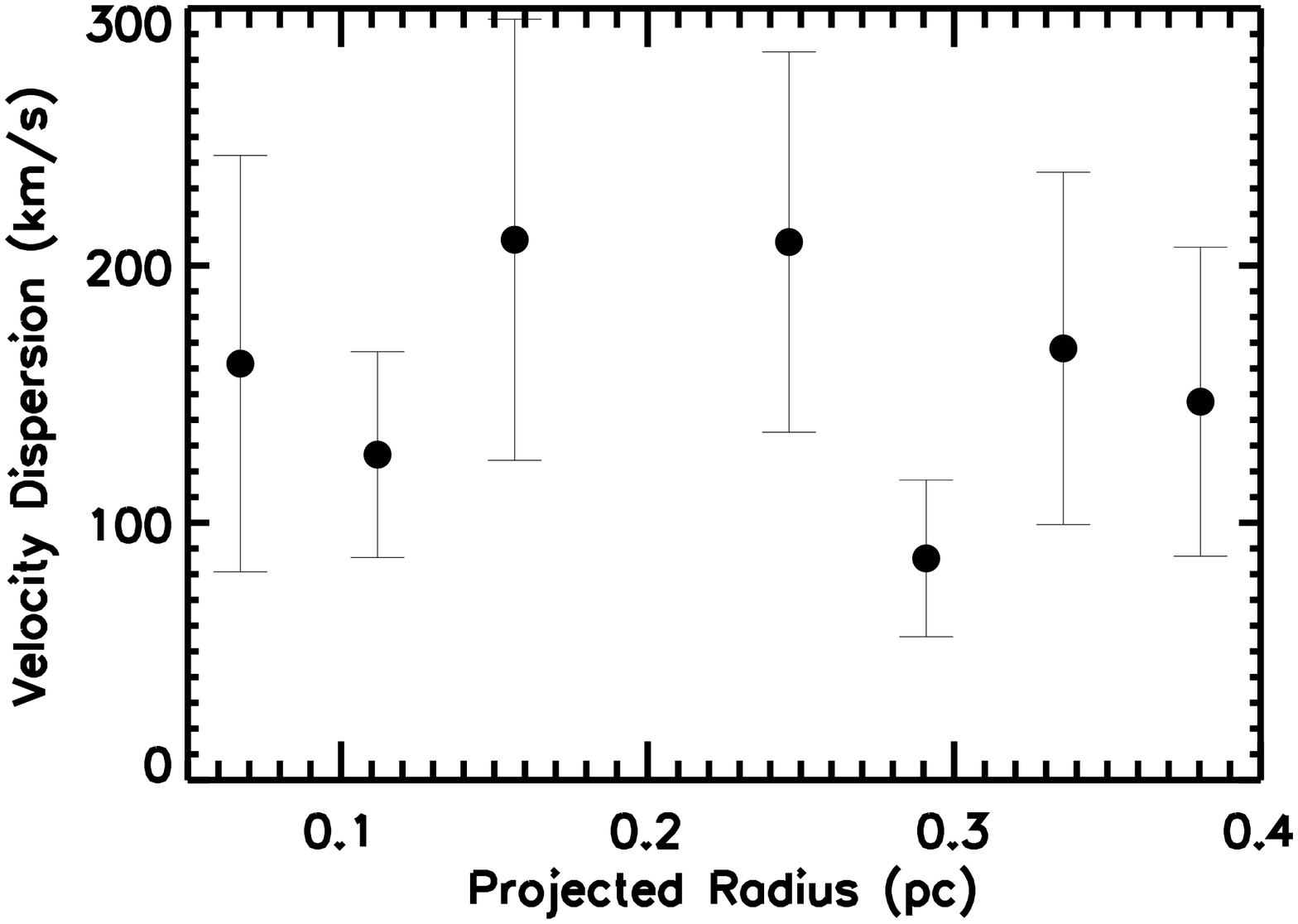} \caption{Surface number
density (left) and radial velocity dispersion (right) of early type
stars as functions of the projected distance from Sgr~A$^*$.
\label{annulushot}}
\end{figure}

\begin{figure}
\epsscale{1.0} \plottwo{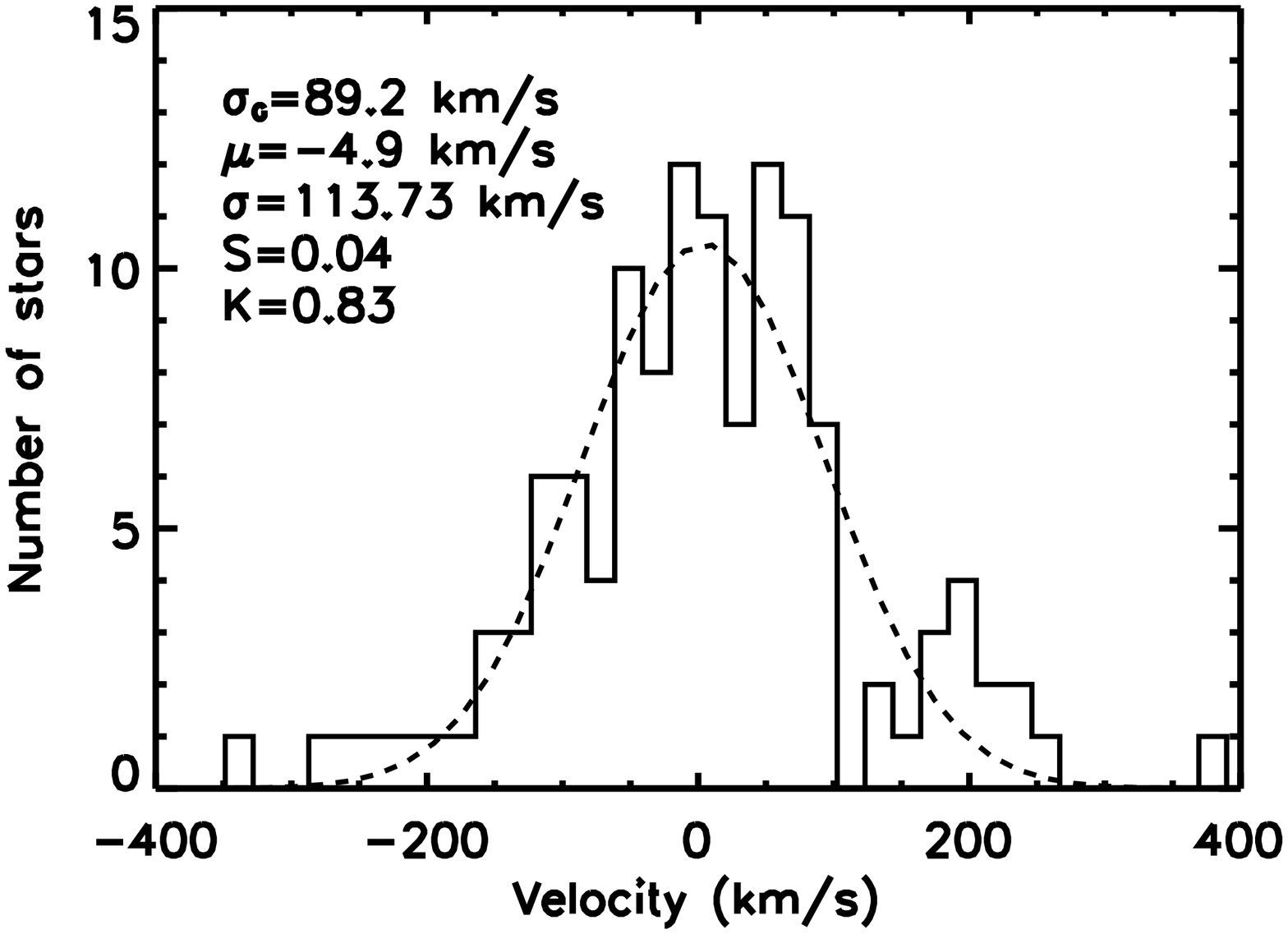}{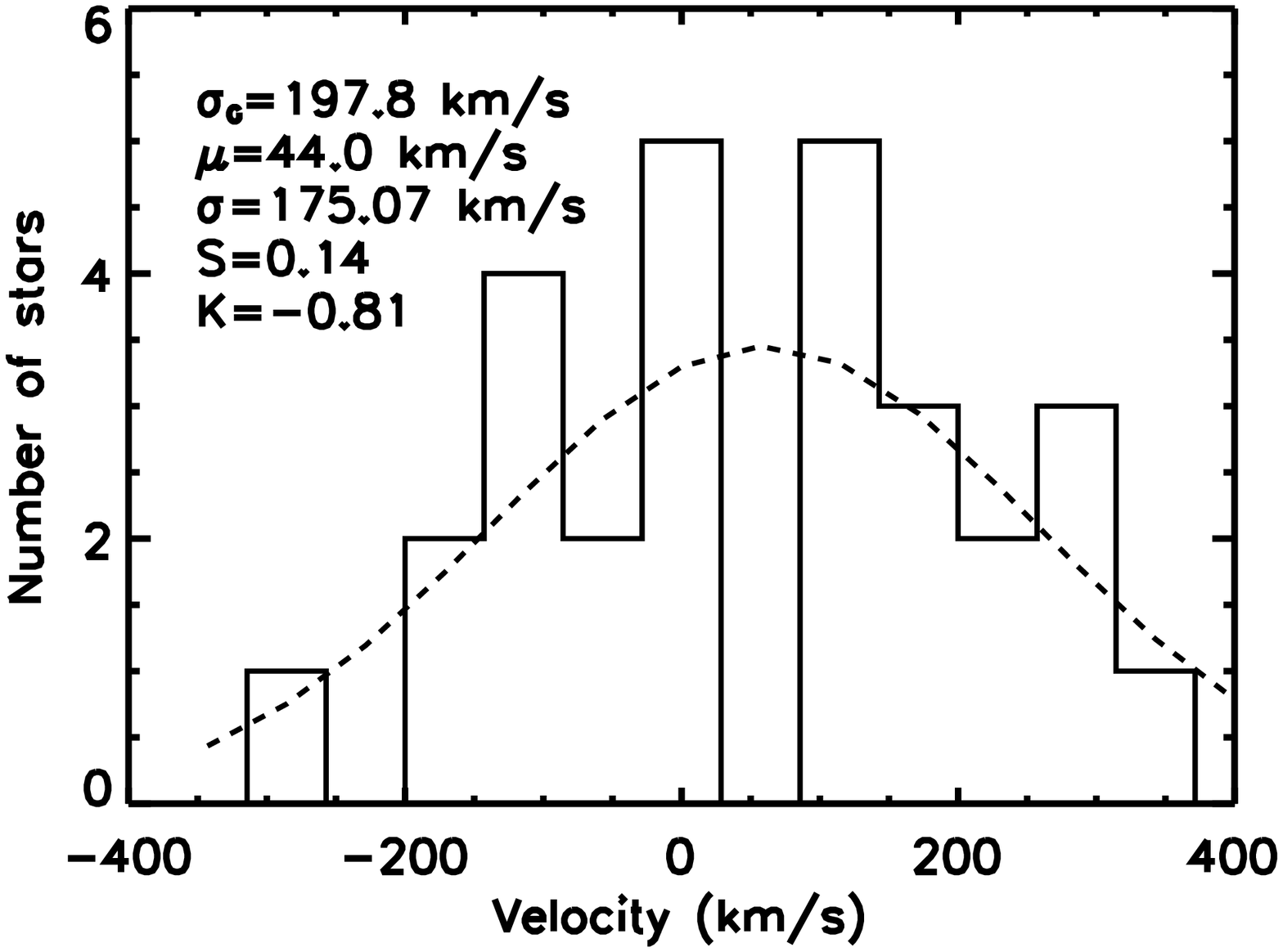} \caption{Distributions
(histograms) and the best Gaussian fittings (dashed lines) of radial
velocities of late type (left panel) and early type (right panel)
stars. The standard deviations $\sigma_G$ of the fitting functions
are indicated. The statistical mean $\mu$, the standard deviation
$\sigma$, the skewness S and the excess kurtosis K of different
samples are also shown. Note that a normal distribution should have
a zero skewness and a zero kurtosis. The radial velocity
measurements from different data sets for the same object are
averaged before the computation of the distributions.
\label{star_vdist}}
\end{figure}

\begin{figure}
\epsscale{1.0} \plottwo{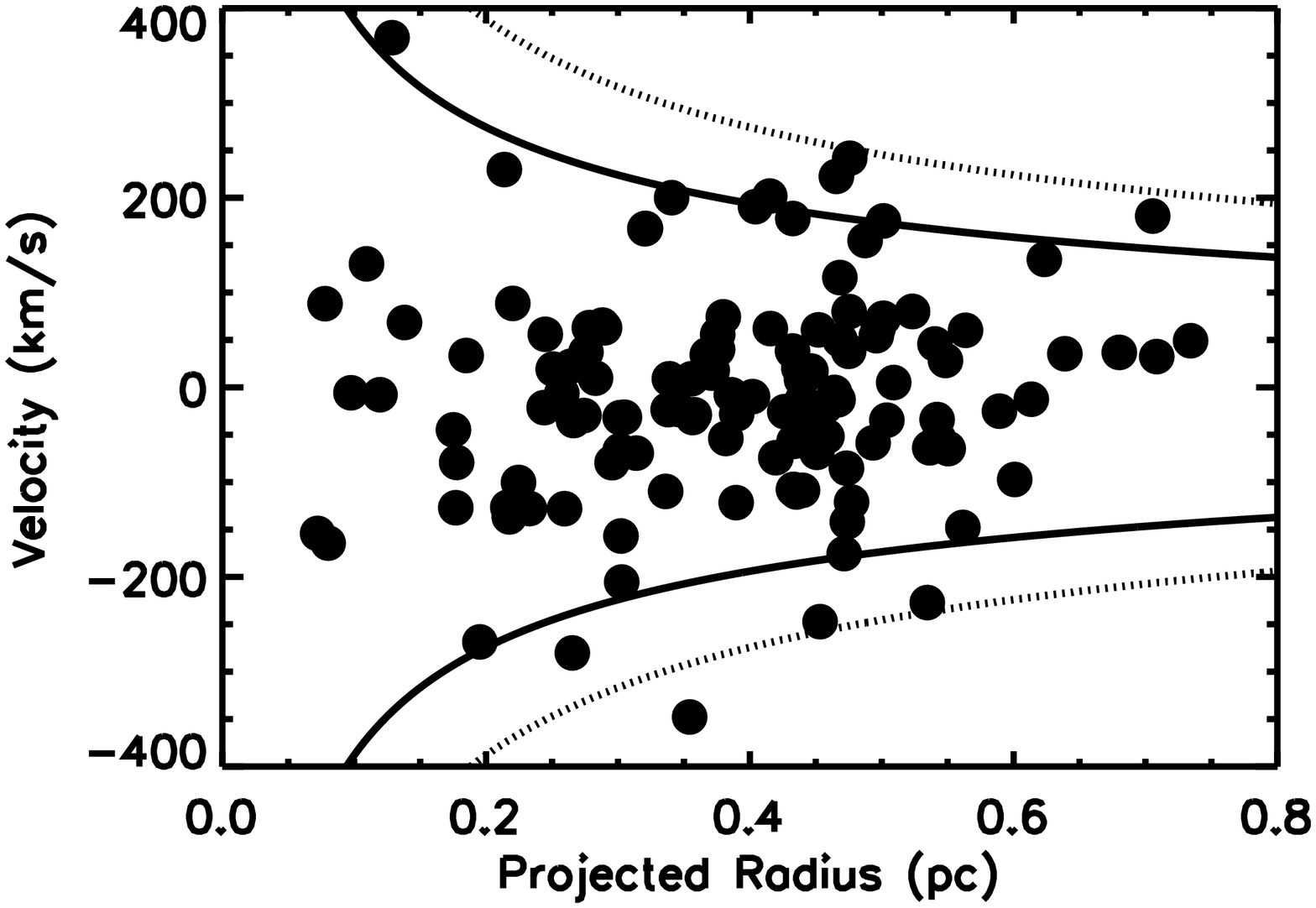}{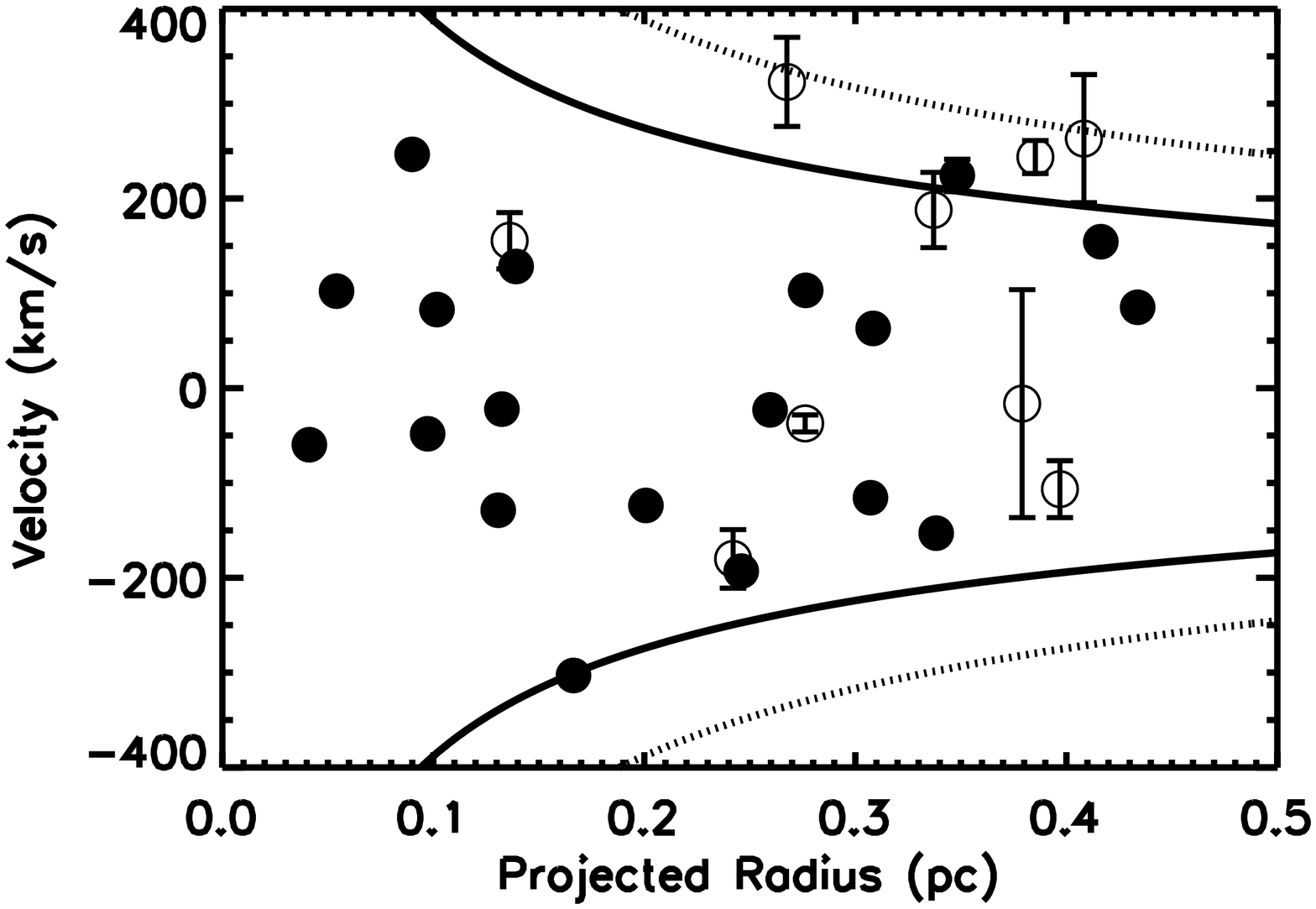} \caption{Radial
velocities of stars as a function of projected Galactic radius. The
left panel is for cool stars and the right panel is for hot stars.
In the right panel, filled circles indicate stars showing narrow
He~I line absorption and open circles indicates stars showing broad
He~I line emission. Note that for some stars the size of the error
bar is smaller than the symbol size. Solid curves in the figures
show velocities for ideal circular orbits around a central mass of
3.5$\times$10$^{6}$~M$_{\odot}$. Dashed curves show the escape
velocities at the corresponding radii. \label{star_vr}}
\end{figure}

\begin{figure}
\epsscale{1.0} \plottwo{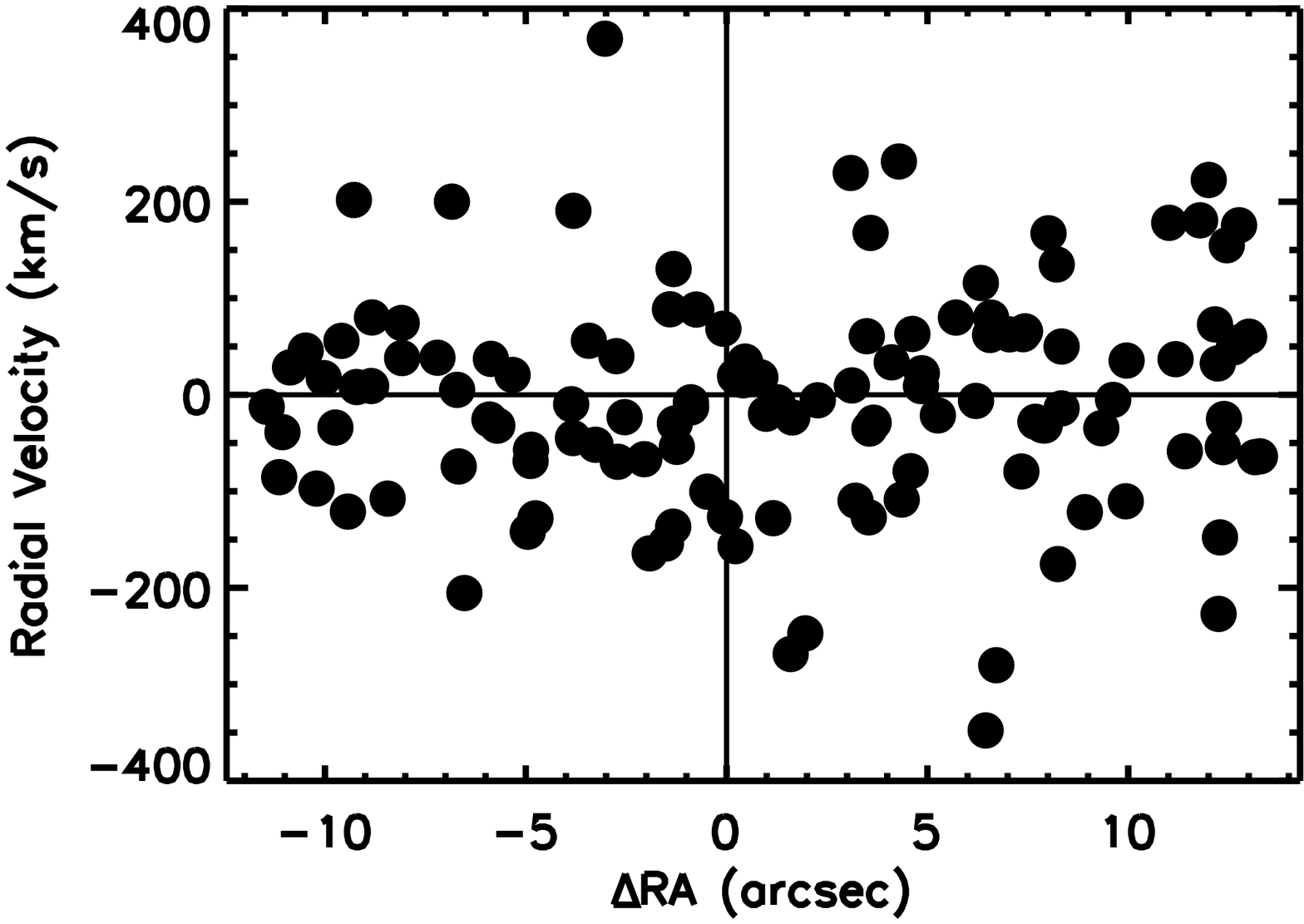}{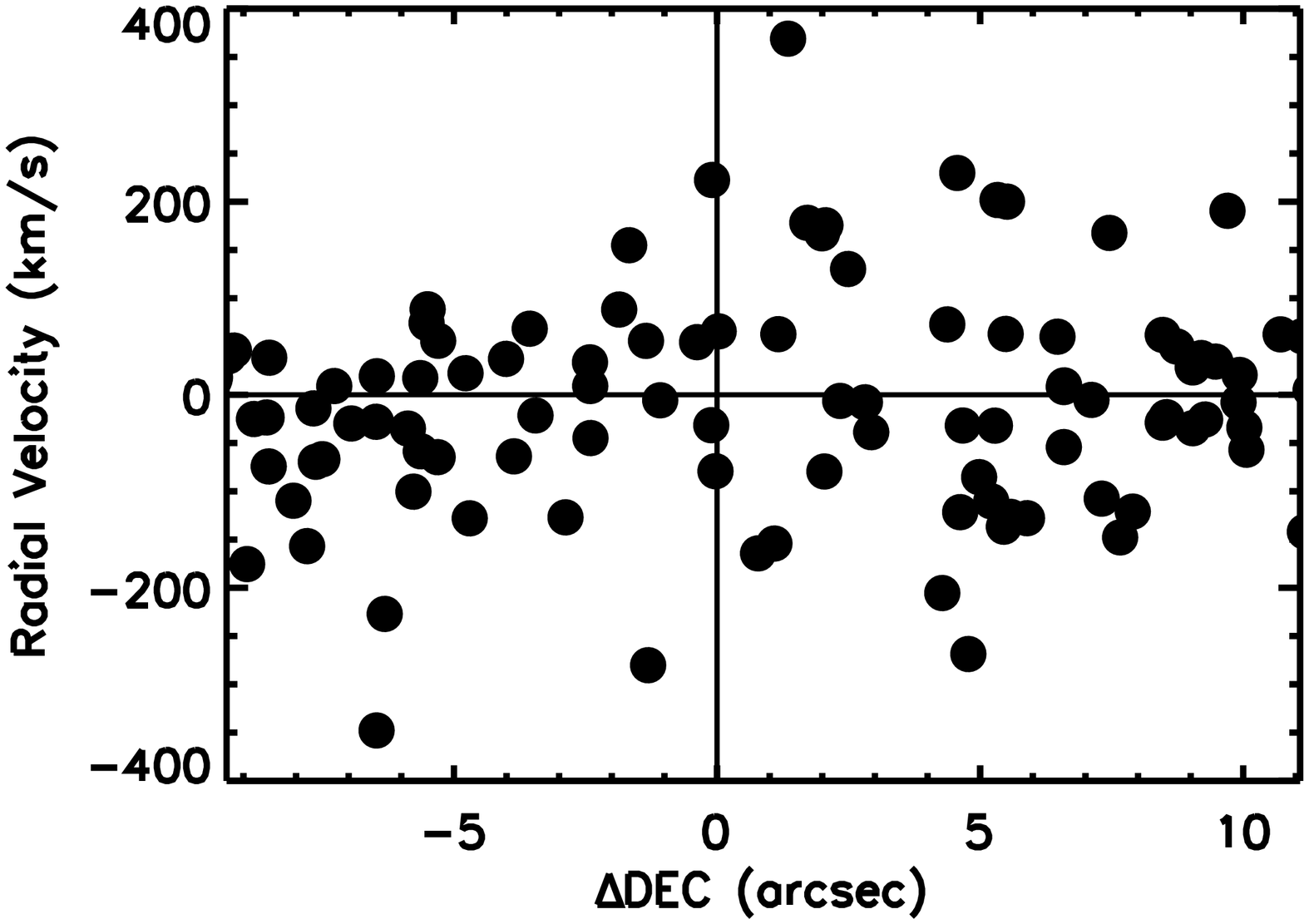} \caption{Radial
velocities of cool stars vs. the RA and DEC offsets from Sgr~A$^*$.
Note the sizes of error bars are smaller than the symbol size in
these plots. \label{coolstar_disk}}
\end{figure}

\begin{figure}
\epsscale{1.0} \plottwo{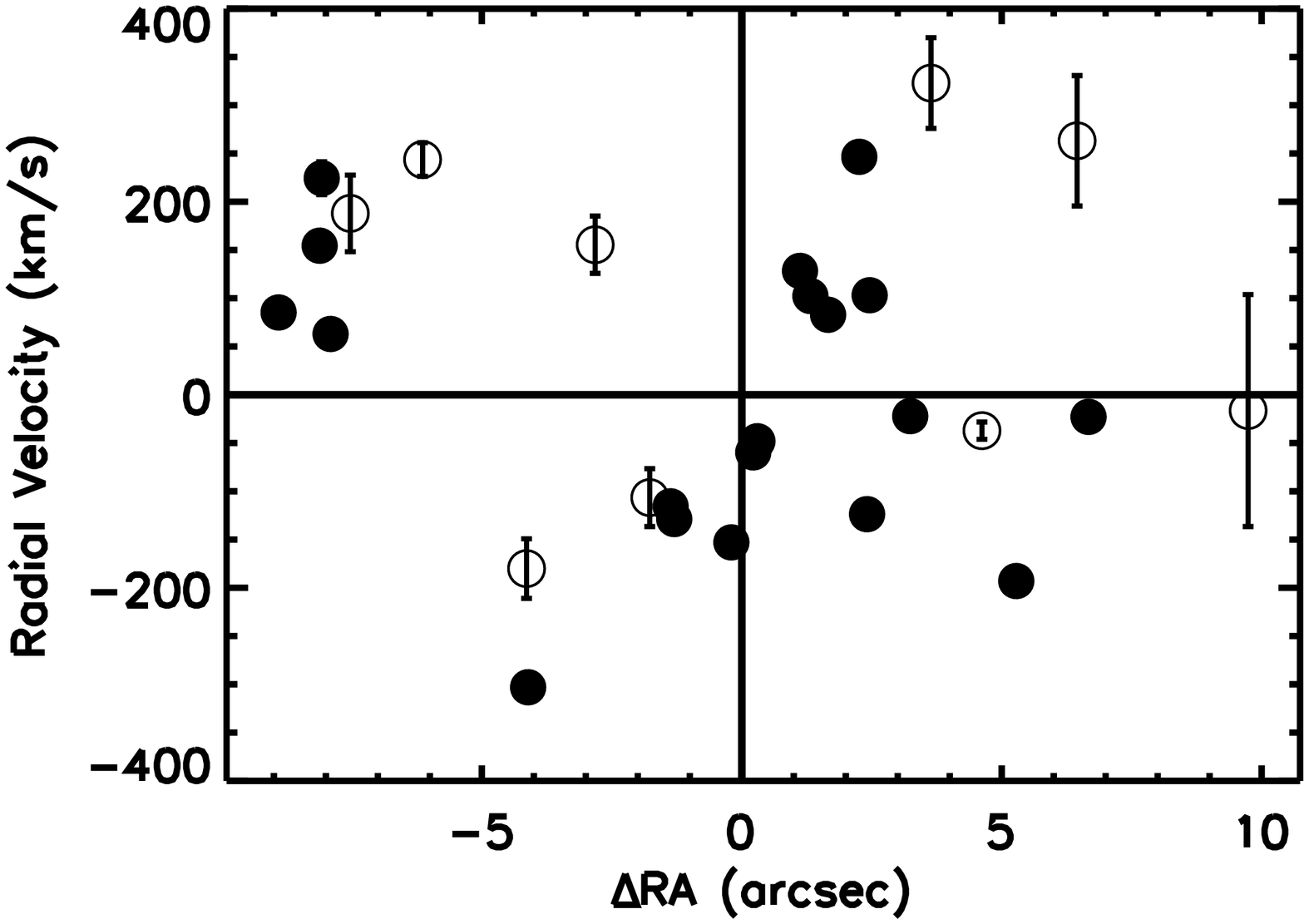}{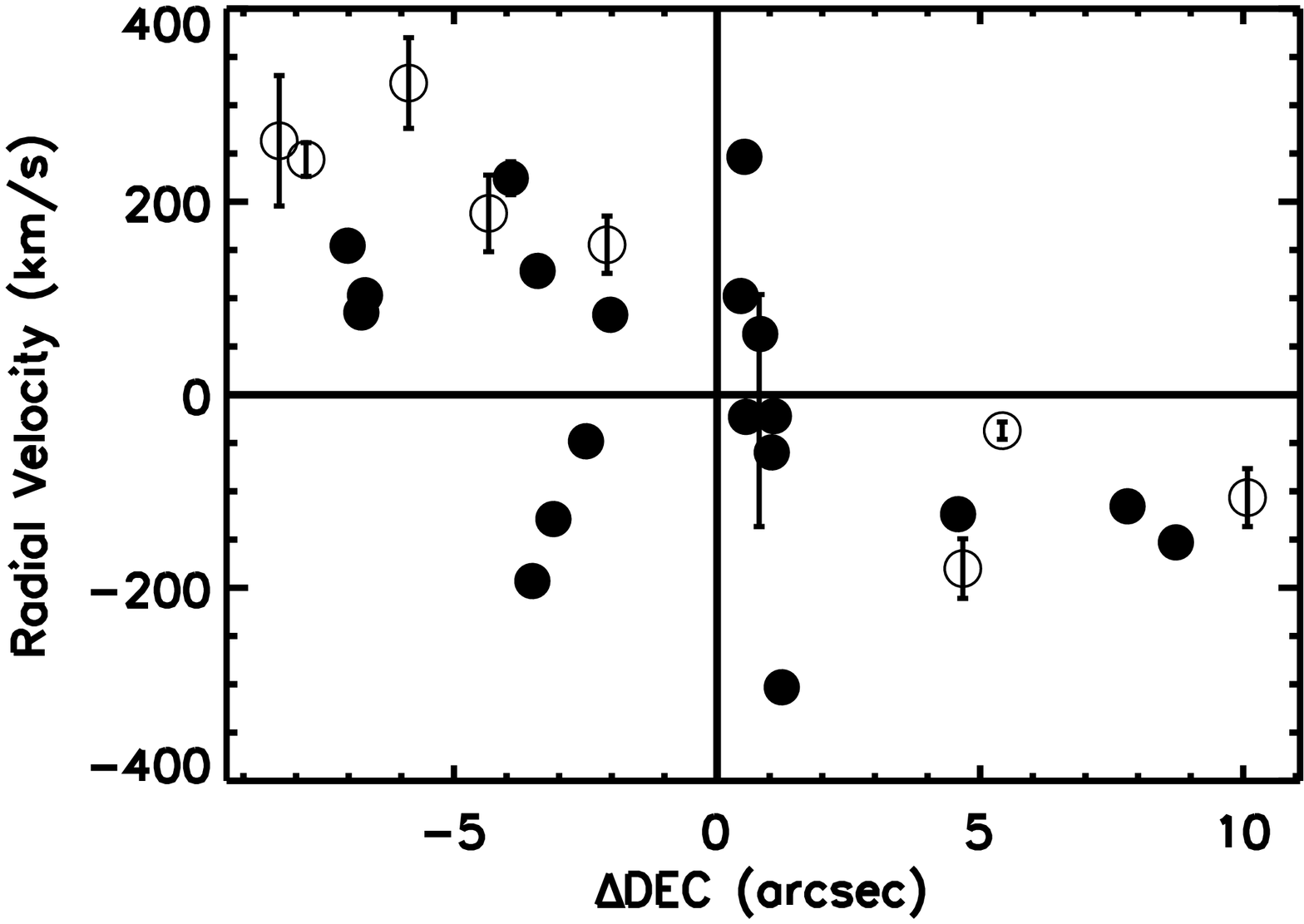} \caption{Radial
velocities of hot stars vs. the RA and DEC offsets from Sgr~A$^*$.
Narrow type stars are plotted with solid circles and broad type
stars are plotted with open circles. \label{hotstar_disk}}
\end{figure}

\begin{figure}
\epsscale{1.0} \plottwo{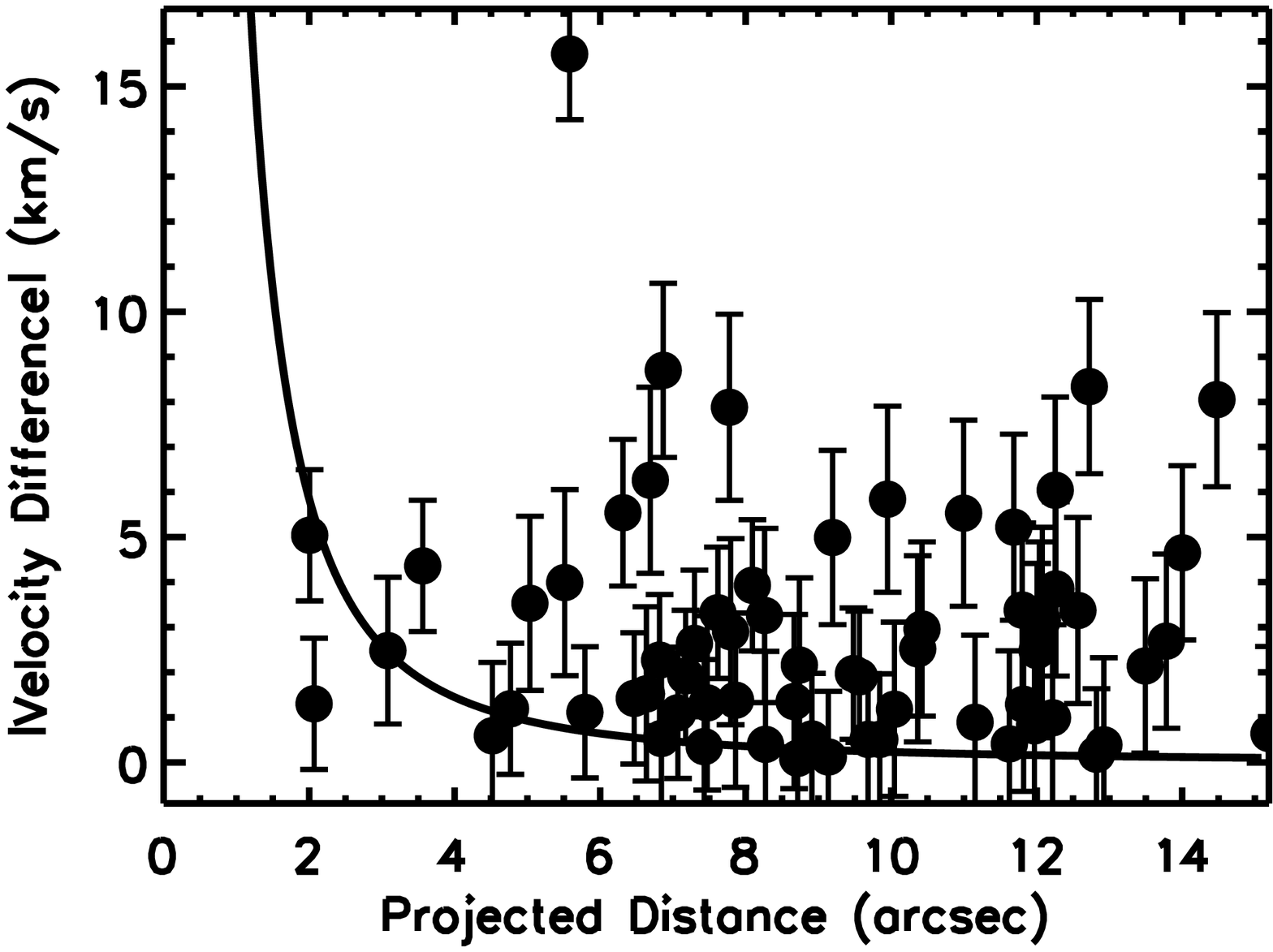}{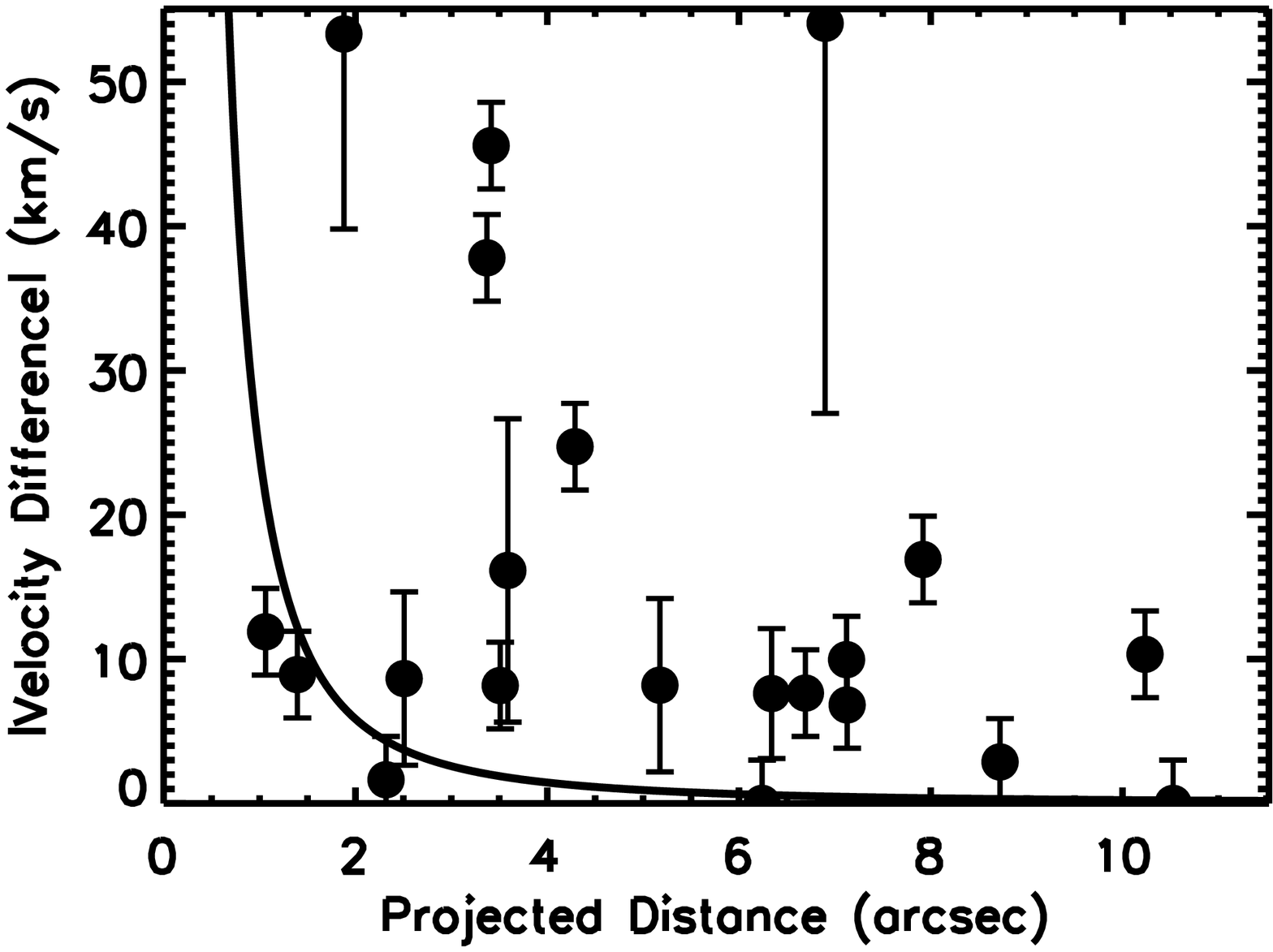} \caption{The difference
of the radial velocities between 1999 and 2005 for cool stars (left
panel) and hot stars (right panel) as a function of the projected
Galactic radius. The solid curve shows the expected maximum radial
velocity changes at different projected distances for an edge-on
circular orbit system around a gravitationally dominant object with
mass of 3.5$\times$10$^6$~M$_{\odot}$. \label{star_vaccl}}
\end{figure}

\clearpage
\begin{figure}
\epsscale{1.0} \plottwo{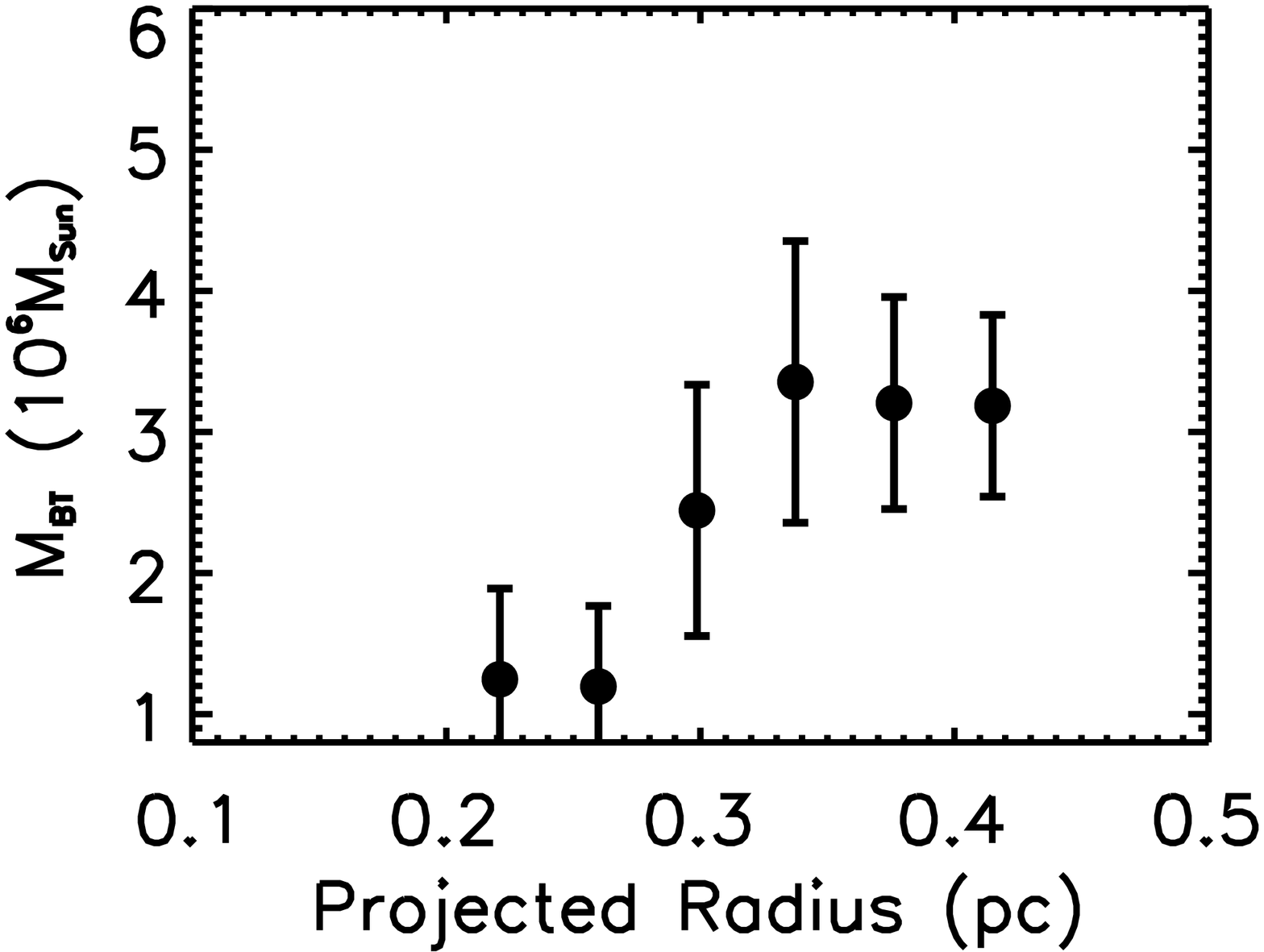}{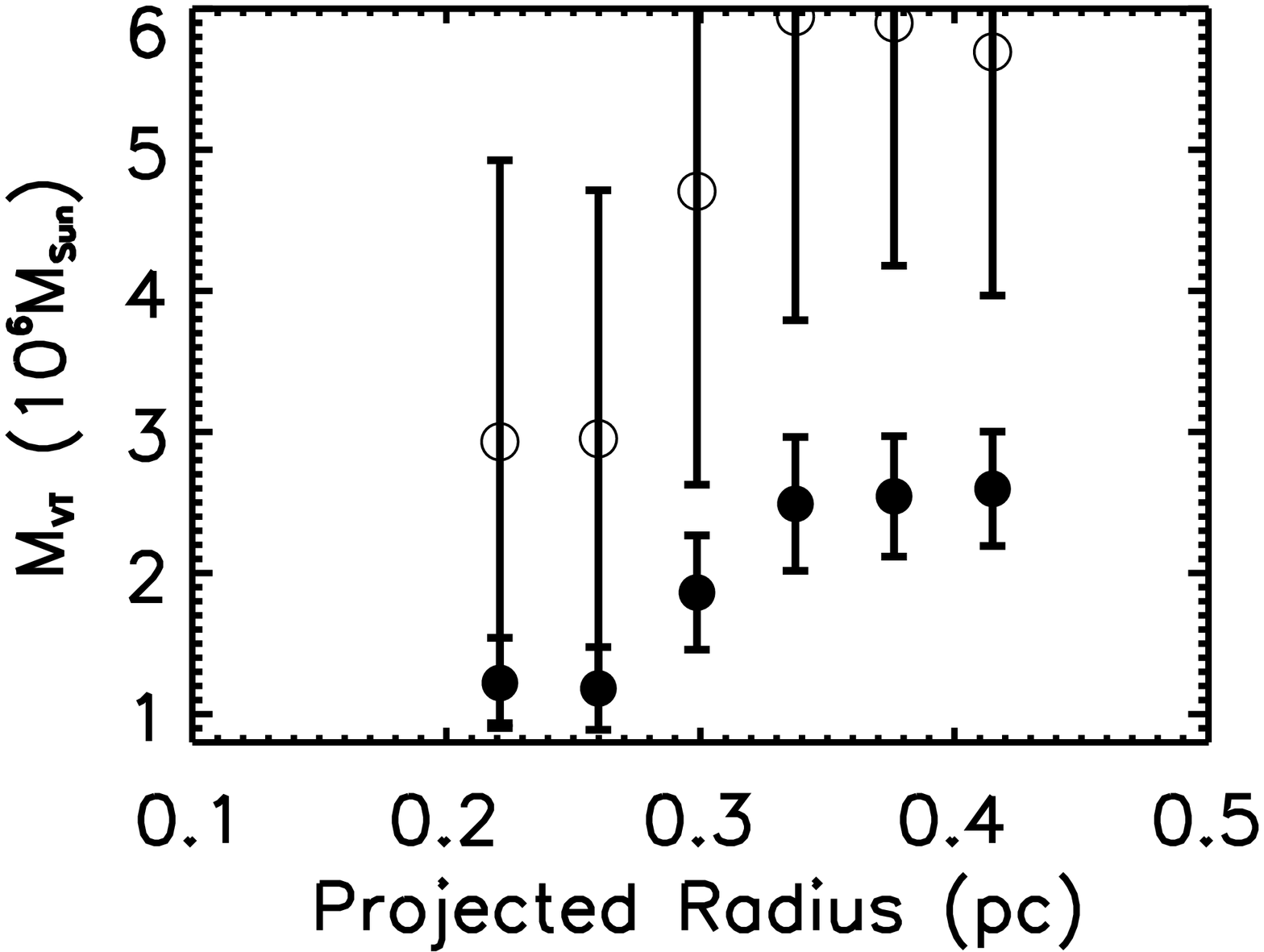} \caption{The estimated
enclosed mass as a function of radius. The left panel shows the
results from using the Bahcall-Tremaine estimator
(Eq.~\ref{enclosedmasseq}) and the right panel shows the results
using the Virial Theorem estimators (Eq.~\ref{enclosedmasseqvt1} and
\ref{enclosedmasseqvt2}). The Bahcall-Tremaine estimator assumes the
stellar velocities are isotropic and all mass is in a central point.
The Virial Theorem estimators do not assume any anisotropy about the
stellar velocities, but the assumption of all mass in a central
point gives a lower limit (filled circles) and the assumption of
mass following stars gives an upper limit (open circles) to the
estimated black mass. Bigger error bars in the case of the Virial
Theorem are due to the r$^{-1}_{ij}$ term. In calculation, the
maximum projected radius is set to keep the data set circularly
symmetric about the GC. Both figures are based on the combined data
set of observations in 1999 and 2005. \label{enclosedmass}}
\end{figure}




\clearpage
\begin{deluxetable}{lcccccc}
\tabletypesize{\scriptsize} \tablewidth{0pt}
\tablecaption{Observational Parameters \label{obspara}} %
\tablehead{\colhead{Name\tablenotemark{a}} &\colhead{Date}&
\colhead{Resolution\tablenotemark{b}} &
\colhead{Filter\tablenotemark{c}}          & \colhead{IT}  &
\colhead{NFrame} & \colhead{Slit Size} \\
&  &
 \colhead{(R=$\frac{\lambda}{\Delta\lambda}$ )}&     & \colhead{(sec)}  &
& \colhead{(arcsec${\times}$arcsec)}   } %
\startdata
    GC1&  1999 JUN 04&        14,000&        NIRSPEC-7&       60&     40&       0.72 $\times$ 24\\
    GC2&  1999 JUL 04&        23,300&        NIRSPEC-6/7&     100&    67&       0.43 $\times$ 24\\
    GC3&  2005 APR 13&        16,700&        NIRSPEC-7&       60&     36&       0.58 $\times$ 24\\
    GC4&  2005 APR 14&        16,700&        NIRSPEC-7&       60&     37&       0.58 $\times$ 24\\
\enddata
\tablecomments{Parameters of individual observations. Integration
time (IT) at each slit position and the total number of slit
positions (NFrame)
are shown in the table. The slit size is in units of arcsecs $\times$ arcsecs.} %
\tablenotetext{a}{Names referring different slit
scans in the order of the dates when observations were carried out.} %
\tablenotetext{b}{The resolution is
$\lambda$/${\Delta}{\lambda}_{FWHM}$,
where${\Delta}{\lambda}_{FWHM}$ is the half-power line width of arc
lamp lines. The values are measured at $\lambda=2.171407~\mu$m.
$\Delta\lambda\simeq$1.561$\times10^{-4}$, 9.286$\times10^{-5}$,
1.296$\times10^{-4}$ and 1.301$\times10^{-4}~{\mu}m$ for GC1, GC2,
GC3 and GC4 respectively. The slit width was 5 pixels in GC1, 3
pixels in GC2 and
4 pixels in GC3 and GC4.} %
\tablenotetext{c}{NIRSPEC-7 has half-power points of 1.85 and
2.62~$\mu$m. NIRSPEC-6 has half-power points of 1.56 and
2.30~$\mu$m. Half-power points are taken from \cite{figGKMBM03}. }
\end{deluxetable}

\begin{deluxetable}{ccccccccc}
\tabletypesize{\scriptsize} \tablewidth{0pt}
\tablecaption{Wavelength Ranges of Echelle Orders ($\mu$m)
\label{orderwave}} \tablehead{ \colhead{Echelle}      &
\multicolumn{2}{c}{GC1} & %
\multicolumn{2}{c}{GC2} & %
\multicolumn{2}{c}{GC3} & %
\multicolumn{2}{c}{GC4} \\
\colhead{Order}      &
\colhead{Min}          & \colhead{Max}  &
\colhead{Min}          & \colhead{Max}  &
\colhead{Min}          & \colhead{Max}  &
\colhead{Min}          & \colhead{Max}
}
\startdata
32&  -     & -    & -    & -    & 2.353&2.388 & 2.353&2.388  \\
33&  2.281 &2.315 &2.290 &2.324 & 2.282&2.316 & 2.282&2.316 \\
34&  2.214 &2.248 &2.223 &2.256 & 2.215&2.249 & 2.215&2.249 \\
35&  2.152 &2.184 &2.160 &2.192 & 2.152&2.185 & 2.152&2.185 \\
36&  2.092 &2.124 &2.100 &2.131 & 2.093&2.124 & 2.093&2.124 \\
37&  2.036 &2.067 &2.044 &2.074 & 2.037&2.067 & 2.037&2.067 \\
\enddata
\tablecomments{Wavelength Coverage of different data sets. Because
the orders are longer than the width of the detector, the spectra
are not contiguous in wavelength. The wavelengths are in microns.}
\end{deluxetable}

\begin{deluxetable}{lrrrrrrrr}
\tabletypesize{\scriptsize}
\tablewidth{0pt}
\tablecaption{Cool Star Velocity \label{coolstarvelo}}
\tablehead{
\colhead{Name}           & \colhead{RA}      &
\colhead{DEC}          & \colhead{V99}  &
\colhead{V05} &
\colhead{V$_{avg}$} &
\colhead{Temp} & \colhead{Spectral} &
\colhead{V$_{Figer}$} \\
& \colhead{($''$)} & \colhead{($''$)}  &
\colhead{(km~s$^{-1}$)}  &
\colhead{(km~s$^{-1}$)}  &
\colhead{(km~s$^{-1}$)}  &
\colhead{(K)} & \colhead{Type}&
\colhead{(km~s$^{-1}$)}
}
\startdata
     ID327&    -1.51&     1.09&  -154.1 $\pm$     1.8&     ...              &  -154.1 $\pm$     1.8&  5161&$<$G8III&  -156.0 $\pm$     1.1\\
     ID259&    -0.75&    -1.86&    85.8 $\pm$     1.3&    90.8 $\pm$     0.7&    88.3 $\pm$     1.5&  3714&   M1III&    96.5 $\pm$     0.8\\
    IRS29S&    -1.91&     0.78&  -165.0 $\pm$     1.3&  -163.7 $\pm$     0.7&  -164.3 $\pm$     1.5&  4338&   K2III&  -160.0 $\pm$     1.1\\
 2.27-1.08&     2.27&    -1.08&    -5.7 $\pm$     1.8&     ...              &    -5.7 $\pm$     1.8&  4996&$<$G8III&     ...              \\
     G288?&    -1.32&     2.49&   130.3 $\pm$     1.3&     ...              &   130.3 $\pm$     1.3&  4167&   K3III&   133.7 $\pm$     1.1\\
     ID365&     1.26&     2.81&    -6.6 $\pm$     1.3&    -9.1 $\pm$     1.0&    -7.8 $\pm$     1.6&  4206&   K3III&     1.9 $\pm$     1.1\\
-3.03+1.35&    -3.03&     1.35&     ...              &   368.9 $\pm$     1.0&   368.9 $\pm$     1.0&  5004&$<$G8III&     ...              \\
    IRS33W&    -0.08&    -3.56&    66.2 $\pm$     1.3&    70.6 $\pm$     0.7&    68.4 $\pm$     1.5&  3697&   M2III&    73.5 $\pm$     0.8\\
    IRS13W&    -3.83&    -2.40&   -44.6 $\pm$     1.3&   -45.2 $\pm$     1.0&   -44.9 $\pm$     1.6&  3865&   M0III&   -40.0 $\pm$     1.1\\
 3.54-2.88&     3.54&    -2.88&  -127.0 $\pm$     1.8&     ...              &  -127.0 $\pm$     1.8&  4482&   K1III&     ...              \\
 4.58-0.03&     4.58&    -0.03&   -79.2 $\pm$     1.8&     ...              &   -79.2 $\pm$     1.8&  4620&   K0III&     ...              \\
     ID244&     4.11&    -2.41&    33.0 $\pm$     1.3&    34.2 $\pm$     0.7&    33.6 $\pm$     1.5&  3628&   M2III&    40.3 $\pm$     0.8\\
     ID398&     1.60&     4.77&  -270.4 $\pm$     1.8&  -266.9 $\pm$     0.7&  -268.6 $\pm$     1.9&  4194&   K3III&  -262.0 $\pm$     1.1\\
 3.09+4.56&     3.09&     4.56&   231.8 $\pm$     1.8&   227.8 $\pm$     1.0&   229.8 $\pm$     2.1&  4430&   K1III&     ...              \\
      IRS7&    -0.04&     5.58&  -118.7 $\pm$     1.3&  -134.4 $\pm$     0.7&  -126.5 $\pm$     1.5&  3364&   M7III&  -109.0 $\pm$     0.8\\
-1.33+5.45&    -1.33&     5.45&     ...              &  -136.7 $\pm$     1.0&  -136.7 $\pm$     1.0&  4380&   K1III&     ...              \\
-1.41-5.50&    -1.41&    -5.50&    88.6 $\pm$     1.8&     ...              &    88.6 $\pm$     1.8&  4310&   K2III&     ...              \\
     IRS20&    -0.48&    -5.77&   -99.8 $\pm$     1.3&  -100.9 $\pm$     0.7&  -100.4 $\pm$     1.5&  3246&$>$M7III&   -90.9 $\pm$     0.8\\
 1.16+5.89&     1.16&     5.89&     ...              &  -127.9 $\pm$     1.0&  -127.9 $\pm$     1.0&  4316&   K2III&     ...              \\
 5.26-3.45&     5.26&    -3.45&     ...              &   -21.4 $\pm$     0.7&   -21.4 $\pm$     0.7&  5123&$<$G8III&     ...              \\
-3.43-5.30&    -3.43&    -5.30&    53.2 $\pm$     1.3&    58.7 $\pm$     1.0&    56.0 $\pm$     1.6&  3492&   M4III&     ...              \\
    G577A?&     0.21&    -6.47&    18.5 $\pm$     1.3&    19.9 $\pm$     0.7&    19.2 $\pm$     1.5&  3687&   M2III&    28.5 $\pm$     0.8\\
 6.21+2.34&     6.21&     2.34&    -5.6 $\pm$     1.8&    -7.1 $\pm$     0.7&    -6.4 $\pm$     1.9&  4053&   K4III&     ...              \\
-4.76-4.70&    -4.76&    -4.70&  -124.9 $\pm$     1.8&  -131.1 $\pm$     1.0&  -128.0 $\pm$     2.1&  4106&   K4III&     ...              \\
     ID189&     4.86&    -4.78&    23.4 $\pm$     1.3&    21.2 $\pm$     0.7&    22.3 $\pm$     1.5&  4105&   K4III&    26.9 $\pm$     1.1\\
 6.71-1.31&     6.72&    -1.31&  -280.0 $\pm$     1.3&  -280.5 $\pm$     0.7&  -280.2 $\pm$     1.5&  4364&   K1III&     ...              \\
     GCHe2&     3.56&    -5.88&   -39.4 $\pm$     1.8&   -30.7 $\pm$     0.7&   -35.0 $\pm$     1.9&  5033&$<$G8III&     ...              \\
     ID144&    -1.29&    -6.95&   -30.3 $\pm$     1.3&   -29.2 $\pm$     0.7&   -29.8 $\pm$     1.5&  3610&   M3III&   -22.3 $\pm$     0.8\\
     ID213&    -5.87&    -4.01&    37.2 $\pm$     1.8&     ...              &    37.2 $\pm$     1.8&  4118&   K3III&    41.1 $\pm$     1.1\\
     ID415&     4.63&     5.48&    62.0 $\pm$     1.3&    64.0 $\pm$     0.7&    63.0 $\pm$     1.5&  4271&   K2III&    68.2 $\pm$     0.8\\
     ID430&     3.12&     6.60&    11.0 $\pm$     1.3&     8.4 $\pm$     1.0&     9.7 $\pm$     1.6&  3887&   M0III&    16.0 $\pm$     1.1\\
     ID297&     7.43&     0.03&    65.8 $\pm$     1.8&    65.5 $\pm$     0.7&    65.7 $\pm$     1.9&  3873&   M0III&    71.1 $\pm$     1.1\\
 IRS1NE(3)&     7.38&     1.16&    62.1 $\pm$     1.8&    63.4 $\pm$     0.7&    62.8 $\pm$     1.9&  3824&   M0III&    72.7 $\pm$     1.1\\
 IRS1NE(2)&     7.34&     2.04&   -77.8 $\pm$     1.3&   -81.2 $\pm$     0.7&   -79.5 $\pm$     1.5&  3623&   M3III&   -68.8 $\pm$     1.1\\
-2.05-7.50&    -2.05&    -7.50&     ...              &   -66.8 $\pm$     1.0&   -66.8 $\pm$     1.0&  3649&   M2III&     ...              \\
     BHA4E&    -5.71&     5.28&   -35.9 $\pm$     1.8&   -28.0 $\pm$     1.0&   -32.0 $\pm$     2.1&  3679&   M2III&   -28.2 $\pm$     1.1\\
     ID126&     0.22&    -7.79&  -158.3 $\pm$     1.8&  -155.4 $\pm$     1.0&  -156.8 $\pm$     2.1&  4415&   K1III&  -146.0 $\pm$     0.8\\
-6.53+4.28&    -6.53&     4.28&  -205.4 $\pm$     1.8&     ...              &  -205.4 $\pm$     1.8&  4208&   K3III&     ...              \\
 7.86-0.11&     7.86&    -0.11&   -30.9 $\pm$     1.8&   -32.2 $\pm$     0.7&   -31.5 $\pm$     1.9&  3562&   M3III&     ...              \\
    IRS12N&    -2.71&    -7.63&   -67.5 $\pm$     1.3&   -71.4 $\pm$     0.7&   -69.4 $\pm$     1.5&  3851&   M0III&   -60.7 $\pm$     0.8\\
 IRS1NE(1)&     8.02&     2.00&   168.9 $\pm$     1.8&   165.6 $\pm$     0.7&   167.3 $\pm$     1.9&  3437&   M5III&   -67.5 $\pm$     1.1\\
 3.58+7.45&     3.59&     7.45&   167.9 $\pm$     1.8&   167.4 $\pm$     1.0&   167.6 $\pm$     2.1&  4905&$<$G8III&     ...              \\
     G849?&     3.21&    -8.05&  -110.4 $\pm$     1.8&  -109.1 $\pm$     0.7&  -109.8 $\pm$     1.9&  4073&   K4III&  -107.0 $\pm$     1.1\\
    IRS14N&     1.63&    -8.56&   -23.5 $\pm$     1.3&   -23.5 $\pm$     0.7&   -23.5 $\pm$     1.5&  3313&$>$M7III&   -13.6 $\pm$     0.8\\
 4.84-7.27&     4.84&    -7.27&    10.1 $\pm$     1.8&     7.9 $\pm$     0.7&     9.0 $\pm$     1.9&  4283&   K2III&     ...              \\
     BHA4W&    -6.84&     5.51&   200.1 $\pm$     1.8&     ...              &   200.1 $\pm$     1.8&  3346&   M7III&     ...              \\
     ID467&    -2.54&     8.54&   -23.0 $\pm$     1.3&   -23.5 $\pm$     0.7&   -23.2 $\pm$     1.5&  3394&   M6III&   -13.8 $\pm$     0.8\\
      IRS9&     6.45&    -6.47&  -347.8 $\pm$     1.3&  -347.7 $\pm$     0.7&  -347.8 $\pm$     1.5&  4012&   K4III&  -341.0 $\pm$     0.8\\
-8.85-2.41&    -8.85&    -2.41&     9.1 $\pm$     1.8&     ...              &     9.1 $\pm$     1.8&  3529&   M4III&     ...              \\
 7.92+4.67&     7.92&     4.67&   -29.3 $\pm$     1.8&   -34.3 $\pm$     0.7&   -31.8 $\pm$     1.9&  3765&   M1III&     ...              \\
 3.66+8.47&     3.66&     8.47&   -28.9 $\pm$     1.8&     ...              &   -28.9 $\pm$     1.8&  4906&$<$G8III&     ...              \\
      G904&     0.46&     9.47&    35.3 $\pm$     1.3&    33.4 $\pm$     0.7&    34.4 $\pm$     1.5&  3314&$>$M7III&    43.8 $\pm$     0.8\\
   IRS14SW&     0.41&    -9.56&    14.1 $\pm$     1.3&    15.9 $\pm$     0.7&    15.0 $\pm$     1.5&  3853&   M0III&    24.5 $\pm$     0.8\\
 0.84-9.54&     0.84&    -9.54&     ...              &    18.1 $\pm$     1.0&    18.1 $\pm$     1.0&  3235&$>$M7III&     ...              \\
    IRS12S&    -2.74&    -9.28&    40.2 $\pm$     1.3&    39.7 $\pm$     0.7&    40.0 $\pm$     1.5&  3498&   M4III&    48.8 $\pm$     0.8\\
      G986&    -9.59&    -1.35&    55.8 $\pm$     1.8&     ...              &    55.8 $\pm$     1.8&  3747&   M1III&    62.7 $\pm$     1.1\\
-8.09-5.52&    -8.09&    -5.52&    74.4 $\pm$     1.8&     ...              &    74.4 $\pm$     1.8&  4349&   K2III&     ...              \\
      ID87&    -1.24&    -9.76&   -54.2 $\pm$     1.3&   -53.7 $\pm$     0.7&   -54.0 $\pm$     1.5&  3874&   M0III&   -46.0 $\pm$     0.8\\
     ID498&    -0.89&     9.91&    -4.9 $\pm$     1.8&   -10.7 $\pm$     1.0&    -7.8 $\pm$     2.1&  3761&   M1III&    -3.5 $\pm$     0.8\\
   IRS10EE&     8.92&     4.62&  -121.0 $\pm$     1.8&  -122.2 $\pm$     0.7&  -121.6 $\pm$     1.9&  3738&   M1III&  -113.0 $\pm$     1.1\\
 7.69-6.48&     7.69&    -6.48&     ...              &   -28.0 $\pm$     1.0&   -28.0 $\pm$     1.0&  4329&   K2III&     ...              \\
      ID94&    -3.87&    -9.61&   -11.1 $\pm$     1.8&    -8.5 $\pm$     1.0&    -9.8 $\pm$     2.1&  4144&   K3III&    -7.5 $\pm$     1.1\\
     G1044&    -3.82&     9.70&   192.1 $\pm$     1.8&   189.1 $\pm$     0.7&   190.6 $\pm$     1.9&  3348&   M7III&   199.9 $\pm$     0.8\\
-9.28+5.33&    -9.28&     5.33&   201.8 $\pm$     1.8&     ...              &   201.8 $\pm$     1.8&  4110&   K4III&     ...              \\
 6.56+8.47&     6.56&     8.47&    62.0 $\pm$     1.8&     ...              &    62.0 $\pm$     1.8&  4469&   K1III&     ...              \\
-6.67-8.52&    -6.67&    -8.52&   -74.3 $\pm$     1.8&     ...              &   -74.3 $\pm$     1.8&  4708&   G9III&     ...              \\
     G1130&    -5.90&     9.27&   -28.4 $\pm$     1.8&   -22.9 $\pm$     1.0&   -25.6 $\pm$     2.1&  3679&   M2III&   -20.6 $\pm$     1.1\\
-7.20-8.51&    -7.20&    -8.51&    38.4 $\pm$     1.8&     ...              &    38.4 $\pm$     1.8&  3861&   M0III&     ...              \\
    G1061?&    11.02&     1.71&   178.6 $\pm$     1.8&   177.7 $\pm$     0.7&   178.1 $\pm$     1.9&  3324&   M7III&   184.8 $\pm$     1.1\\
     ID460&    -8.44&     7.31&  -107.7 $\pm$     1.8&     ...              &  -107.7 $\pm$     1.8&  3916&   K5III&  -100.0 $\pm$     1.1\\
-4.87+10.0&    -4.87&    10.06&   -56.9 $\pm$     1.8&     ...              &   -56.9 $\pm$     1.8&  4608&   K0III&     ...              \\
     ID410&     9.94&     5.21&  -110.4 $\pm$     1.8&     ...              &  -110.4 $\pm$     1.8&  4611&   K0III&  -107.0 $\pm$     1.1\\
-5.33+9.93&    -5.33&     9.93&    20.5 $\pm$     1.8&     ...              &    20.5 $\pm$     1.8&  3828&   M0III&     ...              \\
     G1150&    -9.21&     6.59&     8.0 $\pm$     1.8&     ...              &     8.0 $\pm$     1.8&  3348&   M7III&    15.6 $\pm$     1.1\\
     ID134&     8.34&    -7.67&     ...              &   -14.1 $\pm$     0.7&   -14.1 $\pm$     0.7&  5156&$<$G8III&    -9.3 $\pm$     1.1\\
4.36-10.47&     4.36&   -10.47&     ...              &  -108.6 $\pm$     1.0&  -108.6 $\pm$     1.0&  4527&   K0III&     ...              \\
     ID366&   -11.06&     2.93&   -38.7 $\pm$     1.8&     ...              &   -38.7 $\pm$     1.8&  4207&   K3III&   -32.9 $\pm$     1.1\\
     ID171&   -10.03&    -5.64&    17.3 $\pm$     1.8&     ...              &    17.3 $\pm$     1.8&  4067&   K4III&    24.5 $\pm$     1.1\\
      ID68&    -4.88&   -10.55&   -68.3 $\pm$     1.8&   -68.8 $\pm$     1.0&   -68.6 $\pm$     2.1&  4230&   K3III&   -64.8 $\pm$     1.1\\
3.49+11.12&     3.49&    11.12&    60.6 $\pm$     1.8&     ...              &    60.6 $\pm$     1.8&  4492&   K1III&     ...              \\
     ID534&     1.96&    11.52&  -244.6 $\pm$     1.8&  -249.8 $\pm$     1.0&  -247.2 $\pm$     2.1&  4187&   K3III&  -240.0 $\pm$     1.1\\
    IRS15N&     0.98&    11.76&   -18.0 $\pm$     1.8&   -21.3 $\pm$     0.7&   -19.6 $\pm$     1.9&  3579&   M3III&   -11.2 $\pm$     0.8\\
     G1152&    -3.26&    11.36&   -51.2 $\pm$     1.8&   -52.4 $\pm$     0.7&   -51.8 $\pm$     1.9&  3648&   M2III&   -43.5 $\pm$     1.1\\
    G1174?&     9.63&     7.11&    -4.9 $\pm$     1.8&    -5.7 $\pm$     0.7&    -5.3 $\pm$     1.9&  4112&   K3III&    -0.1 $\pm$     1.1\\
    G1124?&    12.01&    -0.09&   223.8 $\pm$     1.8&   221.3 $\pm$     0.7&   222.5 $\pm$     1.9&  4476&   K1III&   224.8 $\pm$     1.1\\
     G1183&    -0.88&    12.00&   -11.6 $\pm$     1.8&   -14.5 $\pm$     0.7&   -13.1 $\pm$     1.9&  3578&   M3III&    -4.6 $\pm$     1.1\\
6.33-10.28&     6.33&   -10.28&   115.8 $\pm$     1.8&     ...              &   115.8 $\pm$     1.8&  4363&   K1III&     ...              \\
    G1118?&     8.34&     8.74&    48.4 $\pm$     1.8&    51.5 $\pm$     1.0&    50.0 $\pm$     2.1&  4151&   K3III&    47.9 $\pm$     1.1\\
 8.25-8.93&     8.25&    -8.93&     ...              &  -175.4 $\pm$     1.0&  -175.4 $\pm$     1.0&  5857&$<$G8III&     ...              \\
     ID404&   -11.14&     4.98&   -85.4 $\pm$     1.8&     ...              &   -85.4 $\pm$     1.8&  3900&   K5III&   -77.1 $\pm$     1.1\\
     G1198&    -4.95&    11.17&  -141.4 $\pm$     1.8&  -142.4 $\pm$     1.0&  -141.9 $\pm$     2.1&  3709&   M1III&  -133.0 $\pm$     1.1\\
     ID491&    -8.08&     9.20&    38.0 $\pm$     1.8&     ...              &    38.0 $\pm$     1.8&  4388&   K1III&    38.2 $\pm$     1.1\\
     G1248&     6.59&   -10.34&    76.9 $\pm$     1.8&    82.9 $\pm$     1.0&    79.9 $\pm$     2.1&  4106&   K4III&    80.6 $\pm$     1.1\\
4.29+11.49&     4.29&    11.49&   243.6 $\pm$     1.8&   239.8 $\pm$     0.7&   241.7 $\pm$     1.9&  4579&   K0III&     ...              \\
     G1266&    -9.43&     7.90&  -121.0 $\pm$     1.8&     ...              &  -121.0 $\pm$     1.8&  4502&   K0III&  -113.0 $\pm$     1.1\\
     ID261&    12.45&    -1.67&   156.6 $\pm$     1.8&   153.2 $\pm$     1.0&   154.9 $\pm$     2.1&  4199&   K3III&   158.8 $\pm$     1.1\\
     IRS28&    11.41&    -5.63&   -54.4 $\pm$     1.8&   -62.7 $\pm$     0.7&   -58.6 $\pm$     1.9&  3387&   M6III&   -47.6 $\pm$     1.1\\
12.78-0.38&    12.78&    -0.38&     ...              &    54.5 $\pm$     0.7&    54.5 $\pm$     0.7&  3764&   M1III&     ...              \\
    G1187?&     7.05&    10.71&    62.5 $\pm$     1.3&    62.7 $\pm$     0.7&    62.6 $\pm$     1.5&  3568&   M3III&    71.0 $\pm$     0.8\\
    G1202?&    12.16&     4.38&    73.1 $\pm$     1.8&    72.7 $\pm$     0.7&    72.9 $\pm$     1.9&  3414&   M6III&    79.9 $\pm$     1.1\\
12.76+2.06&    12.76&     2.06&     ...              &   175.6 $\pm$     1.0&   175.6 $\pm$     1.0&  3261&$>$M7III&     ...              \\
 9.33+9.04&     9.33&     9.04&   -34.6 $\pm$     1.8&     ...              &   -34.6 $\pm$     1.8&  4566&   K0III&     ...              \\
     G1306&    -6.71&    11.28&     5.1 $\pm$     1.8&     ...              &     5.1 $\pm$     1.8&  4037&   K4III&    11.2 $\pm$     1.1\\
-8.83-10.2&    -8.83&   -10.20&    80.1 $\pm$     1.8&     ...              &    80.1 $\pm$     1.8&  4280&   K2III&     ...              \\
     G1311&     5.71&    12.23&    81.1 $\pm$     1.8&    78.9 $\pm$     0.7&    80.0 $\pm$     1.9&  3682&   M2III&    84.5 $\pm$     0.8\\
    G1390?&    12.25&    -6.32&  -225.7 $\pm$     1.8&  -228.4 $\pm$     0.7&  -227.1 $\pm$     1.9&  3798&   M0III&  -218.0 $\pm$     1.1\\
13.28-3.86&    13.28&    -3.86&     ...              &   -63.8 $\pm$     1.0&   -63.8 $\pm$     1.0&  4539&   K0III&     ...              \\
    G1438?&   -10.48&    -9.18&    45.8 $\pm$     1.8&     ...              &    45.8 $\pm$     1.8&  3522&   M4III&    53.9 $\pm$     1.1\\
     ID506&    -9.74&    10.02&   -34.0 $\pm$     1.8&     ...              &   -34.0 $\pm$     1.8&  4326&   K2III&   -27.7 $\pm$     1.1\\
    G1271?&    12.35&     6.59&   -56.5 $\pm$     1.8&   -51.9 $\pm$     0.7&   -54.2 $\pm$     1.9&  4083&   K4III&   -53.3 $\pm$     1.1\\
     G1458&   -10.87&     9.04&    28.3 $\pm$     1.8&     ...              &    28.3 $\pm$     1.8&  3504&   M4III&    36.8 $\pm$     1.1\\
13.16-5.31&    13.16&    -5.31&     ...              &   -64.7 $\pm$     1.0&   -64.7 $\pm$     1.0&  2948&$>$M7III&     ...              \\
    G1334?&    12.28&     7.66&  -151.8 $\pm$     1.8&  -143.8 $\pm$     0.7&  -147.8 $\pm$     1.9&  4323&   K2III&  -144.0 $\pm$     1.1\\
13.01+6.47&    13.01&     6.47&     ...              &    60.2 $\pm$     1.0&    60.2 $\pm$     1.0&  3485&   M4III&     ...              \\
     ID109&    12.39&    -8.80&   -25.5 $\pm$     1.8&   -24.8 $\pm$     0.7&   -25.1 $\pm$     1.9&  3531&   M4III&   -17.9 $\pm$     1.1\\
   IRS11SW&   -10.21&    11.65&   -97.2 $\pm$     1.8&     ...              &   -97.2 $\pm$     1.8&  3575&   M3III&   -91.6 $\pm$     1.1\\
-11.45-10.&   -11.45&   -10.91&   -12.5 $\pm$     1.8&     ...              &   -12.5 $\pm$     1.8&  4335&   K2III&     ...              \\
     ID575&     8.22&    13.81&     ...              &   135.0 $\pm$     1.0&   135.0 $\pm$     1.0&  4819&   G8III&   143.3 $\pm$     1.1\\
9.96+13.12&     9.96&    13.12&     ...              &    35.4 $\pm$     1.0&    35.4 $\pm$     1.0&  4547&   K0III&     ...              \\
     ID566&    11.18&    13.51&     ...              &    36.9 $\pm$     1.0&    36.9 $\pm$     1.0&  4380&   K1III&    44.1 $\pm$     1.1\\
     ID580&    11.79&    13.85&     ...              &   180.7 $\pm$     1.0&   180.7 $\pm$     1.0&  4889&   G8III&   182.8 $\pm$     1.1\\
12.23+13.5&    12.23&    13.57&     ...              &    32.4 $\pm$     1.0&    32.4 $\pm$     1.0&  4691&   G9III&     ...              \\
12.63+14.1&    12.63&    14.10&     ...              &    49.2 $\pm$     1.0&    49.2 $\pm$     1.0&  4449&   K1III&     ...              \\
\enddata
\tablecomments{Radial velocity of individual cool stars derived from the Dopplor shift of the CO~(2-1) band head feature.
Designations of the stars are taken from \cite{figGKMBM03} and \cite{tanFNKGM06}. For those objects which were not included in the previous works,
we use their coordinates as the designation. Velocity measurements in two epochs 1999 and 2005 are listed when available.
The temperatures and spectral types are calculated with measured average equivalent widths of the CO feature using the method describe in \cite{tanFNKGM06}.
Values in the last two columns are taken from \cite{figGKMBM03}, which was based on the same two data sets obtained in 1999.
}
\end{deluxetable}

\begin{deluxetable}{cccccccccc}
\tabletypesize{\scriptsize}
\tablewidth{0pt}
\tablecaption{Cool Star Velocity Statistics \label{coolstarvelodis}}
\tablehead{
\colhead{GC}& 
\colhead{1}&
\colhead{2}&
\colhead{3}&
\colhead{4}&
\colhead{1 $\cup$ 2}&
\colhead{1 $\cap$ 2}&
\colhead{3 $\cup$ 4}&
\colhead{3 $\cap$ 4}&
\colhead{1 $\cup$ 2  $\cup$ 3 $\cup$ 4}
}
\startdata
                               Number of stars  &     98  &     30  &     68  &     68  &    102  &     26  &     86  &     50  &    123 \\
                   V$_{average}$ (km~s$^{-1}$)  &   -9.6  &  -23.2  &  -18.0  &   -1.6  &   -9.2  &  -27.0  &   -6.6  &  -15.3  &   -4.9 \\
                    V$_{median}$ (km~s$^{-1}$)  &   -5.7  &   11.6  &  -20.1  &   -7.4  &   -5.7  &   11.0  &  -10.7  &  -14.5  &   -7.8 \\
                    V$_{stddev}$ (km~s$^{-1}$)  &  111.6  &  107.8  &  110.8  &  130.3  &  111.3  &  108.4  &  122.3  &  119.1  &  113.7 \\
                  V$_{skewness}$ (km~s$^{-1}$)  &   -0.2  &   -1.3  &   -0.2  &   -0.1  &   -0.2  &   -1.3  &    0.0  &   -0.4  &    0.0 \\
                  V$_{kurtosis}$ (km~s$^{-1}$)  &    0.5  &    1.4  &    0.7  &    0.6  &    0.5  &    1.5  &    0.7  &    0.6  &    0.8 \\
             (V2-V1)$_{average}$ (km~s$^{-1}$)  &         &         &         &         &         &    3.1  &         &    3.6 \\
              (V2-V1)$_{stddev}$ (km~s$^{-1}$)  &         &         &         &         &         &    2.5  &         &    1.5 \\
           Gaussian V$_{Center}$ (km~s$^{-1}$)  &    1.1  &   31.7  &   -2.2  &   11.7  &    4.7  &   27.2  &    3.7  &    4.5  &    4.2 \\
           Gaussian ${\sigma}_V$ (km~s$^{-1}$)  &   81.9  &   69.2  &   78.1  &   86.1  &   86.1  &   62.8  &   88.2  &   74.1  &   89.2 \\
\enddata
\tablecomments{Statistical results based on the velocity distribution of cool star samples.
}
\end{deluxetable}

\begin{deluxetable}{cccrrrrrrcl}
\tabletypesize{\scriptsize}
\rotate
\tablewidth{0pt}
\tablecaption{Radial Velocities of Narrow He~I Line Stars \label{hotstarvelo2}}
\tablehead{
\colhead{Name}      &
\colhead{RA}          & \colhead{Dec}  & \colhead{V1}&
\colhead{V2}&
\colhead{V3}  &
\colhead{V4} & \colhead{V$_{avg}$}  &\colhead{$\Delta$V}& \colhead{EW} &\colhead{Ref.}\\
&
\colhead{arcsec}          & \colhead{arcsec}  & \colhead{km~s$^{-1}$} &
\colhead{km~s$^{-1}$}&
\colhead{km~s$^{-1}$}  &\colhead{km~s$^{-1}$}  &
\colhead{km~s$^{-1}$} & \colhead{km~s$^{-1}$} & \colhead{\AA} &
}
\startdata
                       IRS16NW&        0.22&        1.04&       -55.6$\pm$        13.9&       -55.4$\pm$         7.9&       -63.9$\pm$        11.7&       -64.1$\pm$        16.2&       -59.7$\pm$         6.4&       -11.9$\pm$         3.0&        1.87&          (1),(2),(3),(6) \\
                        IRS16C&        1.32&        0.45&       113.5$\pm$        13.4&        89.4$\pm$         2.6&       104.3$\pm$        15.7&       102.1$\pm$        16.6&       102.3$\pm$         6.7&        -8.9$\pm$         3.0&        1.53&          (1),(2),(3),(6) \\
                    IRS16SW(W)&        1.43&       -1.22&       465.5$\pm$        16.9&       433.6$\pm$         2.4&       424.8$\pm$         2.8&       380.4$\pm$         3.3&       426.1$\pm$         4.4&       -53.3$\pm$        13.5&        1.47&      (1),(2),(3),(4),(6) \\
                       IRS16CC&        2.26&        0.52&       248.3$\pm$         6.1&       241.1$\pm$         3.0&       241.5$\pm$         4.2&       255.1$\pm$         4.0&       246.5$\pm$         2.2&         1.6$\pm$         3.0&        1.68&          (1),(3),(4),(6) \\
                        IRS33N&        0.30&       -2.49&         ...                 &      -135.1$\pm$        14.5&        39.5$\pm$         3.2&        37.8$\pm$         0.3&       -19.2$\pm$         5.0&        -8.6$\pm$         6.0&        2.43&              (3),(5),(6) \\
                   MPE+1.0-7.4&        1.66&       -2.03&         ...                 &        82.9$\pm$         9.3&         ...                 &         ...                 &        82.9$\pm$         9.3&         ...                 &        2.75&                      (6) \\
                   MPE-2.0-8.5&       -1.30&       -3.11&      -131.9$\pm$         3.6&         ...                 &         ...                 &      -125.7$\pm$         2.5&      -128.8$\pm$         2.2&        37.8$\pm$         3.0&        2.05&                          \\
                       IRS16NE&        3.24&        1.08&        12.8$\pm$        18.6&        -2.4$\pm$        12.2&       -35.6$\pm$        23.7&       -63.4$\pm$         2.7&       -22.2$\pm$         8.2&       -45.6$\pm$         3.0&        1.11&          (1),(2),(3),(6) \\
                        IRS33E&        1.12&       -3.41&       148.0$\pm$        15.5&       129.0$\pm$        14.1&       125.2$\pm$         2.6&       111.1$\pm$         2.6&       128.3$\pm$         5.3&       -16.1$\pm$        10.5&        1.46&          (1),(2),(3),(6) \\
                        IRS34W&       -4.11&        1.23&      -316.2$\pm$         1.6&         ...                 &      -289.9$\pm$         3.2&         ...                 &      -303.1$\pm$         1.8&        24.7$\pm$         3.0&        1.11&              (1),(2),(6) \\
                        IRS7SE&        2.41&        4.58&      -128.1$\pm$         4.2&         ...                 &      -122.5$\pm$         4.6&      -116.2$\pm$         2.6&      -122.3$\pm$         2.2&         8.2$\pm$         6.0&        1.87&                  (1),(6) \\
                           A21&        5.28&       -3.51&      -185.7$\pm$         2.7&      -232.4$\pm$        22.7&      -174.5$\pm$         3.9&      -179.5$\pm$         1.9&      -193.0$\pm$         5.8&         7.6$\pm$         4.5&        1.77&                          \\
        ID308\tablenotemark{a}&        6.67&        0.55&        -3.9$\pm$         5.8&       -24.8$\pm$         2.7&       -44.0$\pm$         5.5&       -19.1$\pm$         5.6&       -22.9$\pm$         2.5&        -7.6$\pm$         3.0&        1.52&                      (6) \\
                 MPE+1.41-12.2&        2.46&       -6.69&       121.4$\pm$        16.2&        97.6$\pm$        21.3&        96.3$\pm$         6.0&        97.1$\pm$        11.8&       103.1$\pm$         7.5&        -6.8$\pm$         3.0&        2.66&                      (6) \\
                         IRS26&       -1.37&        7.80&      -109.2$\pm$        10.6&      -124.0$\pm$         2.2&      -118.0$\pm$        20.3&      -111.2$\pm$        14.8&      -115.6$\pm$         6.8&       -16.9$\pm$         3.0&        1.75&                      (6) \\
                         IRS6W&       -7.91&        0.82&        62.9$\pm$         3.0&         ...                 &         ...                 &         ...                 &        62.9$\pm$         3.0&         ...                 &        1.52&                  (1),(3) \\
                           A12&       -0.20&        8.72&      -127.4$\pm$        16.3&      -151.2$\pm$         1.1&         ...                 &      -166.7$\pm$         3.3&      -148.4$\pm$         5.6&        -2.9$\pm$         3.0&        1.91&                  (3),(6) \\
                         AFNWB&       -8.08&       -3.92&       224.3$\pm$        17.0&         ...                 &         ...                 &         ...                 &       224.3$\pm$        17.0&         ...                 &        2.02&                      (3) \\
                         G1138&       -8.12&       -7.02&       154.4$\pm$         5.5&         ...                 &         ...                 &         ...                 &       154.4$\pm$         5.5&         ...                 &        4.12&                      (3) \\
                    -8.91-6.76&       -8.91&       -6.76&        85.3$\pm$         5.9&         ...                 &         ...                 &         ...                 &        85.3$\pm$         5.9&         ...                 &        1.90&                      (3) \\
\enddata
\tablecomments{Radial velocity measurements of narrow type He~I stars from four data sets.
The eqivarlent width for each object is calculated from the average width of the fitted Gaussian componets.
}
\tablenotetext{a}{Source name is taken from \cite{figGKMBM03}}
\tablenotetext{b}{Source name is taken from \cite{ottGES03}}
\tablerefs{
(1) \cite{genPEGO00};
(2) \cite{pauMMR01};
(3) \cite{tanFNKGM06};
(4) \cite{pauMS03};
(5) \cite{pauMM04};
(6) \cite{pauGMNB06}.
}
\end{deluxetable}

\begin{deluxetable}{cccrrrrrrcl}
\tabletypesize{\scriptsize}
\rotate
\tablewidth{0pt}
\tablecaption{Radial Velocities of Broad He~I Line Stars \label{hotstarvelo3}}
\tablehead{
\colhead{Name}      &
\colhead{RA}          & \colhead{Dec}  & \colhead{V1} &
\colhead{V2}&
\colhead{V3}  &
\colhead{V4} & \colhead{V$_{avg}$} & \colhead{$\Delta$V} & \colhead{EW} &\colhead{Ref.}\\
&
\colhead{arcsec}          & \colhead{arcsec}  & \colhead{km~s$^{-1}$} &
\colhead{km~s$^{-1}$}&
\colhead{km~s$^{-1}$}  &\colhead{km~s$^{-1}$}  &
\colhead{km~s$^{-1}$} &  \colhead{km~s$^{-1}$} &\colhead{\AA} &
}
\startdata
                        IRS13E&       -2.82&       -2.09&       143.4$\pm$        61.4&       168.9$\pm$        37.3&       164.8$\pm$        60.1&       144.6$\pm$        73.0&       155.4$\pm$        29.7&         8.2$\pm$         3.0&       23.33&                  (2),(6) \\
                         IRS7W&       -4.14&        4.67&      -193.9$\pm$        27.1&      -199.1$\pm$        24.1&      -163.7$\pm$        59.1&         ...                 &      -185.6$\pm$        23.1&         0.0$\pm$         3.0&      203.05&          (1),(2),(3),(6) \\
                         GCHe2&        3.64&       -5.86&       253.1$\pm$       100.8&       479.8$\pm$       144.6&       305.4$\pm$        51.7&       253.4$\pm$        40.2&       322.9$\pm$        47.0&       -54.1$\pm$        27.0&      265.75&              (2),(3),(6) \\
        ID415\tablenotemark{a}&        4.62&        5.42&       -23.2$\pm$         8.1&         ...                 &       -48.5$\pm$        21.8&       -53.9$\pm$        23.3&       -41.9$\pm$        11.0&       -10.0$\pm$         3.0&       71.56&          (1),(2),(4),(6) \\
                          AFNW&       -7.53&       -4.34&       187.9$\pm$        39.7&         ...                 &         ...                 &         ...                 &       187.9$\pm$        39.7&         ...                 &      228.30&          (1),(2),(3),(6) \\
        ID180\tablenotemark{b}&        9.74&        0.80&         ...                 &         ...                 &       -16.4$\pm$       120.2&         ...                 &       -16.4$\pm$       120.2&         ...                 &       77.04&                  (2),(6) \\
                            AF&       -6.14&       -7.81&       243.7$\pm$        17.4&         ...                 &         ...                 &         ...                 &       243.7$\pm$        17.4&         ...                 &      290.19&          (1),(2),(3),(6) \\
                       IRS15SW&       -1.77&       10.08&      -119.6$\pm$        24.2&         ...                 &       -62.1$\pm$       105.0&      -124.8$\pm$        32.4&      -102.2$\pm$        37.5&        10.3$\pm$         3.0&      270.48&          (1),(2),(3),(6) \\
                         IRS9S&        6.45&       -8.32&       347.2$\pm$        33.7&        54.9$\pm$       267.5&       325.9$\pm$        18.0&       324.5$\pm$         5.9&       263.1$\pm$        67.6&         0.0$\pm$         3.0&       41.31&              (1),(3),(6) \\
\enddata
\tablecomments{Radial velocity measurements of broad type He~I stars from four data sets.
The eqivarlent width for each object is calculated by integrating the area below the normalized
spectrum within the $\pm$1000~km~s$^{-1}$ wavelength range of the corresponding spectral line emission peak.
}
\tablenotetext{a}{Source name is taken from \cite{figGKMBM03}}
\tablenotetext{b}{Source name is taken from \cite{ottGES03}}
\tablerefs{
(1) \cite{genPEGO00};
(2) \cite{pauMMR01};
(3) \cite{tanFNKGM06};
(4) \cite{pauMS03};
(5) \cite{pauMM04}.
}
\end{deluxetable}

\begin{deluxetable}{ccccccr}
\tabletypesize{\scriptsize}
\tablewidth{0pt}
\tablecaption{Spectral Types of He~I Line Stars \label{hotspty}}
\tablehead{
\colhead{Name}      &
\colhead{RA}          & \colhead{Dec}  & \colhead{P2006$^{a}$} & \colhead{Alternative} & \colhead{He~I Line} & \colhead{Spectral$^{a}$}\\
&
\colhead{arcsec}          & \colhead{arcsec}  &\colhead{Designation} &\colhead{Name} & \colhead{Width} &\colhead{Type}
}
\startdata
        IRS16NW&   0.22&   1.04&       E19&               &    Narrow&       Ofpe/WN9\\
         IRS16C&   1.32&   0.45&       E20&               &    Narrow&       Ofpe/WN9\\
     IRS16SW(W)&   1.43&  -1.22&       E23&       IRS16SW,&    Narrow&       Ofpe/WN9\\
        IRS16CC&   2.26&   0.52&       E27&               &    Narrow&    O9.5-B0.5~I\\
         IRS33N&   0.30&  -2.49&       E33&            A11&    Narrow&       B0.5-1~I\\
    MPE+1.0-7.4&   1.66&  -2.03&       E34&         IRS16S&    Narrow&       B0.5-1 I\\
    MPE-2.0-8.5&  -1.30&  -3.11&       E43&               &    Narrow&     O8.5-9.5 I\\
        IRS16NE&   3.24&   1.08&       E39&               &    Narrow&       Ofpe/WN9\\
         IRS33E&   1.12&  -3.41&       E41& ID221, IRS33SE&    Narrow&       Ofpe/WN9\\
         IRS34W&  -4.11&   1.23&       E56&               &    Narrow&       Ofpe/WN9\\
         IRS7SE&   2.41&   4.58&       E62&               &    Narrow&         B0-3 I\\
            A21&   5.28&  -3.51&          &               &    Narrow&               \\
          ID308&   6.67&   0.55&       E67&          IRS1E&    Narrow&         B1-3~I\\
  MPE+1.41-12.2&   2.46&  -6.69&       E69&               &    Narrow&               \\
          IRS26&  -1.37&   7.80&       E73&               &    Narrow&         O9-B~I\\
          IRS6W&  -7.91&   0.82&          &               &    Narrow&               \\
            A12&  -0.20&   8.72&       E75&               &    Narrow&         O9-B~I\\
          AFNWB&  -8.08&  -3.92&          &               &    Narrow&               \\
          G1138&  -8.12&  -7.02&          &               &    Narrow&               \\
     -8.91-6.76&  -8.91&  -6.76&          &               &    Narrow&               \\
         IRS13E&  -2.82&  -2.09&    E51(?)&               &     Broad&         WN8(?)\\
          IRS7W&  -4.14&   4.67&    E68(?)&               &     Broad&         WC9(?)\\
          GCHe2&   3.64&  -5.86&       E65&          IRS9W&     Broad&            WN8\\
          ID415&   4.62&   5.42&       E70&         IRS7E2&     Broad&       Ofpe/WN9\\
           AFNW&  -7.53&  -4.34&       E74&          AHHNW&     Broad&            WN8\\
          ID180&   9.74&   0.80&       E78&         HeI~N1&     Broad&            WC9\\
             AF&  -6.14&  -7.81&       E79&            AHH&     Broad&       Ofpe/WN9\\
        IRS15SW&  -1.77&  10.08&       E83&               &     Broad&        WN8/WC9\\
          IRS9S&   6.45&  -8.32&       E80&         IRS9SE&     Broad&            WC9\\
\enddata
\tablenotetext{a}{Designations and spectral types are taken from  \cite{pauGMNB06}}
\end{deluxetable}




\end{document}